\documentclass{vgtc}                          

\newcommand{\hot}[1]{{\color{black} #1}}
\newcommand{\prevhot}[1]{{\color{black} #1}}

\ifpdf
  \pdfoutput=1\relax                   
  \usepackage{graphicx}                
  \DeclareGraphicsExtensions{.pdf,.png,.jpg,.jpeg} 
\else
  \usepackage{graphicx}                
  \DeclareGraphicsExtensions{.eps}     
\fi%

\graphicspath{{figures/}{pictures/}{images/}{./}} 

\usepackage{microtype}                 
\PassOptionsToPackage{warn}{textcomp}  
\usepackage{textcomp}                  
\usepackage{mathptmx}                  
\usepackage{times}                     
\usepackage{cite}                      
\usepackage{tabu}                      
\usepackage{booktabs}                  
\usepackage{amsmath}
\usepackage{amsfonts}
\usepackage[cal=cm]{mathalfa}

\DeclareMathOperator*{\sh}{SH}
\DeclareMathOperator*{\p}{P}
\DeclareMathOperator*{\rec}{REC}
\DeclareMathOperator*{\lone}{L1}
\DeclareMathOperator*{\tv}{TV}

\newenvironment{myitemize}{
\begin{itemize}
 \setlength{\itemsep}{1pt}
 \setlength{\parskip}{0pt}
 \setlength{\parsep}{0pt}}{\end{itemize}
}

\onlineid{9330}

\vgtccategory{IEEE PacificVis Conference Track}

\vgtcinsertpkg

\usepackage{caption}
\usepackage{subcaption}
\usepackage{comment}

\title{ViSNeRF: Efficient Multidimensional Neural Radiance Field Representation for Visualization Synthesis of Dynamic Volumetric Scenes}

\author{Siyuan Yao\thanks{e-mail: syao2@nd.edu} %
\and Yunfei Lu\thanks{e-mail: ylu25@nd.edu} %
\and Chaoli Wang\thanks{e-mail: chaoli.wang@nd.edu}}
\affiliation{\scriptsize University of Notre Dame}

\abstract{
Domain scientists often face I/O and storage challenges when keeping raw data from large-scale simulations. Saving visualization images, albeit practical, is limited to preselected viewpoints, transfer functions, and simulation parameters. Recent advances in scientific visualization leverage deep learning techniques for visualization synthesis by offering effective ways to infer unseen visualizations when only image samples are given during training. However, due to the lack of 3D geometry awareness, existing methods typically require many training images and significant learning time to generate novel visualizations faithfully. To address these limitations, we propose ViSNeRF, a novel 3D-aware approach for visualization synthesis using neural radiance fields. Leveraging a multidimensional radiance field representation, ViSNeRF efficiently reconstructs visualizations of dynamic volumetric scenes from a sparse set of labeled image samples with flexible parameter exploration over transfer functions, isovalues, timesteps, or simulation parameters. Through qualitative and quantitative comparative evaluation, we demonstrate ViSNeRF's superior performance over several representative baseline methods, positioning it as the state-of-the-art solution. \hot{The code is available at \url{https://github.com/JCBreath/ViSNeRF}.}
} 

\begin{document}



\vspace{-0.1in}
\firstsection{Introduction}

\maketitle

In scientific research, large-scale simulations are essential for modeling complex phenomena across diverse science and engineering fields. 
However, huge raw data generated from these simulations, especially those involving time sequences and ensemble runs, presents significant challenges for domain scientists.
Specifically, I/O operations and storage capacity limitations can lead to cumbersome and inefficient data retrieval and visualization processes, hindering timely analysis and slowing the overall scientific discovery.

A common practice to mitigate these challenges is preserving visualization images rendered from simulation data. 
While this approach offers a practical solution by reducing the data footprint, it has inherent limitations. 
Saving visualization images restricts scientists to preselected viewpoints, transfer functions (TFs), and simulation parameters. 
Such constraints limit the flexibility required for exploratory data analysis, where researchers often need to investigate multiple perspectives and vary visualization settings to uncover hidden patterns or anomalies.

Over the past few years, deep learning has become viable for addressing diverse generation tasks in scientific visualization~\cite{Wang-TVCG}. 
Most studies are dedicated to {\em data generation} for scalar fields~\cite{Han-VIS20, Yao-CG23, Han-VI22, Tang-CG24, Tang-PVIS24} and vector fields~\cite{Gu-PVIS22, Han-CGA19, Gu-CGA21, Guo-PVIS20, Han-CG22} across various tasks such as super-resolution generation, data translation, reconstruction, and completion. 
In contrast, only a few have focused on {\em visualization generation}~\cite{Berger-TFGAN-TVCG19, Hong-DNN-VolVis-PVIS19, He-InSituNet-TVCG20, Han-CoordNet-TVCG}. 
Examples such as InSituNet~\cite{He-InSituNet-TVCG20} and CoordNet~\cite{Han-CoordNet-TVCG} support post hoc analysis on scientific simulations without accessing raw simulation data.
However, these existing methods do not develop {\em 3D awareness} of {\em dynamic} volumetric scenes from 2D visualization images.
As a result, they constrain the viewpoints to interpolated positions between fixed camera angles, inhibiting the ability to freely rotate and scale the scene or utilize the full six degrees of freedom for camera movement.
This limitation is significant because freely navigating visualizations in 3D is essential for thoroughly exploring complex data, identifying intricate structures, and gaining deeper scientific insights.
Moreover, although these methods eliminate the need to handle simulation data, they still require over a hundred views of each scene frame and numerous intermediate timesteps and ensemble runs to achieve smooth interpolation.
Additionally, the training can take from hours to days, further complicating their usability. 
These dilemmas not only amplify the costs of preparing training data but also constrain the practicality of these techniques for real-world scientific applications.

To address the limitations of existing methods, we present ViSNeRF, a deep learning framework for \underline{\textbf{Vi}}sualization \underline{\textbf{S}}ynthesis using \underline{\textbf{Ne}}ural \underline{\textbf{R}}adiance \underline{\textbf{F}}ields.
ViSNeRF introduces an innovative factorization approach that allows a single NeRF model to represent dynamic scenes with multiple controllable parameters.
By incorporating this multidimensional NeRF representation, ViSNeRF enables us to explore dynamic volumetric scenes interactively.
This includes adjusting parameters such as TFs for {\em direct volume rendering} (DVR), isovalues for {\em isosurface rendering} (IR), timesteps, and simulation parameters, as well as examining the scene from any desired viewpoint with confidence.
Moreover, combining an explicit feature grid and a pair of {\em multilayer perceptron} (MLP) decoders for color and density, the hybrid NeRF representation allows ViSNeRF to accelerate training while improving generation quality. 
The contributions of ViSNeRF are as follows:
\begin{myitemize}
\vspace{-0.05in}
  \item ViSNeRF efficiently synthesizes high-resolution and high-quality visualizations from novel viewpoints via a hybrid radiance field representation.
  \item Leveraging a generalized factorization strategy, ViSNeRF accomplishes dynamic \hot{scene} generation with plausible interpolation over the parameter space, including TFs, isovalues, timesteps, or simulation parameters.
  \item Our comprehensive study validates the effectiveness of ViSNeRF and demonstrates its superior performance compared to state-of-the-art methods.
\vspace{-0.05in}
\end{myitemize}

\vspace{-0.1in}
\section{Related Work}

{\bf Deep learning for visualization generation.} 
Deep learning has been applied in scientific visualization to solve data generation, visualization generation, prediction, objection detection and segmentation, and feature learning and extraction tasks~\cite{Wang-TVCG}.
We restrict our attention to works that have succeeded in visualization generation.

Berger et al.\ \cite{Berger-TFGAN-TVCG19} designed GAN-VR, a {\em generative adversarial network} (GAN) framework to synthesize DVR and allow users to explore the space of viewpoints and TFs.
Hong et al.\ \cite{Hong-DNN-VolVis-PVIS19} introduced DNN-VolVis, which applies the rendering effects of the user-defined goal image to the original DVR image without knowing the explicit TFs. 
He et al.\ \cite{He-InSituNet-TVCG20} proposed InSituNet that generates visualization at simulation time and enables post hoc exploration of ensemble simulations.
Shi et al.\ \cite{Shi-VDL-TVCG22} leveraged a view-dependent surrogate model named VDL-Surrogate to infer volume data and generate visualizations with user-defined visual mappings for parameter-space exploration.
Han and Wang~\cite{Han-CoordNet-TVCG} built CoordNet, which leverages {\em implicit neural representation} (INR) to solve various scientific visualization tasks, including view synthesis.

To achieve super-resolution of IR images while maintaining consistent geometric properties, Weiss et al.\ \cite{Weiss-SRNet-TVCG21} proposed a fully {\em convolutional neural network} (CNN) trained with depth and normal maps.
Weiss et al.\ \cite{Weiss-LearningAdaptive-TVCG22} presented a neural rendering approach that employs low-resolution rendering images to predict the density distribution of the volume data and adaptively samples it to produce high-resolution images.
Weiss and Navab~\cite{Weiss-DeepDVR-ArXiv21} developed DeepDVR to model the DVR process by learning the implicit semantics for feature extraction, classification, and visual mapping. 
To save rendering time, Bauer et al.\ \cite{Bauer-FoVolNet-TVCG23} designed the FoVolNet, which reconstructs full-frame renderings from foveated renderings.

{\em Scene representation networks} (SRNs) represent volumetric data for neural volume rendering without requiring direct access to the original volume data. 
Weiss et al.\ \cite{Weiss-CGF22} introduced a technique using a volumetric grid of latent features that allows for effective rendering by encoding essential volumetric information.
Further advancing this direction, Wurster et al.\ \cite{Wurster-TVCG24} developed an adaptive multi-grid SRN that enhances rendering efficiency by representing large-scale data across various resolutions. 
Wu et al.\ \cite{Wu-TVCG24} proposed a compressive neural representation that employs multi-resolution hash encoding, achieving rapid training speeds and optimized memory usage for large-scale volumetric data.
Recently, Tang and Wang~\cite{Tang-VIS24} introduced StyleRF-VolVis, a method leveraging NeRF to perform geometry-aware stylization with only 2D images.

ViSNeRF is designed for visualization synthesis of dynamic volumetric scenes and is benchmarked exclusively against methods that can natively accomplish this task. GAN-VR~\cite{Berger-TFGAN-TVCG19} requires 200,000 rendering images for training dynamic scenes. DNN-VolVis~\cite{Hong-DNN-VolVis-PVIS19} and InSituNet~\cite{He-InSituNet-TVCG20} only work with low-resolution (256$\times$256) images. VDL-Surrogate~\cite{Shi-VDL-TVCG22} needs access to the raw data during training and incurs a long training time (50 hours). While CoordNet~\cite{Han-CoordNet-TVCG} supports high-resolution (1024$\times$1024) image synthesis, it requires up to five days to train with 200 images. In contrast, our ViSNeRF requires only 42 rendering images for training a static scene and several hundred to a few thousand images for training a dynamic scene. It works with high-resolution (1024$\times$1024) images, achieving fast training (up to 35 minutes for a static scene and up to 2.5 hours for a dynamic scene) and superior quality for synthesized images.

\begin{figure*}[htb]
  \begin{center}
    \includegraphics[width=1.0\linewidth]{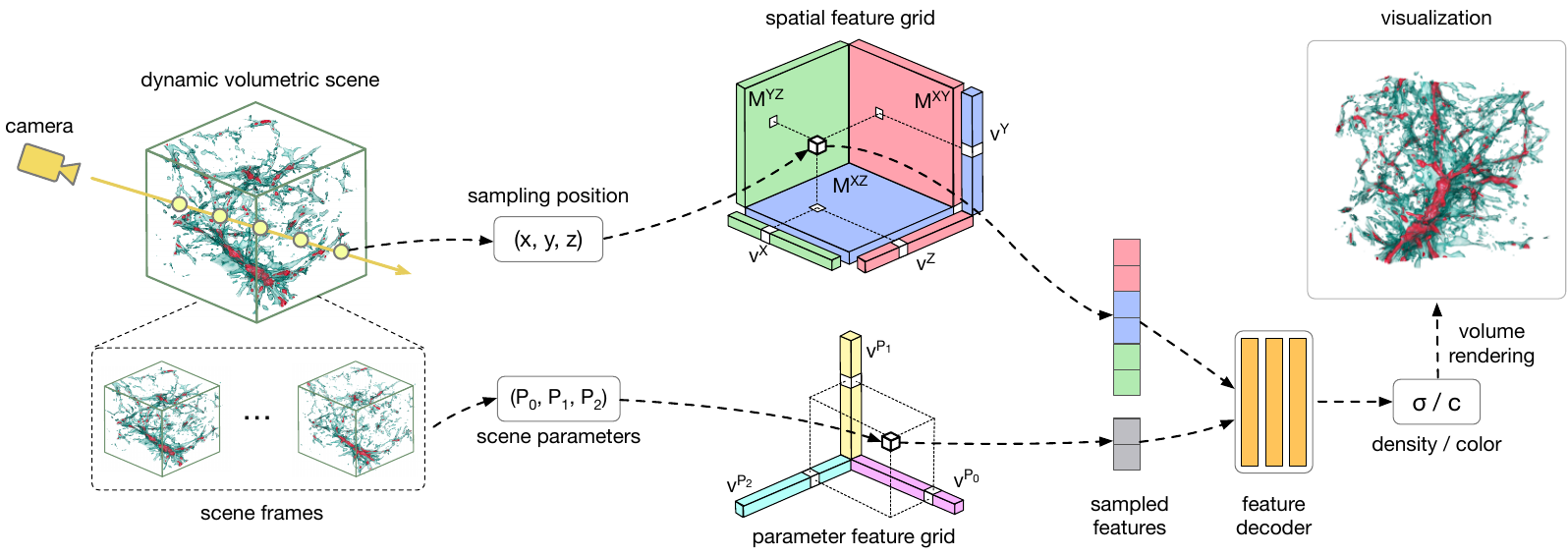}\\
 \end{center}
\vspace{-.25in} 
 \caption{Overview of ViSNeRF using the Nyx dataset as an example of a dynamic volumetric scene. Features are sampled from spatial and parameter feature grids based on the camera ray's sampling position and scene parameters. These features are processed by the decoder to generate density and color values, which are then used in volume rendering to visualize the scene frame from the chosen camera view.}
 \label{fig:overview}
\end{figure*}

{\bf 3D-aware image synthesis.}
Recent works of novel-view image synthesis have moved on to incorporate camera information to enhance the 3D consistency of generated views. 
Early approaches, such as PrGAN~\cite{PrGAN}, VON~\cite{VON}, PlatonicGAN~\cite{PlatonicGAN}, HoloGAN~\cite{HoloGAN}, and BlockGAN~\cite{BlockGAN}, use voxels to represent the scene and generate images based on the voxel shape. 
However, due to the limited voxel resolution, these methods fail to reconstruct fine details of the original scenes. 
Liao et al.\ \cite{ControllableGAN} suggested using 3D primitives to represent the scene for 3D-aware view synthesis. 
While this scheme allows for 3D control, it may be inadequate when reconstructing complex scenes, as primitives provide only limited information.
Consequently, the resulting image synthesis may be suboptimal or of low quality.
Unlike the above approaches, neural field representations are more popular and effective for 3D-aware image synthesis. 
NeRF~\cite{Mildenhall-NeRF-ECCV20} is the seminal work demonstrating the great potential of neural scene representations. 
Numerous variants of NeRF have successfully produced remarkable synthesis results~\cite{Schwarz-GRAF-NeurIPS20,Yu-pixelNeRF-CVPR21,Chan-PiGAN-CVPR21,Jain-DietNeRF-ICCV21,Mildenhall-RawNeRF-CVPR22,Gu-StyleNeRF-ICLR22}, improved the capability of NeRF in many scenarios, and extended its applications from image synthesis to 3D reconstruction~\cite{Oechsle-UNISURF-ICCV21,Wang-NeuS-NeurIPS21,Yariv-VolSDF-NeurIPS21}, 3D content generation~\cite{Niemeyer-GIRAFFE-CVPR21, Jain-DreamField-CVPR22, Wang-CLIP-NeRF-CVPR22, Chan-EG3D-CVPR22}, and dynamic scene representations~\cite{Park-Nerfies-ICCV21, Pumarola-D-NeRF-CVPR21, Singer-Text-To-4D-arXiv23, Cao-HexPlane-CVPR23}.

{\bf Efficient NeRFs.}
Although NeRF can generate realistic results with a compact MLP, slow convergence and long training and inference are common issues among most pure implicit methods.
Later efforts of efficiency improvement~\cite{Liu-NSVF-NeurIPS20, Yu-PlenOctrees-ICCV21, Fridovich-Keil-Plenoxels-CVPR22} focus on space-time tradeoff, which sacrifice memory space to accelerate the rendering process of radiance field methods. 
By factorizing the complex voxel-based feature grid of radiance fields, emerging decomposed hybrid NeRF architectures express exceptionally high efficiency in both computation and memory usage. 
{\em Generative scene network} (GSN)~\cite{Devries-GSN-ICCV21} is the first plane-based work that uses 2D representations of the radiance fields.
Efficient geometry-aware 3D GAN (EG3D)~\cite{Chan-EG3D-CVPR22} enables style-mixing and latent-space interpolation by leveraging StyleGAN2~\cite{Karras-StyleGAN2-CVPR20} to generate features of the triplane representation.
Instant-NGP~\cite{Thomas-InstantNGP} integrates a multiresolution hash table of trainable feature vectors with a compact network, illustrating a hybrid representation approach known for its efficiency in training and inference.
{\em 3D Gaussian splatting} (3DGS)~\cite{Kerbl-TOG23}, which represents 3D scenes with 3D Gaussians, offers an efficient explicit approach that eliminates the need for neural networks, significantly speeding up training and inference.
iVR-GS~\cite{Tang-PVIS25} designs editable 3DGS to achieve inverse volume rendering for explorable visualization of color, opacity, and lighting parameters. 
{\em Tensorial radiance field} (TensoRF)~\cite{Chen-TensoRF-ECCV22} introduces a tensor-based architecture to obtain photorealistic quality with high computational and memory efficiency.

{\bf Factorized NeRFs.}
Recent advancements like K-Planes~\cite{Fridovich-Keil-K-Planes-CVPR23} and HexPlane~\cite{Cao-HexPlane-CVPR23}, inspired by TensoRF, extend to factorized 4D NeRF representations to reconstruct dynamic scenes.
We also acknowledge works, such as Tensor4D~\cite{Shao-Tensor4D-CVPR23}, which factorize signed distance fields to represent 3D geometry. 
Yet, those works do not perform factorization on radiance fields, so they do not fit our scope.
ViSNeRF adopts a generalized factorization strategy for multidimensional NeRF representations to address the demands of visualization synthesis of dynamic volumetric scenes.

\section{ViSNeRF}
\label{section:visnerf-method}

ViSNeRF leverages a NeRF model to learn a volumetric representation of a scene from a set of DVR or IR images, along with camera pose information, which can be provided directly by users or estimated using tools like COLMAP~\cite{Schonberger-COLMAP-ECCV16}.
This NeRF representation allows us to produce 3D-consistent visualizations from any viewpoint.
ViSNeRF uses a hybrid NeRF model combining {\em explicit feature grids} and {\em implicit feature decoders} to improve speed and quality.
To handle the dynamic scenes efficiently, ViSNeRF incorporates a {\em generalized factorization strategy} on the NeRF to achieve a multidimensional representation with minimal increase in model size and GPU memory cost.
This strategy allows us to efficiently represent dynamic scenes with an arbitrary number of adjustable parameters, such as TFs, isovalues, timesteps, and simulation parameters.

As shown in Figure~\ref{fig:overview}, to synthesize a visualization of a dynamic volumetric scene, we first cast rays from screen pixels toward the scene, originating from the camera locations. 
For every point sampled along a ray, we provide ViSNeRF with the sampling point location $(x,y,z)$ and the parameters $(p_0,p_1,...)$ defining a scene frame.
ViSNeRF then retrieves features from spatial matrices and vectors as well as parameter vectors, decoding them into color and density values.
It finally composes a visualization of the scene frame from the camera view by collecting sampled points on the rays.

\vspace{-0.05in}
\subsection{Volume Rendering with Radiance Fields}
\label{subsection:overview-vol-rendering}

Following the classical volume rendering~\cite{Kajiya-RayTracingVolumeDensities-SIGGRAPH84} as NeRF~\cite{Mildenhall-NeRF-ECCV20} suggested, we formulate the rendering function as 
\vspace{-0.05in}
\begin{equation}
    \begin{array}{l}
      \displaystyle \mathcal{C}(\mathbf{r}) = \int_{t_n}^{t_f}  \mathcal{T}(t)\sigma(\mathbf{r}(t))c(\mathbf{r}(t),\mathbf{d}) dt, \\
      \displaystyle \text{where} \;  \mathcal{T}(t)=\exp\biggl(-\int_{t_n}^{t} \sigma(\mathbf{r}(s)) ds\biggr).
      \label{eqn:nerf-color-integral}
\vspace{-0.05in}
    \end{array}
\end{equation}
In Equation~\ref{eqn:nerf-color-integral}, $\mathcal{C}(\mathbf{r})$ denotes the color sampled through the ray $\mathbf{r}(t)=\mathbf{o}+t\mathbf{d}$ from the camera, where $\mathbf{o}$ is the origin, $\mathbf{d}$ is the viewing direction, and $t_n$ and $t_f$ are near and far bounds.
The accumulated transmittance function $\mathcal{T}(t)$ determines the probability that no particle is present along the ray between $t_n$ and $t$.
The volume density function $\sigma(\mathbf{x})$ is the probability that the ray is stopped at location $\mathbf{x}$ by a particle.
The view-dependent color term $c(\mathbf{x}, \mathbf{d})$ belongs to the particle that terminates the ray at location $\mathbf{x}$ with $\mathbf{d}$.

We use stratified sampling along the ray to estimate the integral.
Following the quadrature rule~\cite{Max-Optical-TVCG95}, the resulting color of the ray, $\mathcal{C}(\mathbf{r})$, is estimated using $n$ samples along the ray by 
\vspace{-0.05in}
\begin{equation}
  \begin{array}{l}
    \displaystyle \mathcal{\hat{C}}(\mathbf{r}) = \sum_{i=1}^{n}  \mathcal{T}_i(1-\exp(-\sigma_i\delta_i))c_i, \\
    \displaystyle \text{where} \;  \mathcal{T}_i=\exp\biggl(-\sum_{j=1}^{i-1} \sigma_j\delta_j\biggr).
    \label{eqn:nerf-color}
\vspace{-0.05in}    
  \end{array}
\end{equation}
In Equation~\ref{eqn:nerf-color}, the ray marching step size $\delta_i$ is defined by the distance between the locations of contiguous samples, i.e., $\delta_i=t_{i+1}-t_i$.

\vspace{-0.05in}
\subsection{Factorization of Static Radiance Fields}
\label{subsection:overview-factorization}

In a dynamic scene, individual frames can be treated as static scenes represented by 3D radiance fields.
Like TensoRF~\cite{Chen-TensoRF-ECCV22}, we use the block-term tensor decomposition~\cite{Ye-BlockTerm-CVPR18} to factorize the 3D volume $\mathbf{V}\in\mathbb{R}^{X\times Y\times Z}$ of the static radiance fields as the sum of vector-matrix outer products 
\vspace{-0.05in}
\begin{equation}
  \begin{array}{l}
    \displaystyle \mathbf{V} = \sum_{r=1}^{R_{1}} \mathbf{M}_r^{XY} \circ \mathbf{v}_r^Z + \sum_{r=1}^{R_{2}} \mathbf{M}_r^{XZ} \circ \mathbf{v}_r^Y  + \sum_{r=1}^{R_{3}} \mathbf{M}_r^{YZ} \circ \mathbf{v}_r^X,
    \label{eqn:factorization}
\vspace{-0.05in}
  \end{array} 
\end{equation}
where $\mathbf{M}_r^{XY}$, $\mathbf{M}_r^{XZ}$, and $\mathbf{M}_r^{YZ}$ are the spatial matrices (refer to Figure~\ref{fig:overview}) representing the $XY$, $XZ$, and $YZ$ planes. 
$\mathbf{v}_r^X$, $\mathbf{v}_r^Y$, and $\mathbf{v}_r^Z$ are the spatial vectors along the $X$, $Y$, and $Z$ axes. 
\prevhot{$\mathbf{M}_r^{XY}\circ\mathbf{v}_r^Z$ denotes the outer product of $\mathbf{M}_r^{XY}$ and $\mathbf{v}_r^Z$.} 
$R_1$, $R_2$, and $R_3$ denote the numbers of low-rank components, i.e., $\mathbf{M}_r^{XY} \circ \mathbf{v}_r^Z$, $\mathbf{M}_r^{XZ} \circ \mathbf{v}_r^Y$, and $\mathbf{M}_r^{YZ} \circ \mathbf{v}_r^X$, and the numbers can be updated according to the complexity and dimensions of the effective volume.
By employing factorization, we can reduce the memory complexity for representing the radiance field from $O(N^3)$ to $O(N^2)$ when $N\gg R_1+R_2+R_3$, where $N$ is the spatial resolution.
\prevhot{For simplicity, we assume the same complexity across all three dimensions. Thus, we use the same number of components for each dimension in 3D static radiance fields (i.e., $R=R_1=R_2=R_3$). The factorization is then expressed as
\vspace{-0.05in}
\begin{equation}
  \begin{array}{l}
    \displaystyle \mathbf{V} = \sum_{r=1}^{R} \mathbf{M}_r^{XY} \circ \mathbf{v}_r^Z + \mathbf{M}_r^{XZ} \circ \mathbf{v}_r^Y  + \mathbf{M}_r^{YZ} \circ \mathbf{v}_r^X.
    \label{eqn:factorization-r1}
\vspace{-0.1in}
  \end{array}
\end{equation}}

\vspace{-0.1in}
\subsection{Factorization of Parameter Space}

Allowing NeRF to represent dynamic volumetric scenes is necessary to synthesize visualizations for time-varying or ensemble simulation data with different TFs or isovalues. 
Training multiple NeRF models for visualizing individual volumes is impractical, as isolating these models makes effective interpolation between them impossible.

\prevhot{Existing works such as HexPlane~\cite{Cao-HexPlane-CVPR23} and K-Plane~\cite{Fridovich-Keil-K-Planes-CVPR23} have made strides in factorizing dynamic 4D scenes and achieved commendable results in dynamic scene reconstruction, they still fall short in our (3+$K$)-D scenarios, where $K$ represents the number of changeable parameters defining different scenes. 
A key limitation of these methods is their lack of efficient scalability to multidimensional spaces, primarily due to excessive reliance on planes.
For instance, both HexPlane and K-Plane necessitate the addition of three extra planes when transitioning from 3D to 4D.
Moreover, while HexPlane is limited to 4D, K-Plane suggests adding another four planes to move from 4D to 5D. 
This requirement increases the complexity exponentially, making these methods impractical for direct application in our more complex, higher-dimensional scenarios.}

\begin{figure}[htb]
  \begin{center}
  $\begin{array}{c@{\hspace{0.05in}}c@{\hspace{0.05in}}c@{\hspace{0.05in}}c}
  \includegraphics[height=0.4525in]{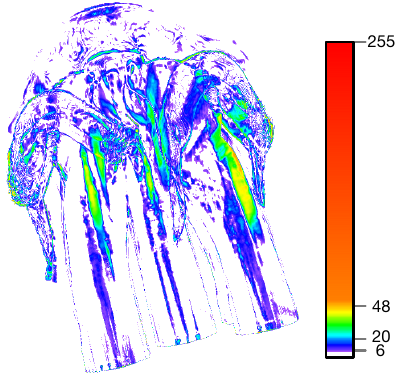}&
  \includegraphics[height=0.4525in]{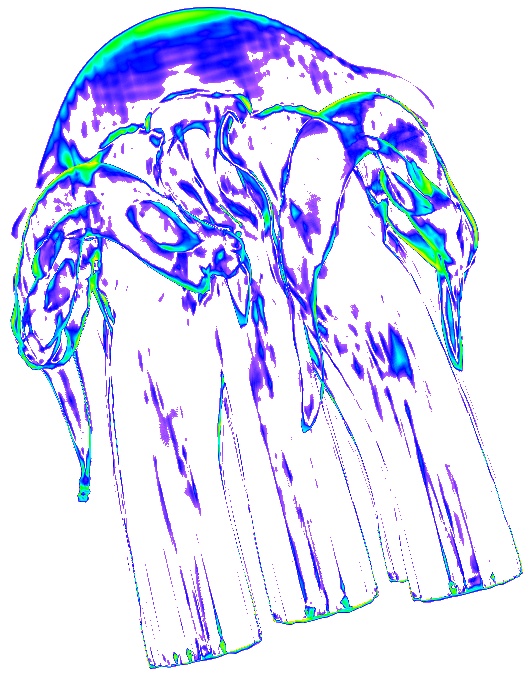}&
  \includegraphics[height=0.4525in]{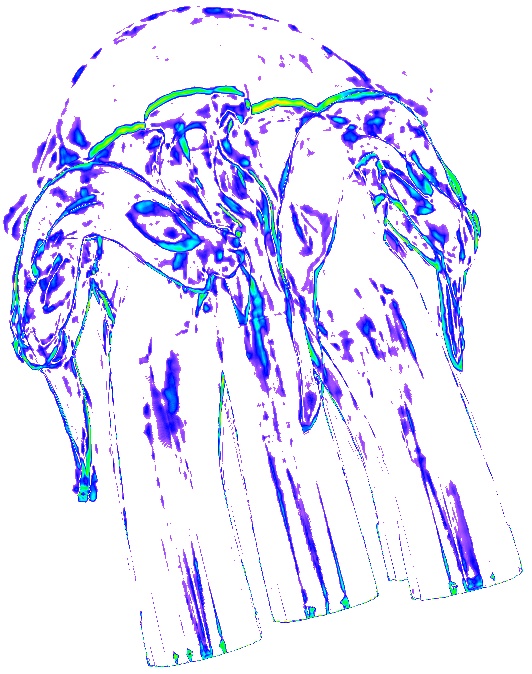}&
 \\
  \includegraphics[height=0.95in]{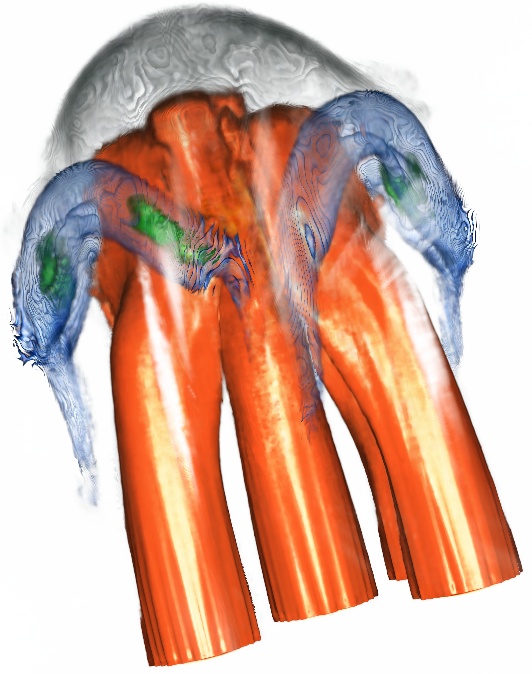}&
  \includegraphics[height=0.95in]{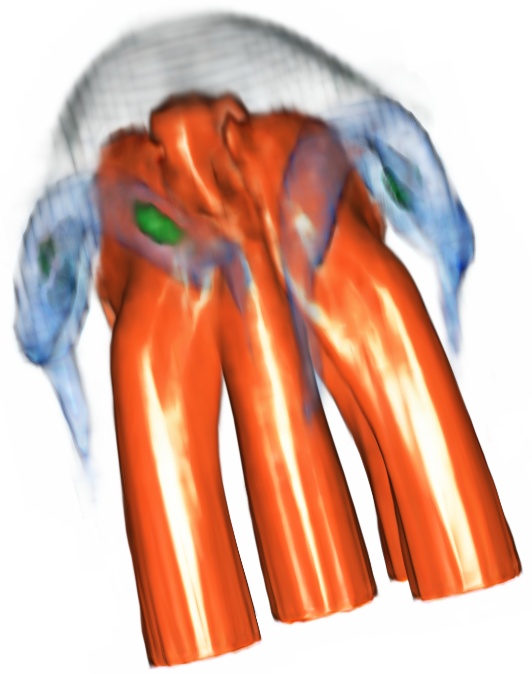}&
  \includegraphics[height=0.95in]{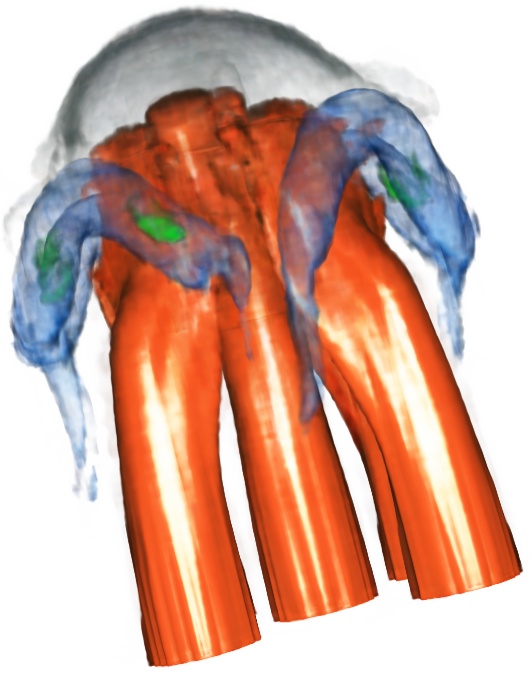}&
  \includegraphics[height=0.95in]{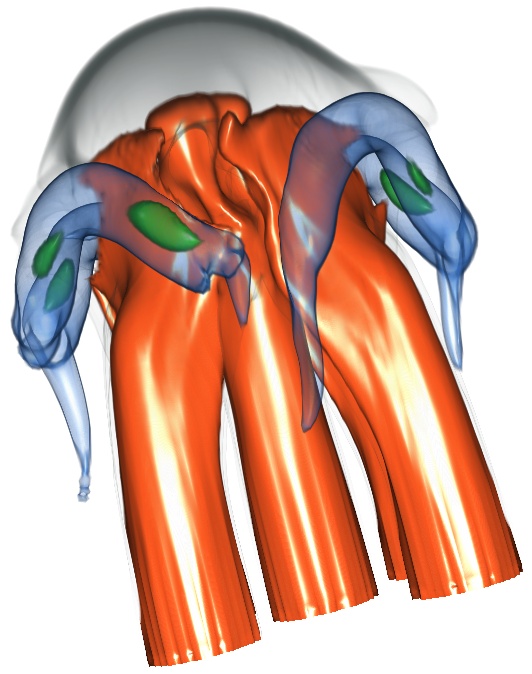}\\
 \mbox{\footnotesize \hot{(a) K-Planes}} & \mbox{\footnotesize \hot{(b) HexPlane}} & \mbox{\footnotesize \hot{(c) ViSNeRF}} & \mbox{\footnotesize \hot{(d) GT}}
 \end{array}$
 \end{center}
\vspace{-.25in} 
 \caption{\prevhot{Inferred five jets (timestep) DVR images generated by K-Planes, HexPlane, and ViSNeRF using 462 views to train the dynamic scene with a full 360-degree view.}} 
 \label{fig:comp-decomp}
 \end{figure}

\prevhot{To address the scalability challenges in our (3+$K$)-D scenarios, shown in Figure~\ref{fig:overview}, we propose to divide the scene into two separate tensors: a 3D tensor $\mathcal{T}^3$ and a $K$-D tensor $\mathcal{T}^K$, each subjected to a distinct factorization strategy. 
While $\mathcal{T}^3$ undergoes vector-matrix decomposition for enhanced expressiveness, $\mathcal{T}^K$ is factorized using the CANDECOMP/PARAFAC (CP) decomposition for higher compactness and scalability.
This technique allows us to factorize the $K$-D tensor into a series of $K$ vectors, effectively managing the complexity inherent in multidimensional parameter spaces. As a result, the factorization of our $K$-D parameter space can be represented as
\vspace{-0.05in}
\begin{equation}
  \begin{array}{l}
    \displaystyle 
    \mathcal{T}^K = \sum_{r=1}^{R} \mathbf{v}_r^1 \circ \mathbf{v}_r^2 \circ \cdots \circ \mathbf{v}_r^K,
    \label{eqn:factorization-r2}
\vspace{-0.05in}
  \end{array}
\end{equation}
where $R$ represents the number of one-rank components in the form of $\mathbf{v}_r^1 \circ \mathbf{v}_r^2 \circ \cdots \circ \mathbf{v}_r^K$. 
This tensor factorization significantly reduces the complexity of the parameter space in arbitrary dimensions from $O(M^K)$ to $O(M)$, where $M$ is the parameter-space grid resolution.
Consequently, we can represent the entire (3+$K$)-D scene in a more manageable and efficient manner. Written in equation
\vspace{-0.05in}
\begin{equation}
  \begin{split}
    \displaystyle 
    \mathcal{T}^{3+K}  = &\mathcal{T}^3 \circ \mathcal{T}^K \\
                                     = &\sum_{r=1}^{R_{s}} \mathbf{M}_r^{XY} \circ \mathbf{v}_r^Z + \mathbf{M}_r^{XZ} \circ \mathbf{v}_r^Y  + \mathbf{M}_r^{YZ} \circ \mathbf{v}_r^X \\
                                       &\circ \sum_{r=1}^{R_{p}} \mathbf{v}_r^1 \circ \mathbf{v}_r^2 \circ \cdots \circ \mathbf{v}_r^K,
    \label{eqn:factorization-full}
\vspace{-0.05in}
  \end{split}
\end{equation}
where $\mathcal{T}^{3+K}$ represents the (3+$K$)-D tensor encapsulating the entire dynamic scene, $R_s$ is the number of low-rank components in $\mathcal{T}^3$, and $R_p$ is the number of one-rank components in $\mathcal{T}^K$. 
This approach significantly reduces the overall complexity from $O(N^3M^K)$ to $O(N^2M)$.
Provided that $R_s \ll N$ and $R_p \ll M^K$, ViSNeRF achieves ultra-compactness in its representation while retaining the capability to handle dynamic scenes with an arbitrary number of parameters. 

In Table~\ref{tab:pse-decomp} and Figure~\ref{fig:comp-decomp}, we demonstrate that ViSNeRF, despite having the smallest model size among the three methods, surpasses both K-Planes and HexPlane in visualization generation quality for 4D scenarios, using the five jets dataset with an additional time dimension. 
In Figure~\ref{fig:comp-decomp}, to help with the comparison, we produce difference images with respect to the ground-truth (GT) result in the CIELUV color space. Noticeable pixel differences are mapped according to the colormap legend as shown in (a). 
For this comparison, K-Plane is configured according to its recommended training settings, and HexPlane is set up with its suggested configurations but matched to the same grid resolution as ViSNeRF. 
All three methods undergo training for 90,000 iterations. We exclude K-Plane and HexPlane from further discussion as these methods are not designed to handle (3+$K$)-D scenarios.}

\begin{table}[htb]
  \caption{\prevhot{Factorization methods for dynamic scenes using five jets (timestep) DVR images with \hot{1024$\times$1024} resolution: average PSNR (dB), SSIM, LPIPS, model size (MS, in MB), training time (TT, in hours), and inference time (IT, in minutes) across all 181 synthesized views. The best ones are highlighted in bold.}}
  \vspace{-0.1in}
  \centering
  {\scriptsize
  \hot{
  \begin{tabular}{c|rrr|rrr}
  method & PSNR$\uparrow$ & SSIM$\uparrow$ & LPIPS$\downarrow$ & MS$\downarrow$ & TT$\downarrow$ & IT$\downarrow$ \\ \hline
   K-Planes  & 26.67 & 0.939 & 0.057 & 417.40 & 3.02 & 22.95 \\
   HexPlane & 26.54 & 0.941 & 0.063  & 22.23 & \bf{1.06} & \bf{18.93} \\
   ViSNeRF   & \bf{27.51} & \bf{0.945} & \bf{0.050}  & \bf{15.43} & 1.43 & 21.20 \\   
  \end{tabular}
  }}
  \label{tab:pse-decomp}
\end{table}

In our implementation, the combination of features sampled from spatial matrices and vectors is achieved through element-wise multiplications between each vector-matrix pair. 
This element-wise multiplication is also applied for combining features sampled from the parameter vectors. 
We concatenate the outcomes of these element-wise multiplications to construct the final feature. 
This involves integrating the products derived from the three pairs of spatial matrix and vector as well as the product of the parameter vectors.

\vspace{-0.05in}
\subsection{Feature Decoder}
\label{subsec:fd}

Unlike fully explicit methods, such as PlenOctrees~\cite{Yu-PlenOctrees-ICCV21} and Plenoxels~\cite{Fridovich-Keil-Plenoxels-CVPR22}, using {\em spherical harmonics} (SH) as the decoder, our hybrid ViSNeRF model leverages two small MLPs to decode the density and color features separately.
TensoRF~\cite{Chen-TensoRF-ECCV22} and K-Planes~\cite{Fridovich-Keil-K-Planes-CVPR23} compare a fully explicit model and a hybrid model.
The results suggest that, although using SH reduces rendering time, the quality of synthesized images gets worse due to the limited expressiveness of SH and the difficulty of SH coefficient optimization. 
Instead of directly employing SH as the decoder, we utilize SH to encode the view directions, which are then fed as input into the color MLP decoder, denoted as $g_c$. We observe that encoding view directions with SH slightly enhances network performance compared to {\em positional encoding} (PE)~\cite{Vaswani-Transformer-NeurIPS17}.

\begin{figure}[htb]
  \begin{center}
  $\begin{array}{c@{\hspace{0.02in}}c@{\hspace{0.02in}}c}
  \includegraphics[height=1.225in]{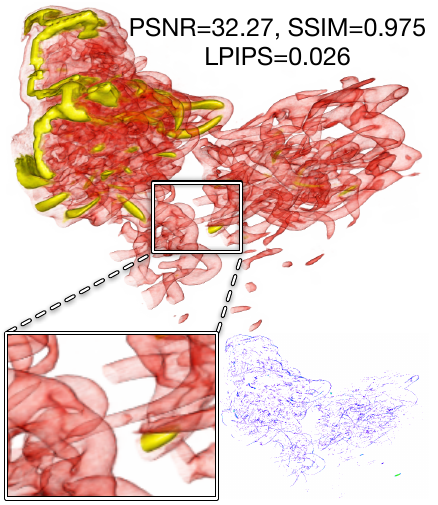}&
  \includegraphics[height=1.225in]{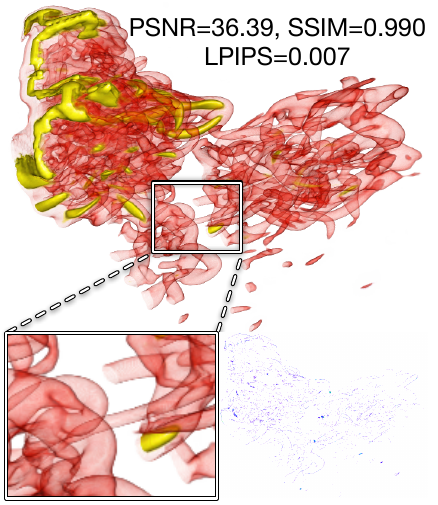}&
  \includegraphics[height=1.225in]{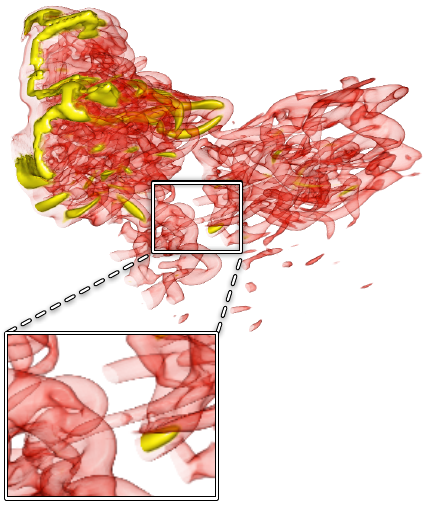}\\
 \mbox{\footnotesize (a) TensoRF} & \mbox{\footnotesize \prevhot{(b) ViSNeRF}} & \mbox{\footnotesize (c) GT}
 \end{array}$
 \end{center}
\vspace{-.25in} 
 \caption{Inferred DVR images of the Tangaroa dataset 
 using 42 views to train the full 360-degree view. Zoom-in and difference images are provided for better comparison. 
 } 
 \label{fig:comp-decoder}
 \end{figure}

In TensoRF~\cite{Chen-TensoRF-ECCV22}, without an implicit MLP decoder for the density features, the explicit density features are not sufficiently expressive to discern the edges of overlapping translucent regions in DVR images, especially when the training images are scarce (e.g., less than 50 images for the full 360-degree view).
Thus, in addition to the MLP $g_c$ for color features $f_c$, we use an extra small MLP $g_\sigma$ to decode the density features $f_\sigma$ at a trivial cost of efficiency, which also helps improve the accuracy of overlapping translucent regions for DVR images. 
Both $g_c$ and $g_\sigma$ have only one hidden layer. They map the feature of a coordinate $\mathbf{x}$ to density and color values as
\vspace{-0.05in}
\begin{equation}
  \begin{array}{l}
    \displaystyle \sigma(\mathbf{x}) = g_\sigma(f_\sigma(\mathbf{x}), E_{\p}(\mathbf{x})), \\
    \displaystyle c(\mathbf{x}) = g_c(f_c(\mathbf{x}), E_{\sh}(\mathbf{d})),
    \label{eqn:mlp}
\vspace{-0.05in}
  \end{array}
\end{equation}
where $E_{\p}$ and $E_{\sh}$ denote PE and SH encoding.
As shown in Figure~\ref{fig:comp-decoder}, the DVR images generated by ViSNeRF exhibit improved clarity in the translucent regions compared with TensoRF. Both methods apply the same L1 and TV regularization \hot{(refer to Section~\ref{subsec:loss})} to ensure fairness. 

 \hot{
\vspace{-0.05in}
\subsection{Loss Functions}
\label{subsec:loss}

To optimize ViSNeRF, a straightforward approach utilizes a loss function that measures the difference between the generated visualization and the GT, known as reconstruction loss.
However, as shown in Figure~\ref{fig:comp-loss}, without any regularization, excessive noises and film grains are apparent in the visualized spatial matrices of ViSNeRF and the generated images.
Therefore, in addition to the reconstruction loss, we apply an L1 norm loss and a total variation (TV) loss as regularization terms in the loss function 
\vspace{-0.05in}
\begin{equation}
  \begin{array}{l}
    \displaystyle \mathcal{L} = \mathcal{L}_{\rec} + \lambda_1 \mathcal{L}_{\lone} + \lambda_2 \mathcal{L}_{\tv},
    \label{eqn:loss}
\vspace{-0.05in}
  \end{array}
\end{equation}
where $\lambda_1$ and $\lambda_2$ are weights of the regularization terms.

{\bf Reconstruction loss.} 
During the training process, in each batch, a set of rays $\mathcal{R}$ is randomly selected from the pool of all pixels from the images. 
As characterized in Equation~\ref{eqn:nerf-color}, we query $n$ samples along each ray $\mathbf{r}$ and predict the color of the corresponding pixel $\mathcal{\hat{C}}(\mathbf{r})$.
The reconstruction loss is computed using the MSE between the predicted colors $\mathcal{\hat{C}}(\mathbf{r})$ and GT colors $\mathcal{C}(\mathbf{r})$ of the set of pixels in each batch, and is defined as
\vspace{-0.05in}
\begin{equation}
  \begin{array}{l}
    \displaystyle \mathcal{L}_{\rec} = \frac{1}{|\mathcal{R}|} \sum_{\mathbf{r}\in\mathcal{R}} \left|\left| \mathcal{C}(\mathbf{r}) - \mathcal{\hat{C}}(\mathbf{r}) \right|\right|^2_2.
    \label{eqn:loss-mse}
\vspace{-0.05in}
  \end{array}
\end{equation}

{\bf L1 norm loss.} 
To reduce overfitting by extracting more relevant features, we employ L1 norm loss as a regularization term
\vspace{-0.05in}
\begin{equation}
  \begin{array}{l}

    \displaystyle \mathcal{L}_{\lone} = \frac{1}{|W|} \sum_{w\in W} 
    
    \begin{cases}
      ||1-w||_1 & \text{if } W\in \mathbf{v}_p \\
      ||w||_1   & \text{otherwise}
    \end{cases}

    \label{eqn:loss-l1-norm}
\vspace{-0.05in}
  \end{array}
\end{equation}
where $W$ are the weights in the vectors and matrices in the radiance field representation, and $\mathbf{v}_p$ is the set of vectors representing input parameter features.
For the vectors containing the parameter features, the weights are initialized and regularized to 1 for no alteration in the spatial content. 
Note that L1 regularization is only applied to the vectors and matrices, not the implicit MLP decoder. 

\begin{figure}[htb]
  \centering
    \begin{subfigure}[b]{0.5\textwidth}
      \centering
  $\begin{array}{c@{\hspace{0.05in}}c@{\hspace{0.05in}}c}
  \includegraphics[width=0.3\linewidth]{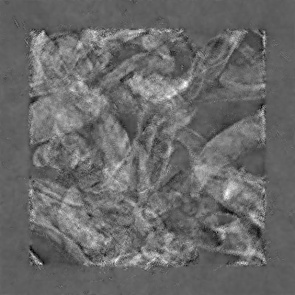}&
  \includegraphics[width=0.3\linewidth]{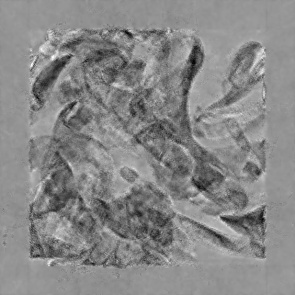}&
  \includegraphics[width=0.3\linewidth]{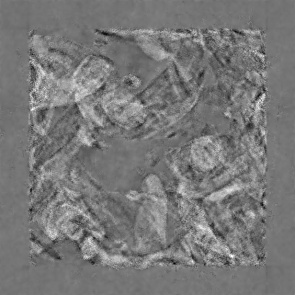}\\
 \end{array}$
  \vspace{-.075in} 
 {\footnotesize \caption{no regularization}}
\end{subfigure}%
\hfill
\begin{subfigure}[b]{0.5\textwidth}
  \centering
  $\begin{array}{c@{\hspace{0.05in}}c@{\hspace{0.05in}}c}
  \includegraphics[width=0.3\linewidth]{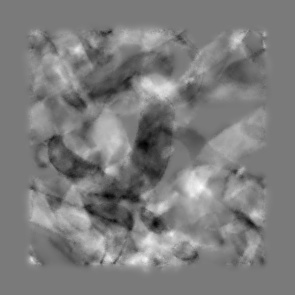}&
  \includegraphics[width=0.3\linewidth]{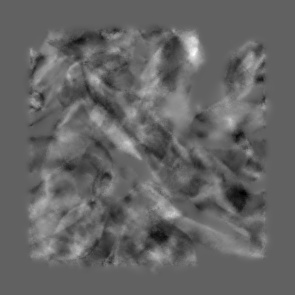}&
  \includegraphics[width=0.3\linewidth]{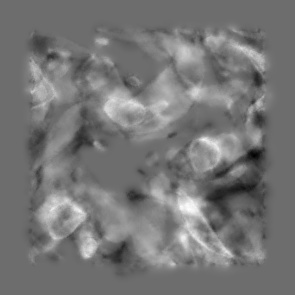}\\
 \end{array}$
  \vspace{-.075in} 
  {\footnotesize \caption{L1+TV regularization}}
\end{subfigure}
\begin{subfigure}[b]{0.5\textwidth}
  \centering
  $\begin{array}{c@{\hspace{0.05in}}c@{\hspace{0.05in}}c}
  \includegraphics[width=0.3\linewidth]{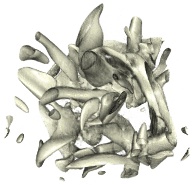}&
  \includegraphics[width=0.3\linewidth]{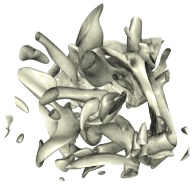}&
  \includegraphics[width=0.3\linewidth]{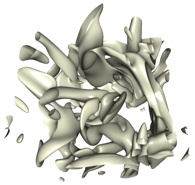}\\
  \mbox{\footnotesize (c) no regularization} & \mbox{\footnotesize (d) L1+TV} & \mbox{\footnotesize GT}
 \end{array}$
\end{subfigure}
\vspace{-.15in} 
 \caption{Spatial matrices and IR images of novel views generated by ViSNeRF using 42 views of the vortex dataset for training. (a) and (b) shows the three spatial matrices of density feature.} 
 \label{fig:comp-loss}
 \end{figure}

{\bf Total variation loss.} 
Although L1 norm loss reduces the artifacts by preventing overfitting, if we have only a few views of the scene available for training, the vectors and matrices would still be noisy since some regions are not seen in these views.
TV loss is a regularization term that aims to denoise the vectors and matrices by penalizing high variation and encouraging smoothness. 
That is, 

\vspace{-0.05in}
\begin{equation}
  \begin{array}{l}
    \displaystyle \mathcal{L}_{\tv} = \mathcal{L}_{\tv_{1}} + \mathcal{L}_{\tv_{2}}, \\
    \displaystyle \mathcal{L}_{\tv_{1}} =  \frac{1}{|\mathcal{V}||\mathcal{Q}|l} \sum_{\mathbf{v},q,i} \bigl( \bigl|\bigl|\mathbf{v}_{q}^{i}-\mathbf{v}_{q}^{i-1}\bigr|\bigr|^2_2 \bigr), \\
    \displaystyle \mathcal{L}_{\tv_{2}} =  \frac{1}{|\mathcal{M}||\mathcal{Q}|hw} \sum_{\mathbf{M},q,i,j} \bigl( \bigl|\bigl|\mathbf{M}_{q}^{i,j}-\mathbf{M}_{q}^{i-1,j}\bigr|\bigr|^2_2 \\
    \displaystyle                                                                   + \bigl|\bigl|\mathbf{M}_{q}^{i,j}-\mathbf{M}_{q}^{i,j-1}\bigr|\bigr|^2_2 \bigr), \\
    \label{eqn:loss-tv}
\vspace{-0.1in}
  \end{array}
\end{equation}
where $\mathbf{v}\in \mathcal{V}$, $\mathbf{M}\in \mathcal{M}$, $q\in \mathcal{Q}$, $\mathcal{V}$ is the set of vectors, $\mathcal{M}$ is the set of matrices, $\mathcal{Q}$ is the channels of the features, $l$ is the length of the vector, and $h$ and $w$ are the height and width of the plane.
}


\begin{table*}[htb]
\caption{Parameter-space exploration of dynamic scenes: numbers of training images sampled and their resolutions. 
}
\vspace{-0.1in}
\centering
{\scriptsize
\begin{tabular}{c|cccccc}
dataset & \prevhot{volume} & \# images & \# parameters  & \# parameter & total \# & image \\ 
(scenario) & \prevhot{resolution} & per scene & \hot{($K$)}  & samples \hot{($M$)} & images & resolution\\ \hline
five jets (timestep) & \prevhot{128$\times$128$\times$128} & 42 & 1 & 11 & 462 & \hot{1024$\times$1024} \\
Tangaroa (isovalue) & \prevhot{300$\times$180$\times$120} & 42 & 1 & 13 & 546 & 1024$\times$1024 \\
vortex (TF-1) & \prevhot{128$\times$128$\times$128} & 42 & 2 & 36 & 1512 & 256$\times$256 \\
vortex (TF-2) & \prevhot{128$\times$128$\times$128} & 42 & 1 & 11 & 462 & 256$\times$256 \\
\prevhot{Nyx-DVR/IR (simulation parameters)} & \prevhot{256$\times$256$\times$256} & \prevhot{42} & \prevhot{3} & \prevhot{45} & \prevhot{1890} & \prevhot{256$\times$256} \\
\end{tabular}
}
\label{tab:pse-dataset}
\end{table*}

\vspace{-0.05in}
\section{Results and Discussion}

For dynamic volumetric scenes, we present the results of ViSNeRF and compare them with four baseline methods (i.e., InSituNet, CoordNet, StyleGAN2, and EG3D) in parameter-space exploration tasks.
For static volumetric scenes, we also compare ViSNeRF with eight baseline and representative methods (i.e., InSituNet, CoordNet, StyleGAN2, EG3D, NeRF, 3DGS, Instant-NGP, and TensoRF). 
The experimental results are furnished in the appendix. 
The appendix also includes optimization schemes, baseline training details, hyperparameter study, and additional results and discussion.

\vspace{-0.05in}
\subsection{Datasets}

As shown in Table~\ref{tab:pse-dataset}, we evaluate ViSNeRF in five dynamic volumetric scenes from four simulation datasets with different controllable parameters.
The image resolutions are determined based on the content present in the data. 
\hot{The vortex and Nyx are relatively simple and can use low-resolution renderings. 
The Tangaroa and five jets need high-resolution renderings to capture the details.}

To evenly distribute the viewpoints around the volume data, the camera positions are determined by the vertices of an icosphere, which approximates a sphere composed of equilateral triangles. 
The number of vertices starts with 12 at the subdivision level 0 (initial icosahedron), followed by 42, 92, 162, and 252 for successive subdivision levels. 
Experiments on static scenes show that 42 vertices at subdivision level 1 are sufficient to create a training set that delivers satisfactory generation quality for ViSNeRF. 
Hence, we also use 42 views per scene frame of dynamic volumetric scenes for training ViSNeRF and the baseline models.
For each viewpoint, we record the camera poses for ViSNeRF and other 3D-aware methods and convert them to the corresponding azimuth ([-180,180]) and elevation ([-90,90]) for 2D-based approaches.

The parameter settings that define each scene frame are evenly sampled within the chosen parameter ranges.
The exact number of sampled parameter settings is provided in Table~\ref{tab:pse-dataset}.
For inference on the full 360-degree view, we synthesize 181 views along a path on a spherical surface from azimuth -180 and elevation -90, through azimuth 0 and elevation 0, to azimuth 180 and elevation 90.
During the inference sequence, as the camera moves, the scene dynamically changes by varying parameters within their defined ranges.

\begin{figure}[htb]
  \begin{center}
  $\begin{array}{c@{\hspace{0.05in}}c}
  \includegraphics[width=0.475\linewidth]{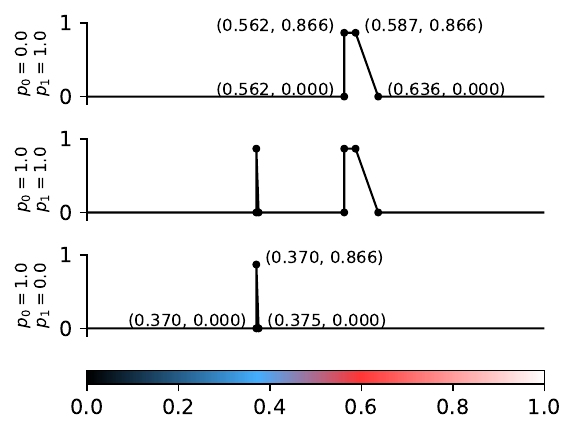}&
  \includegraphics[width=0.475\linewidth]{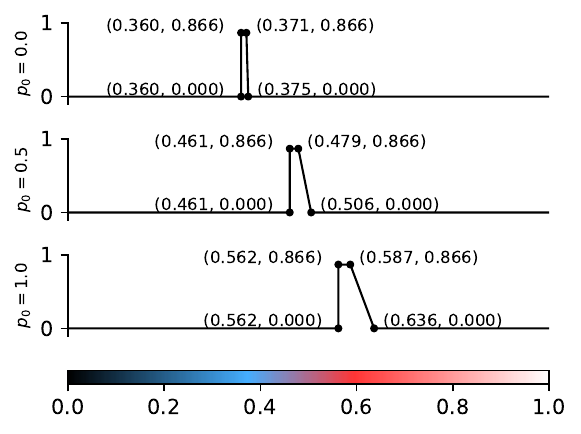}\\
 \mbox{\footnotesize (a) \hot{TF-1}} & \mbox{\footnotesize (b) TF-2}
 \end{array}$
 \end{center}
\vspace{-.25in} 
 \caption{Illustration of two ways of interpolation over TFs of the vortex dataset. The tuple of each control point indicates its scalar value ($x$-axis) and opacity ($y$-axis).
 } 
 \label{fig:tf-illustration}
 \end{figure}

\begin{table}[htb]
  \caption{Parameter-space exploration of dynamic scenes: average PSNR (dB), SSIM, LPIPS, MS (in MB), TT (in hours), and IT (in minutes) across all 181 synthesized views. 
  }
  \vspace{-0.1in}
  \centering
  \resizebox{\columnwidth}{!}{
  \begin{tabular}{c|c|rrr|rrr}
  dataset          & method & PSNR$\uparrow$ & SSIM$\uparrow$ & LPIPS$\downarrow$ & MS$\downarrow$ & TT$\downarrow$ & IT$\downarrow$ \\ \hline
                    & \hot{InSituNet} & \hot{16.37} & \hot{0.850} & \hot{0.176} & \hot{351.74} & \hot{39.22} & \hot{\bf{0.70}} \\
  five jets         & \hot{CoordNet}  & \hot{17.71} & \hot{0.864} & \hot{0.169} & \hot{\bf{5.71}} & \hot{112.23} & \hot{13.38} \\
  timestep  & \hot{StyleGAN2} & \hot{16.41} & \hot{0.854} & \hot{0.180}  & \hot{158.68} & \hot{53.25} & \hot{1.07} \\
  (\hot{1024$\times$1024})	                & \hot{EG3D}      & \hot{22.74} & \hot{0.908} & \hot{0.067} & \hot{152.14} & \hot{75.75} & \hot{1.03}\\
                    & \hot{ViSNeRF}   & \hot{\bf{27.51}} & \hot{\bf{0.945}} & \hot{\bf{0.050}}  & \hot{15.43} & \hot{\bf{1.43}} & \hot{21.20} \\ \hline
                    & InSituNet & 17.08 & 0.841 & 0.230 & 351.73 & 51.97 & \bf{0.42}\\
  Tangaroa          & CoordNet  & 19.83 & 0.876 & 0.178 & \bf{5.72} & 120.90 & 11.84\\
  isovalue & StyleGAN2 & 18.53 & 0.860 & 0.178 & 158.68 & 60.64 & 1.19 \\
  (1024$\times$1024)	                & EG3D      & 22.92 & 0.915 & 0.081  & 152.14 & 89.52 & 1.17 \\
                    & \prevhot{ViSNeRF}   & \prevhot{\bf{29.36}} & \prevhot{\bf{0.967}} & \prevhot{\bf{0.042}}  & \prevhot{69.91} & \prevhot{\bf{2.49}} & \prevhot{31.42}\\ \hline
                    & InSituNet & 17.03 & 0.726 & 0.211 & 199.66 & 17.52 & 0.18\\
  vortex              & CoordNet  & 18.06 & 0.728 & 0.222 & \bf{5.72} & 60.57 & 1.35\\
  TF-1  & StyleGAN2 & 18.38 & 0.823 & 0.201 & 103.62 & 9.07 & \bf{0.10} \\
  (256$\times$256)	                & EG3D      & 26.20 & 0.929 & 0.054  & 150.64 & 82.78 & 0.22\\
                    & \prevhot{ViSNeRF}   & \prevhot{\bf{37.26}} & \prevhot{\bf{0.994}} & \prevhot{\bf{0.005}} & \prevhot{12.79} & \prevhot{\bf{1.68}} & \prevhot{1.55} \\ \hline
                    & InSituNet & 16.75 & 0.779 & 0.199 & 199.66 & 6.13 & \bf{0.08} \\
  vortex              & CoordNet  & 17.10 & 0.793 & 0.212 & \bf{5.72} & 35.71 & 0.64\\
  TF-2  & StyleGAN2 & 17.08 & 0.790 & 0.197  & 103.62 & 4.15 & 0.10 \\
  (256$\times$256)	                & EG3D      & 27.53 & 0.969 & 0.024 & 150.63 & 37.97 & 0.35\\
                    & \prevhot{ViSNeRF}   & \prevhot{\bf{36.58}} & \prevhot{\bf{0.995}} & \prevhot{\bf{0.006}} & \prevhot{12.79} & \prevhot{\bf{1.42}} & \prevhot{1.45} \\ \hline
                    & \prevhot{InSituNet} & \prevhot{14.06} & \prevhot{0.495} & \prevhot{0.294} & \prevhot{199.67} & \prevhot{21.41} & \prevhot{\bf{0.12}} \\
  \prevhot{Nyx-DVR}              & \prevhot{CoordNet}  & \prevhot{15.09} & \prevhot{0.514} & \prevhot{0.370} & \prevhot{\bf{5.71}} & \prevhot{87.80} & \prevhot{0.85}\\
  \prevhot{simulation}  & \prevhot{StyleGAN2} & \prevhot{14.48} & \prevhot{0.508} & \prevhot{0.302}  & \prevhot{103.62} & \prevhot{11.79} & \prevhot{0.15} \\
  \prevhot{parameters}	                & \prevhot{EG3D}      & \prevhot{18.76} & \prevhot{0.759} & \prevhot{0.157} & \prevhot{150.64} & \prevhot{108.34} & \prevhot{0.37}\\
  \prevhot{(256$\times$256)}                  & \prevhot{ViSNeRF}   & \prevhot{\bf{30.02}} & \prevhot{\bf{0.978}} & \prevhot{\bf{0.018}} & \prevhot{12.79} & \prevhot{\bf{1.66}} & \prevhot{1.62} \\ \hline
                    & \prevhot{InSituNet} & \prevhot{16.52} & \prevhot{0.550} & \prevhot{0.234} & \prevhot{199.67} & \prevhot{21.89} & \prevhot{\bf{0.12}} \\
  \prevhot{Nyx-IR}              & \prevhot{CoordNet}  & \prevhot{17.41} & \prevhot{0.468} & \prevhot{0.299} & \prevhot{\bf{5.71}} & \prevhot{78.90} & \prevhot{0.85}\\
  \prevhot{simulation}  & \prevhot{StyleGAN2} & \prevhot{16.88} & \prevhot{0.562} & \prevhot{0.250}  & \prevhot{103.62} & \prevhot{11.85} & \prevhot{0.15} \\
  \prevhot{parameters}	                & \prevhot{EG3D}      & \prevhot{20.38} & \prevhot{0.759} & \prevhot{0.149} & \prevhot{150.64} & \prevhot{101.65} & \prevhot{0.35}\\
  \prevhot{(256$\times$256)}                  & \prevhot{ViSNeRF}   & \prevhot{\bf{28.73}} & \prevhot{\bf{0.950}} & \prevhot{\bf{0.042}} & \prevhot{12.79} & \prevhot{\bf{1.66}} & \prevhot{1.60} \\ 
  \end{tabular}
  }
  \label{tab:psnr-lpips-pse}
\end{table}

\begin{figure*}[htb]
  \begin{center}
  $\begin{array}{c@{\hspace{0.01in}}c@{\hspace{0.01in}}c@{\hspace{0.01in}}c@{\hspace{0.01in}}c@{\hspace{0.01in}}c@{\hspace{0.01in}}c@{\hspace{0.01in}}c@{\hspace{0.01in}}c}
    \raisebox{0.15in}{\rotatebox{90}{InSituNet}}&
    \includegraphics[height=0.8125in]{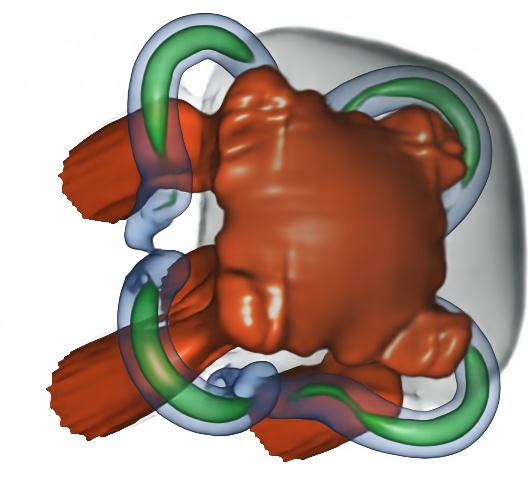}&
    \includegraphics[height=0.8125in]{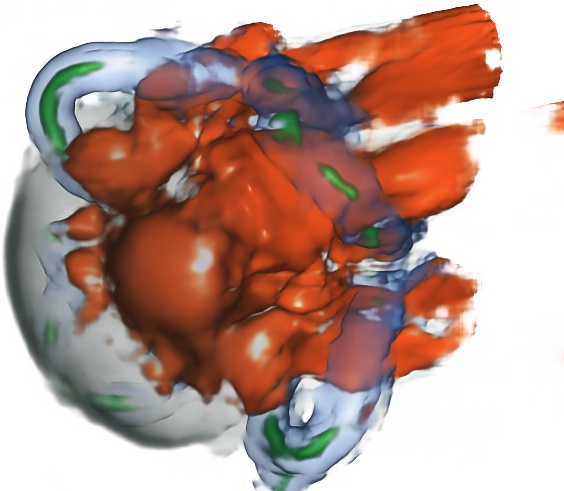}&
    \includegraphics[height=0.8125in]{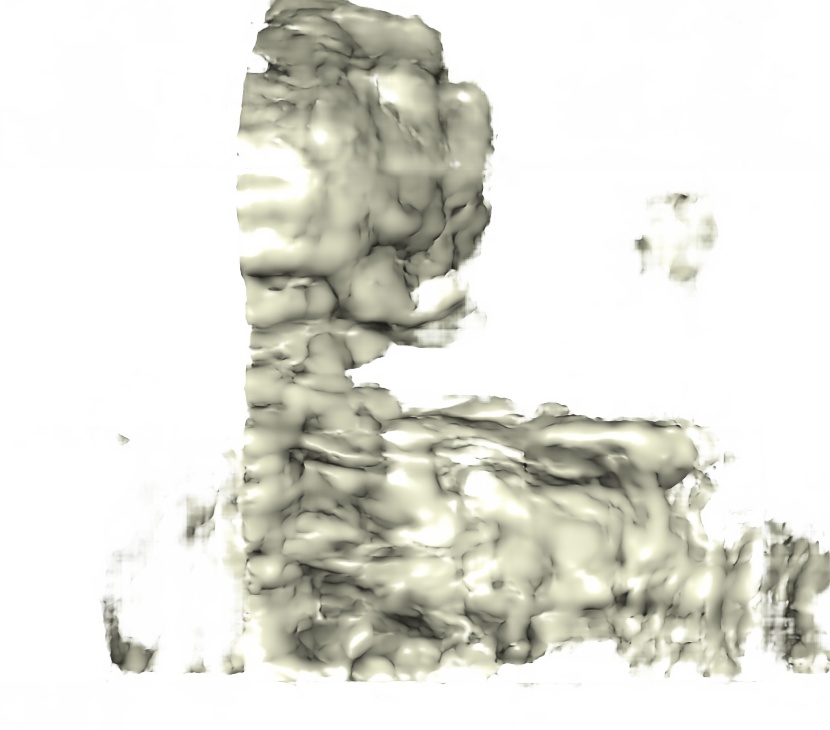}&
    \includegraphics[height=0.8125in]{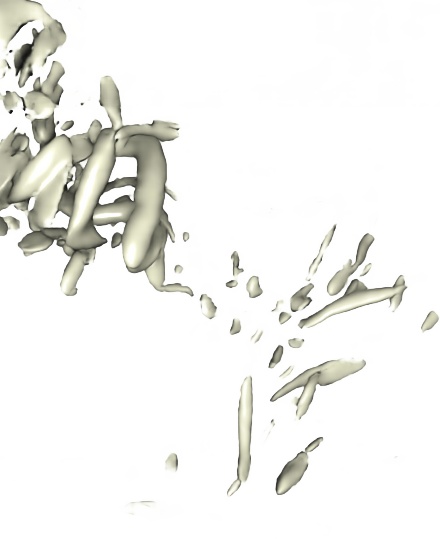}&
    \includegraphics[height=0.8125in]{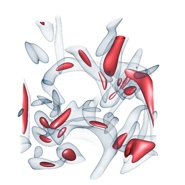}&
    \includegraphics[height=0.8125in]{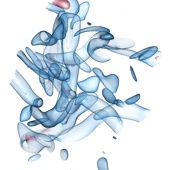}&
    \includegraphics[height=0.8125in]{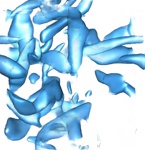}&
    \includegraphics[height=0.8125in]{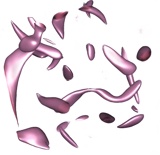}\\
    \raisebox{0.15in}{\rotatebox{90}{CoordNet}}&
    \includegraphics[height=0.8125in]{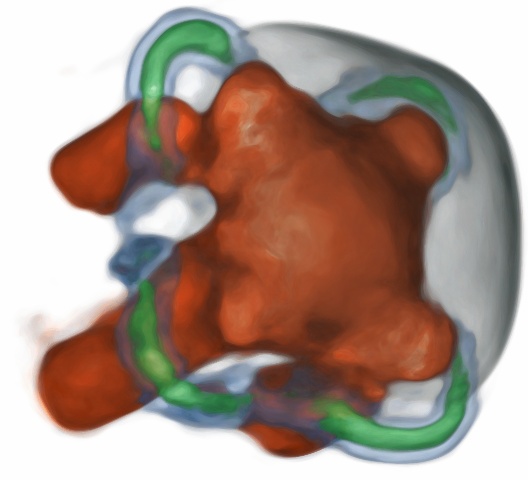}&
    \includegraphics[height=0.8125in]{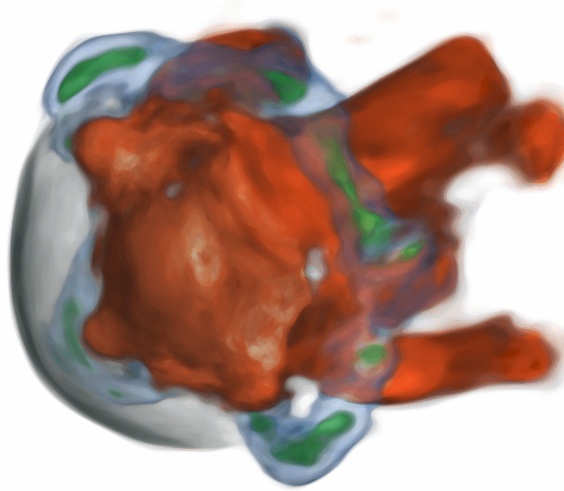}&
    \includegraphics[height=0.8125in]{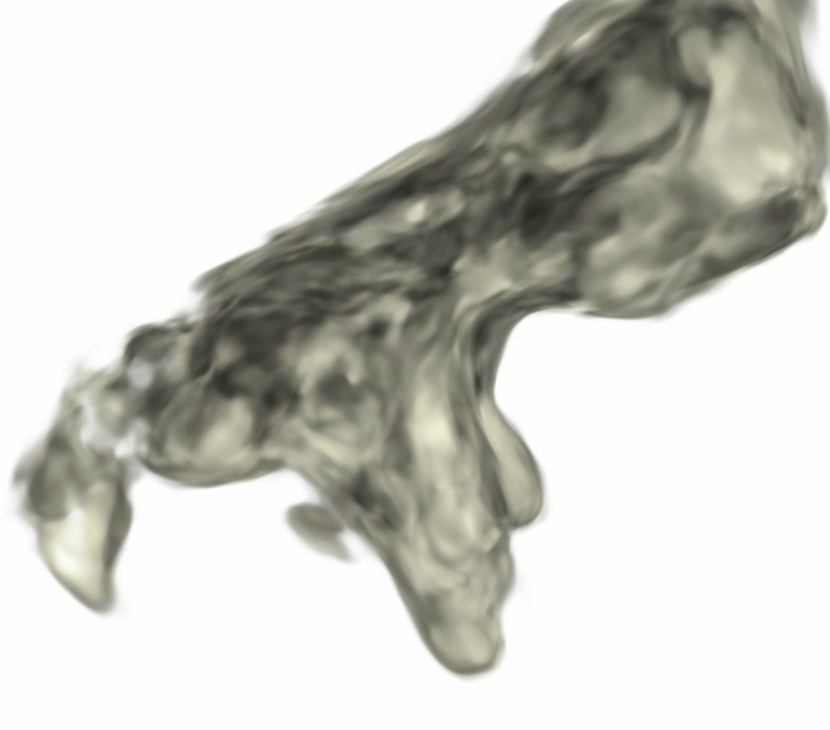}&
    \includegraphics[height=0.8125in]{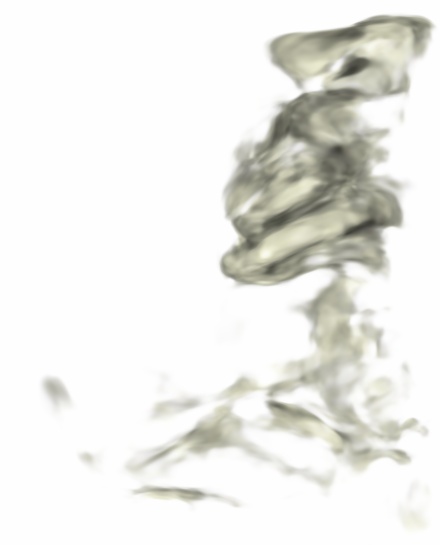}&    
    \includegraphics[height=0.8125in]{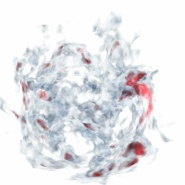}&
    \includegraphics[height=0.8125in]{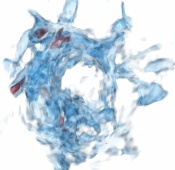}&
    \includegraphics[height=0.8125in]{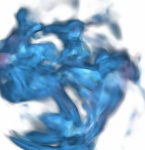}&
    \includegraphics[height=0.8125in]{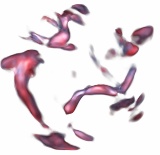}\\
    \raisebox{0.1in}{\rotatebox{90}{StyleGAN2}}&
    \includegraphics[height=0.8125in]{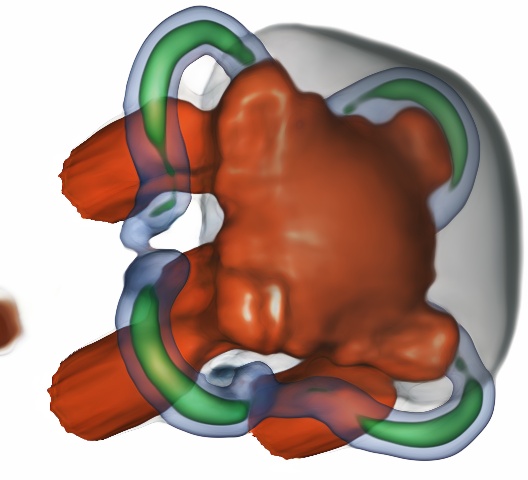}&
    \includegraphics[height=0.8125in]{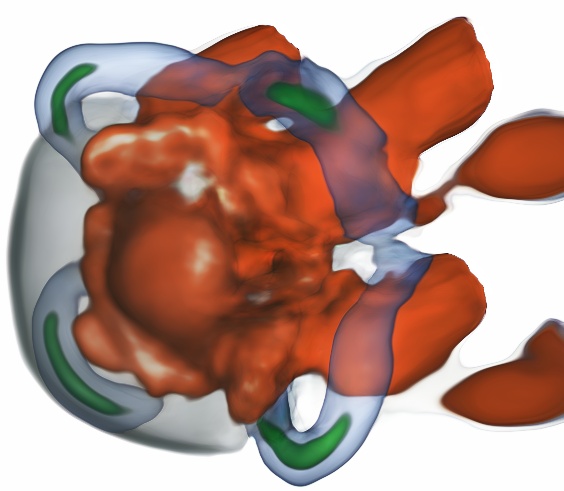}&
    \includegraphics[height=0.8125in]{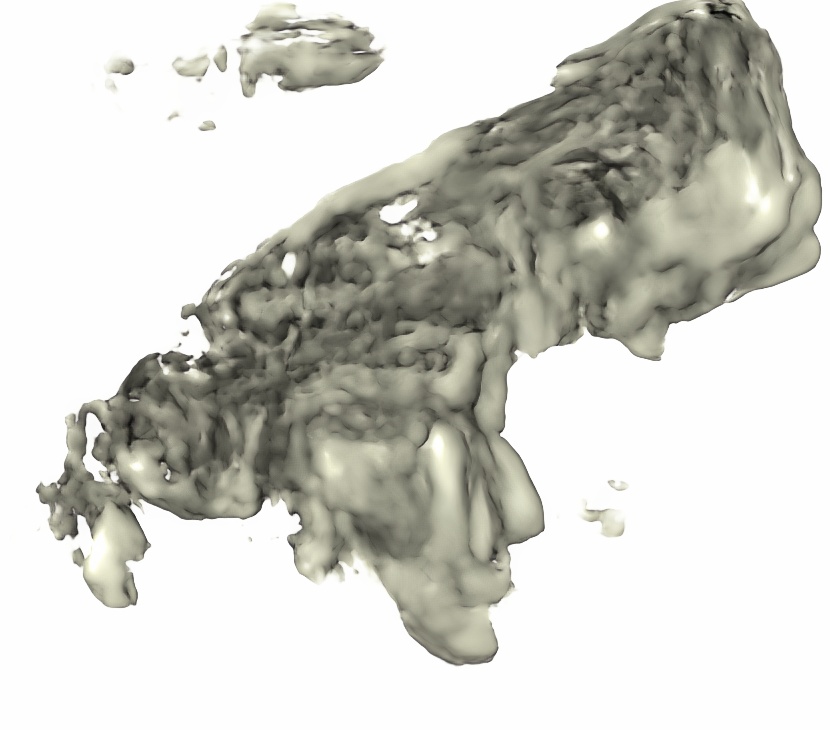}&
    \includegraphics[height=0.8125in]{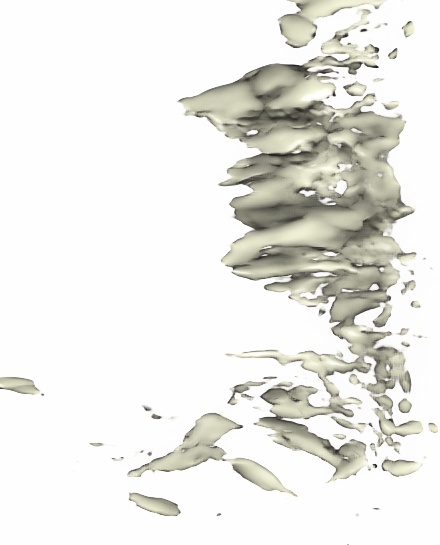}&
    \includegraphics[height=0.8125in]{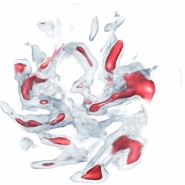}&
    \includegraphics[height=0.8125in]{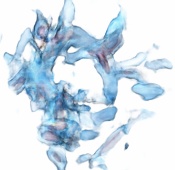}&
    \includegraphics[height=0.8125in]{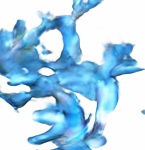}&
    \includegraphics[height=0.8125in]{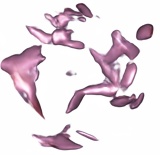}\\
    \raisebox{0.35in}{\rotatebox[origin=c]{90}{EG3D}}&
    \includegraphics[height=0.8125in]{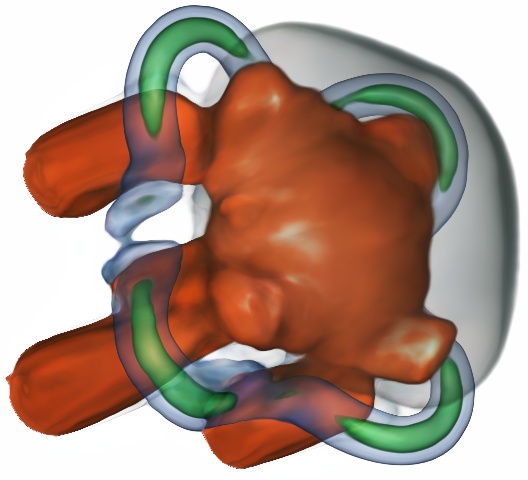}&
    \includegraphics[height=0.8125in]{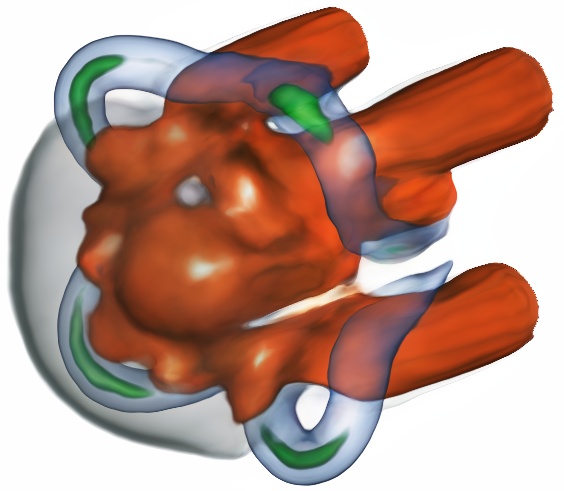}&
    \includegraphics[height=0.8125in]{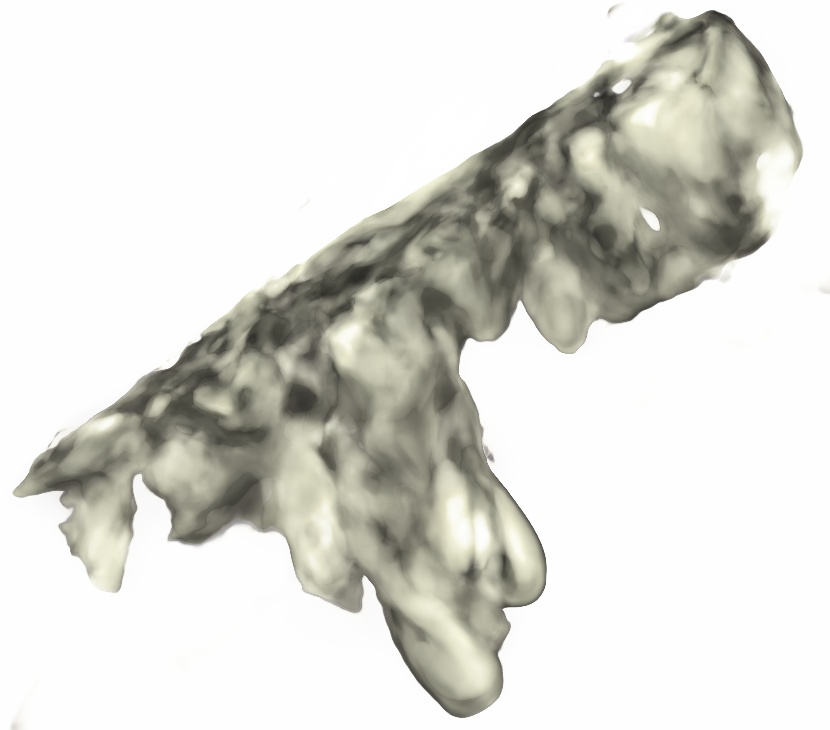}&
    \includegraphics[height=0.8125in]{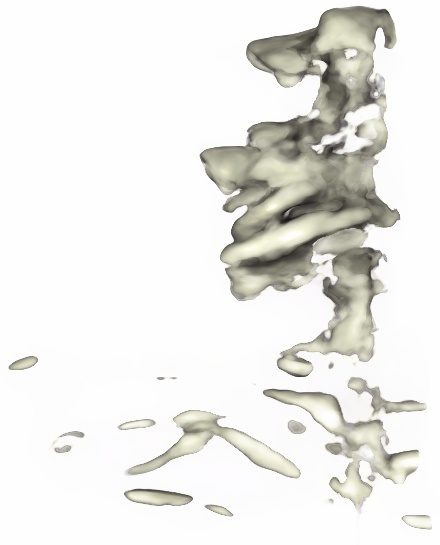}&
    \includegraphics[height=0.8125in]{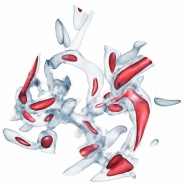}&
    \includegraphics[height=0.8125in]{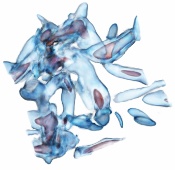}&
    \includegraphics[height=0.8125in]{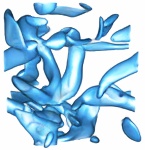}&
    \includegraphics[height=0.8125in]{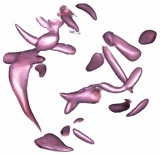}\\
    \raisebox{0.15in}{\rotatebox{90}{ViSNeRF}}&
    \includegraphics[height=0.8125in]{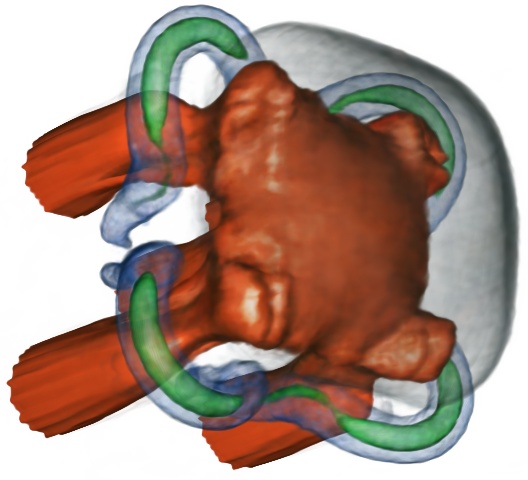}&
    \includegraphics[height=0.8125in]{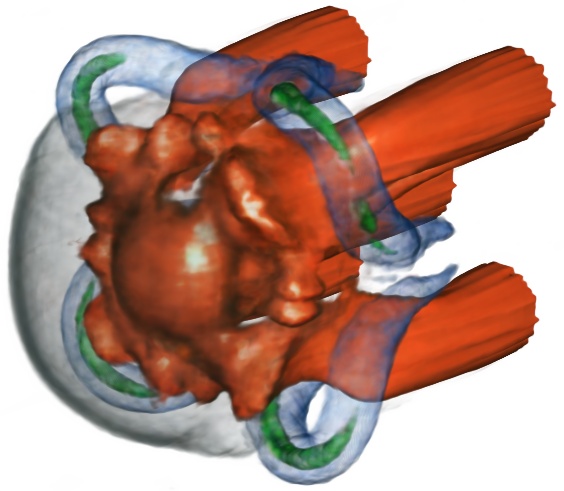}&
    \includegraphics[height=0.8125in]{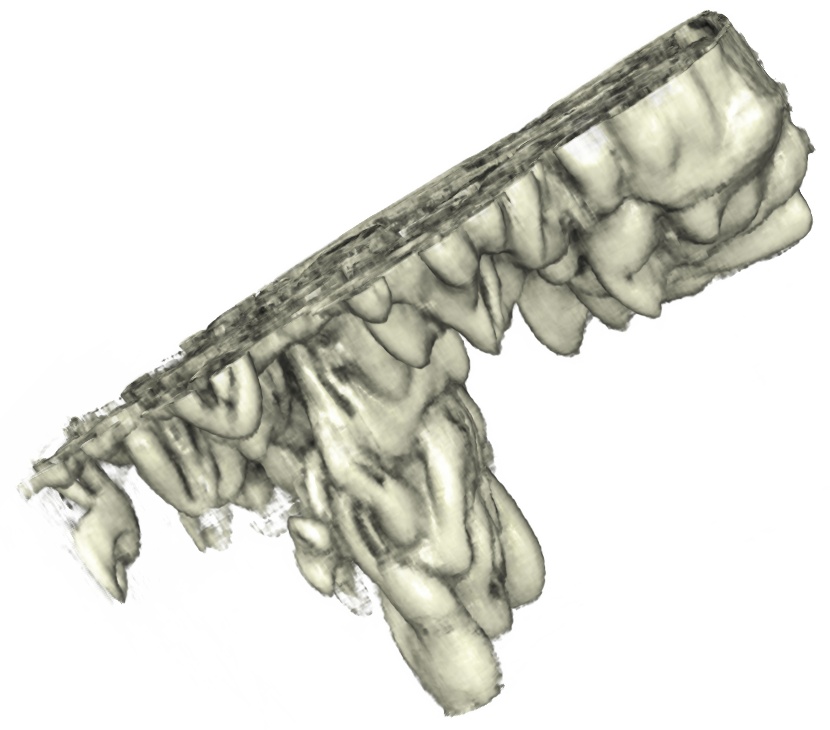}&
    \includegraphics[height=0.8125in]{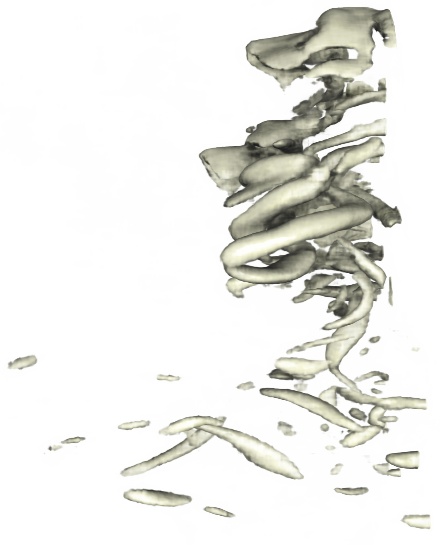}&
    \includegraphics[height=0.8125in]{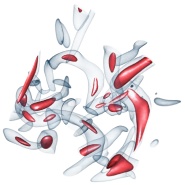}&
    \includegraphics[height=0.8125in]{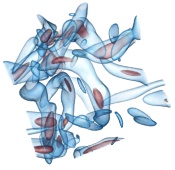}&
    \includegraphics[height=0.8125in]{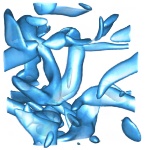}&
    \includegraphics[height=0.8125in]{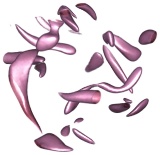}\\
    \raisebox{0.35in}{\rotatebox{90}{GT}}&
    \includegraphics[height=0.8125in]{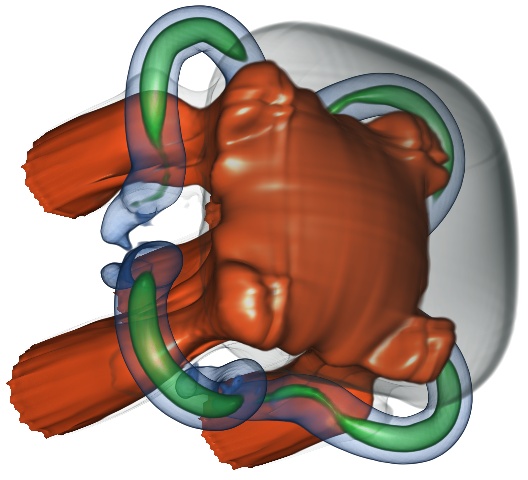}&
    \includegraphics[height=0.8125in]{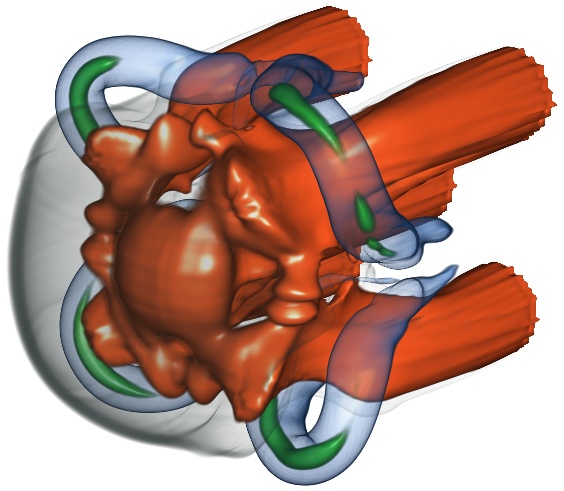}&
    \includegraphics[height=0.8125in]{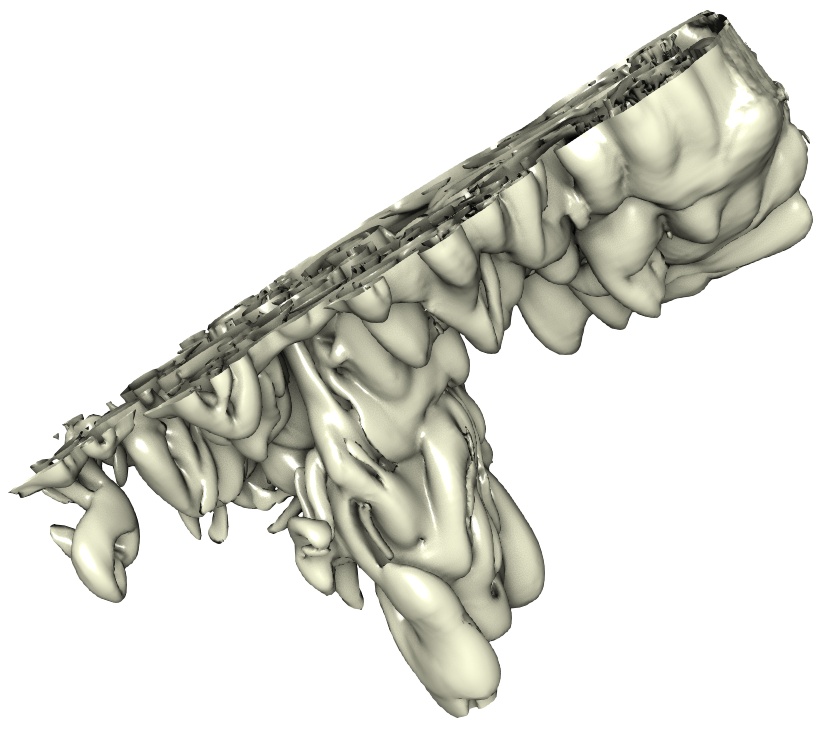}&
    \includegraphics[height=0.8125in]{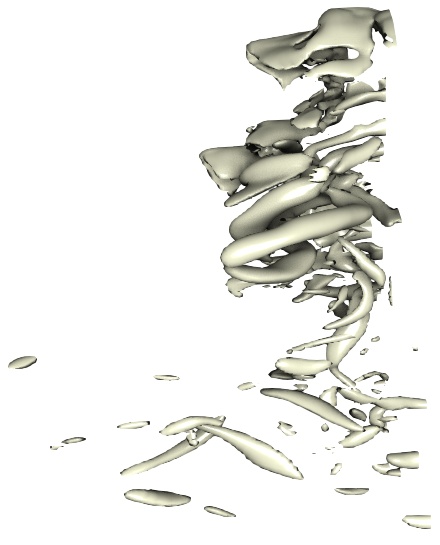}&
    \includegraphics[height=0.8125in]{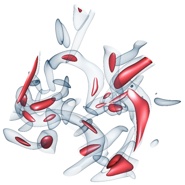}&
    \includegraphics[height=0.8125in]{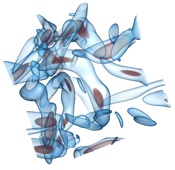}&
    \includegraphics[height=0.8125in]{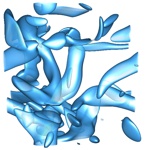}&
    \includegraphics[height=0.8125in]{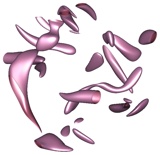}\\    
    & \mbox{\footnotesize (a)} & \mbox{\footnotesize (b)} & \mbox{\footnotesize (c)} & \mbox{\footnotesize (d)} & 
    \mbox{\footnotesize (e)} & \mbox{\footnotesize (f)} & \mbox{\footnotesize (g)} & \mbox{\footnotesize (h)}
  \end{array}$
 \end{center}
 \vspace{-.25in} 
 \caption{Inferred images under different views for parameter-space exploration of dynamic scenes (timesteps, isovalues, and TFs). 
 (a) and (b) DVR images with interpolation over timesteps using the five jets dataset. 
 (c) and (d) IR images with interpolation over isovalues using the Tangaroa dataset.
 (e) to (h) DVR images with interpolation over TFs using the vortex dataset (TF-1: (e) and (f); TF-2: (g) and (h)).
 } 
 \label{fig:comp-pse}
\end{figure*}

\begin{figure*}[htb]
  \begin{center}
    $\begin{array}{c@{\hspace{0.01in}}c@{\hspace{0.01in}}c@{\hspace{0.01in}}c@{\hspace{0.01in}}c@{\hspace{0.01in}}c@{\hspace{0.01in}}c@{\hspace{0.01in}}c@{\hspace{0.01in}}c}
      \raisebox{0.15in}{\rotatebox{90}{InSituNet}}&
      \includegraphics[height=0.8125in]{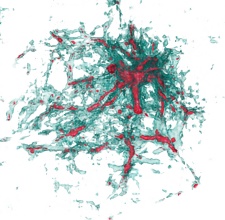}&
      \includegraphics[height=0.8125in]{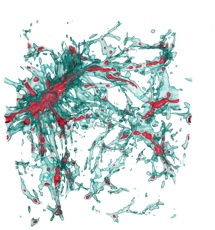}&
      \includegraphics[height=0.8125in]{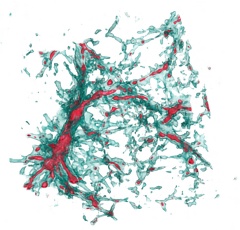}&
      \includegraphics[height=0.8125in]{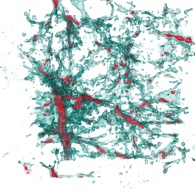}&
      \includegraphics[height=0.8125in]{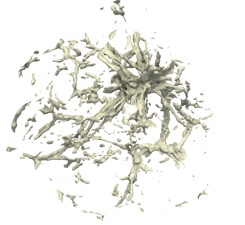}&
      \includegraphics[height=0.8125in]{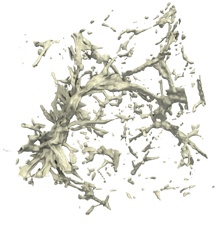}&
      \includegraphics[height=0.8125in]{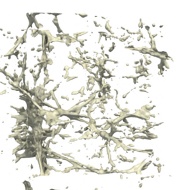}&
      \includegraphics[height=0.8125in]{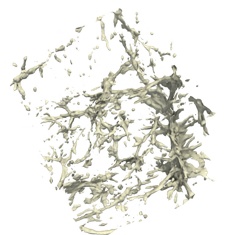}\\
      \raisebox{0.15in}{\rotatebox{90}{CoordNet}}&
      \includegraphics[height=0.8125in]{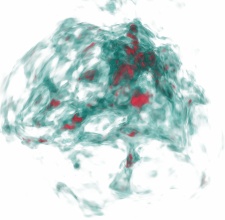}&
      \includegraphics[height=0.8125in]{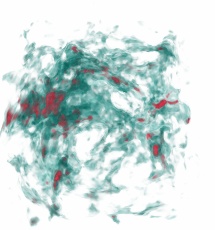}&
      \includegraphics[height=0.8125in]{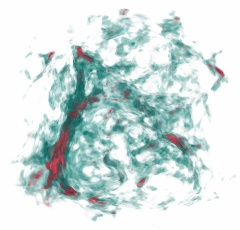}&
      \includegraphics[height=0.8125in]{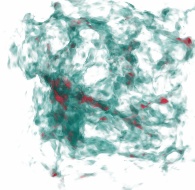}&    
      \includegraphics[height=0.8125in]{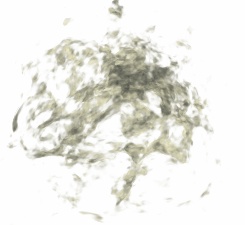}&
      \includegraphics[height=0.8125in]{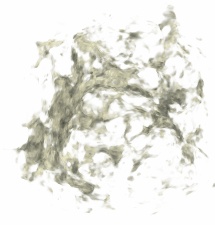}&
      \includegraphics[height=0.8125in]{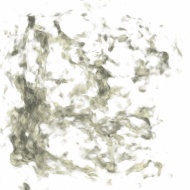}&
      \includegraphics[height=0.8125in]{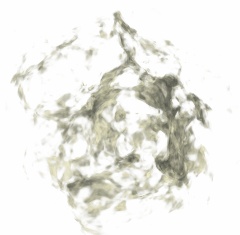}\\
      \raisebox{0.1in}{\rotatebox{90}{StyleGAN2}}&
      \includegraphics[height=0.8125in]{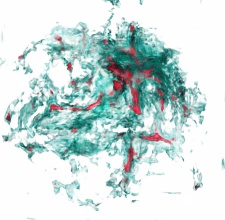}&
      \includegraphics[height=0.8125in]{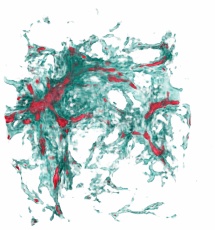}&
      \includegraphics[height=0.8125in]{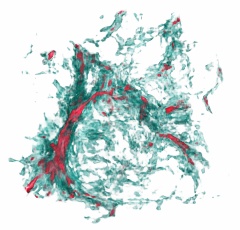}&
      \includegraphics[height=0.8125in]{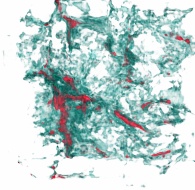}&
      \includegraphics[height=0.8125in]{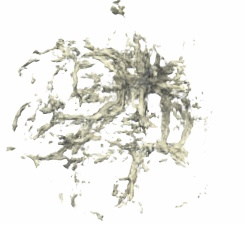}&
      \includegraphics[height=0.8125in]{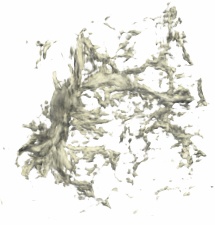}&
      \includegraphics[height=0.8125in]{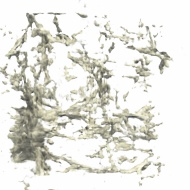}&
      \includegraphics[height=0.8125in]{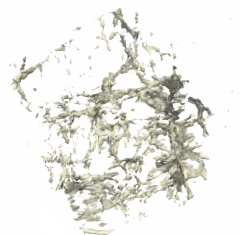}\\
      \raisebox{0.25in}{\rotatebox{90}{EG3D}}&
      \includegraphics[height=0.8125in]{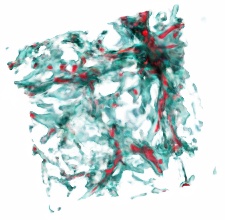}&
      \includegraphics[height=0.8125in]{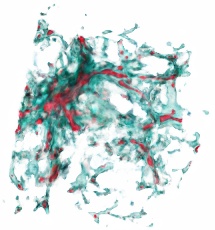}&
      \includegraphics[height=0.8125in]{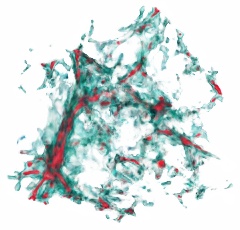}&
      \includegraphics[height=0.8125in]{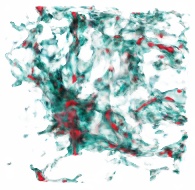}&
      \includegraphics[height=0.8125in]{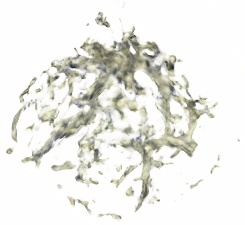}&
      \includegraphics[height=0.8125in]{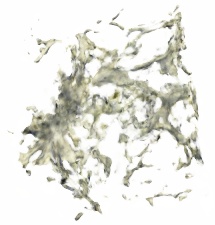}&
      \includegraphics[height=0.8125in]{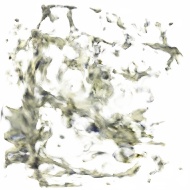}&
      \includegraphics[height=0.8125in]{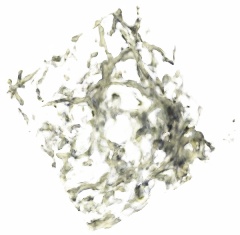}\\
      \raisebox{0.15in}{\rotatebox{90}{ViSNeRF}}&
      \includegraphics[height=0.8125in]{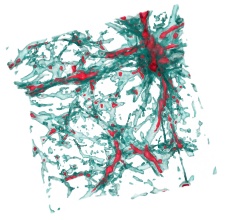}&
      \includegraphics[height=0.8125in]{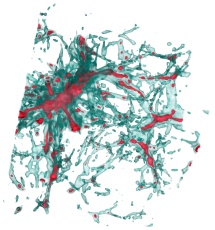}&
      \includegraphics[height=0.8125in]{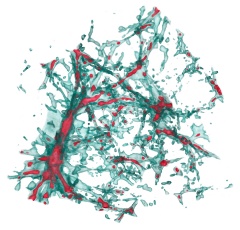}&
      \includegraphics[height=0.8125in]{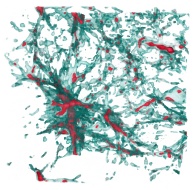}&
      \includegraphics[height=0.8125in]{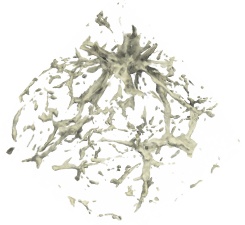}&
      \includegraphics[height=0.8125in]{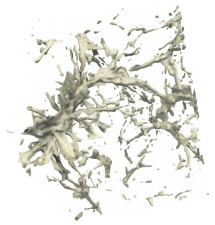}&
      \includegraphics[height=0.8125in]{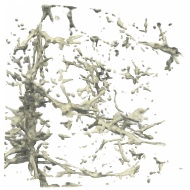}&
      \includegraphics[height=0.8125in]{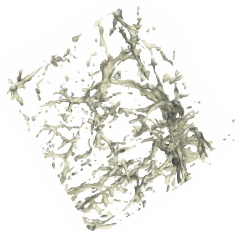}\\
      \raisebox{0.35in}{\rotatebox{90}{GT}}&
      \includegraphics[height=0.8125in]{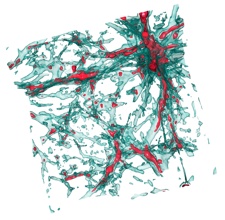}&
      \includegraphics[height=0.8125in]{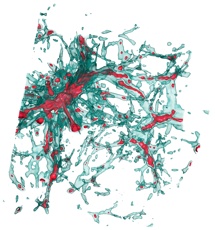}&
      \includegraphics[height=0.8125in]{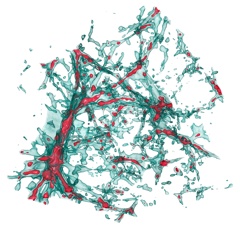}&
      \includegraphics[height=0.8125in]{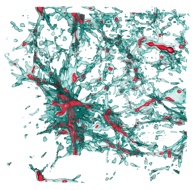}&
      \includegraphics[height=0.8125in]{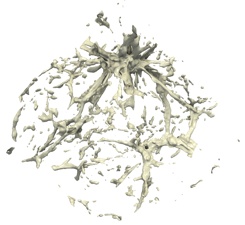}&
      \includegraphics[height=0.8125in]{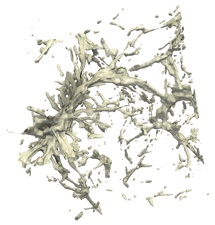}&
      \includegraphics[height=0.8125in]{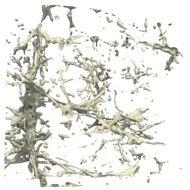}&
      \includegraphics[height=0.8125in]{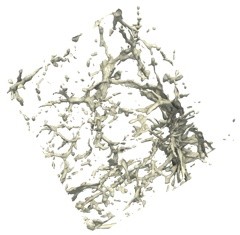}\\    
      &\mbox{\footnotesize (a)} & \mbox{\footnotesize (b)} & \mbox{\footnotesize (c)} & \mbox{\footnotesize (d)} & 
      \mbox{\footnotesize (e)} & \mbox{\footnotesize (f)} & \mbox{\footnotesize (g)} & \mbox{\footnotesize (h)}
    \end{array}$
 \end{center}
\vspace{-.25in} 
 \caption{
 Inferred images under different views for parameter-space exploration of dynamic scenes (simulation parameters).
 Inferred DVR (a to d) and IR (e to h) images with interpolation over simulation parameters using the Nyx dataset. 
 (a), (b), (c), (e), (f), and (g) are single-parameter variation results. 
 (d) and (h) are multiple-parameter variation results. 
 Table~\ref{tab:comp-nyx-params} gives each case's simulation parameter values. 
 } 
 \label{fig:comp-nyx}
\end{figure*}

\begin{figure}[htb]
  \begin{center}
  $\begin{array}{c@{\hspace{0.005in}}c@{\hspace{0.005in}}c@{\hspace{0.005in}}c@{\hspace{0.005in}}c@{\hspace{0.005in}}c@{\hspace{0.005in}}c@{\hspace{0.005in}}c@{\hspace{0.005in}}c}
    \raisebox{0.06in}{\rotatebox{90}{\tiny InSituNet}}&
    \includegraphics[height=0.38in]{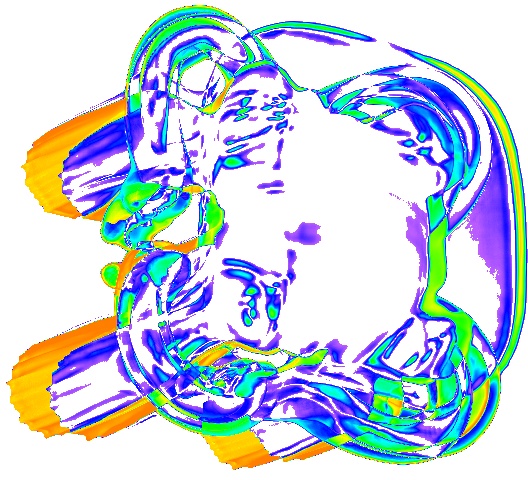}&
    \includegraphics[height=0.38in]{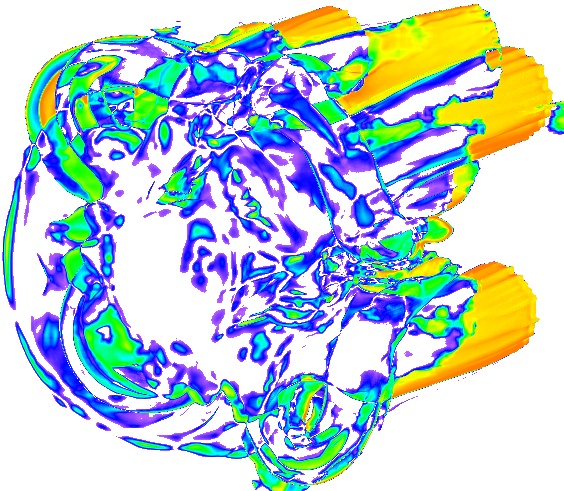}&
    \includegraphics[height=0.38in]{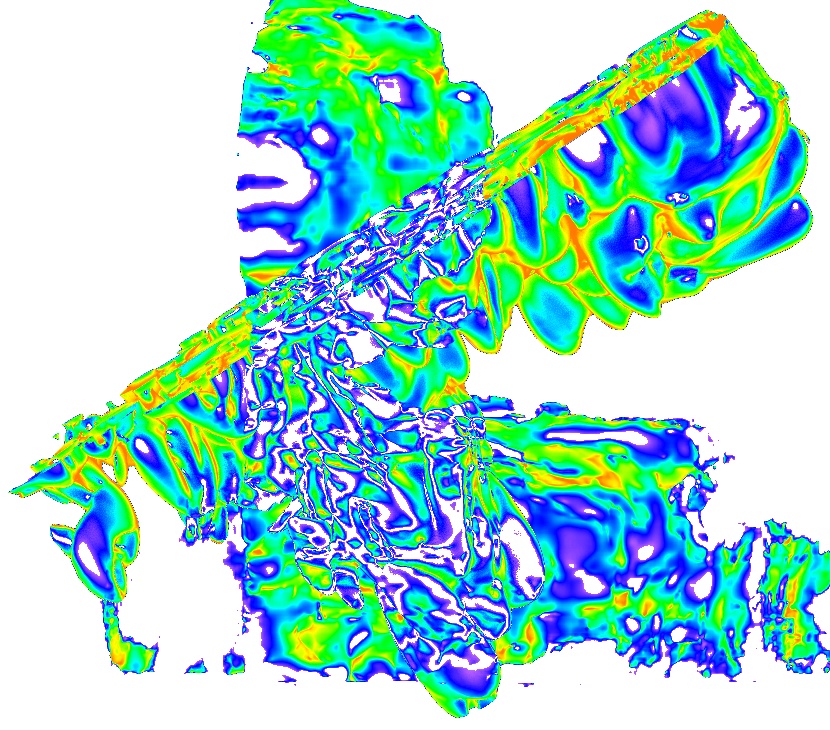}&
    \includegraphics[height=0.38in]{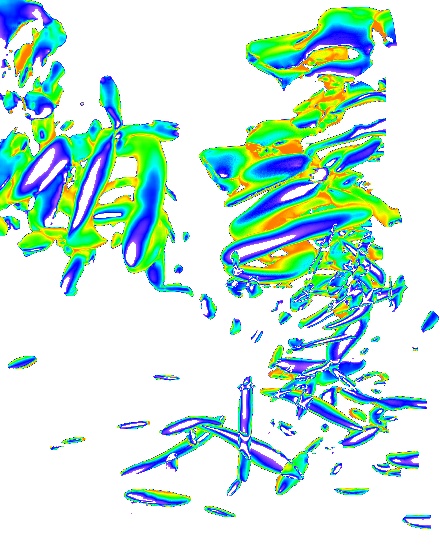}&
    \includegraphics[height=0.38in]{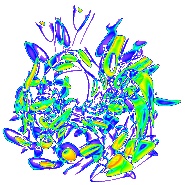}&
    \includegraphics[height=0.38in]{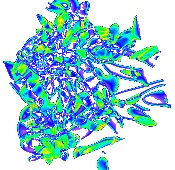}&
    \includegraphics[height=0.38in]{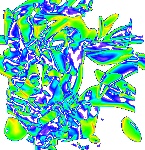}&
    \includegraphics[height=0.38in]{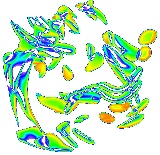}\\
    \raisebox{0.06in}{\rotatebox{90}{\tiny CoordNet}}&
    \includegraphics[height=0.38in]{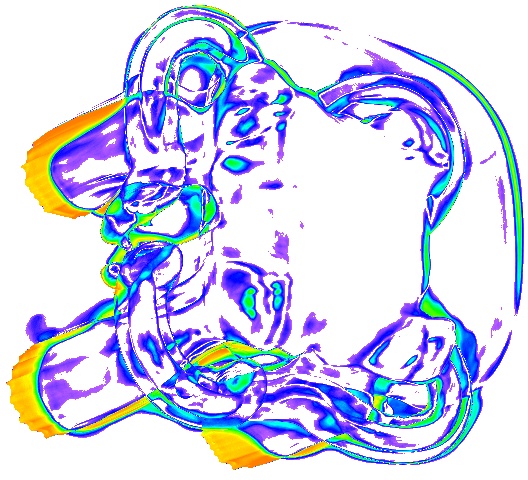}&
    \includegraphics[height=0.38in]{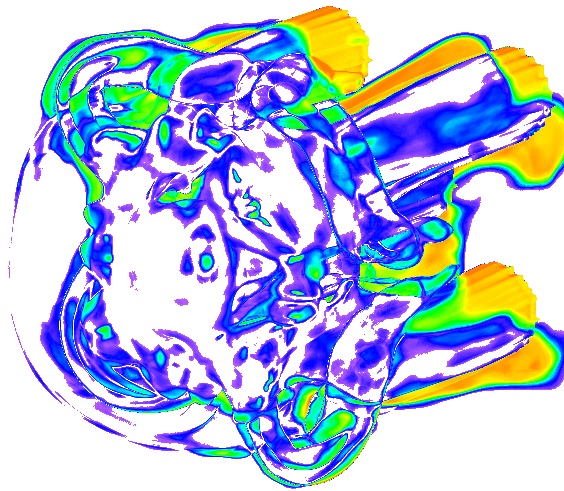}&
    \includegraphics[height=0.38in]{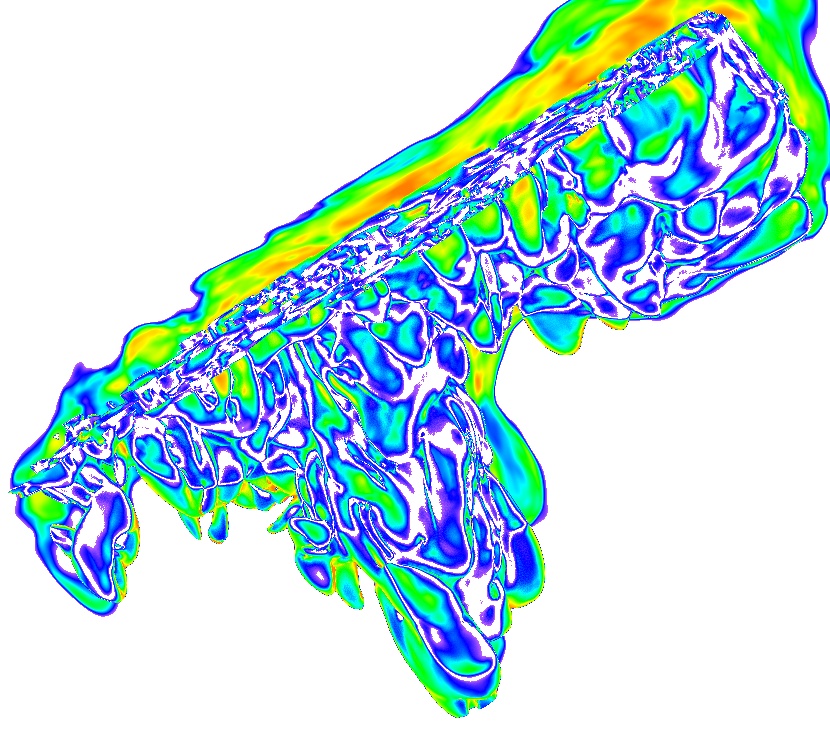}&
    \includegraphics[height=0.38in]{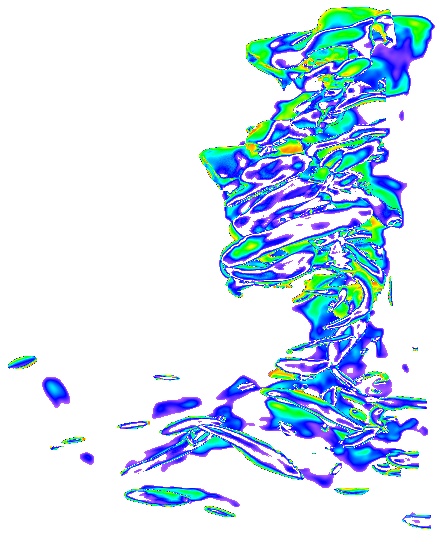}&    
    \includegraphics[height=0.38in]{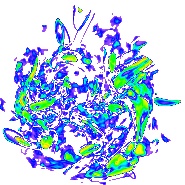}&
    \includegraphics[height=0.38in]{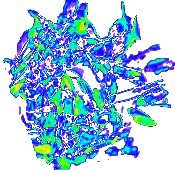}&
    \includegraphics[height=0.38in]{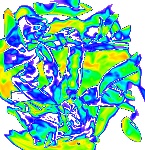}&
    \includegraphics[height=0.38in]{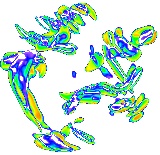}\\
    \raisebox{0.02in}{\rotatebox{90}{\tiny StyleGAN2}}&
    \includegraphics[height=0.38in]{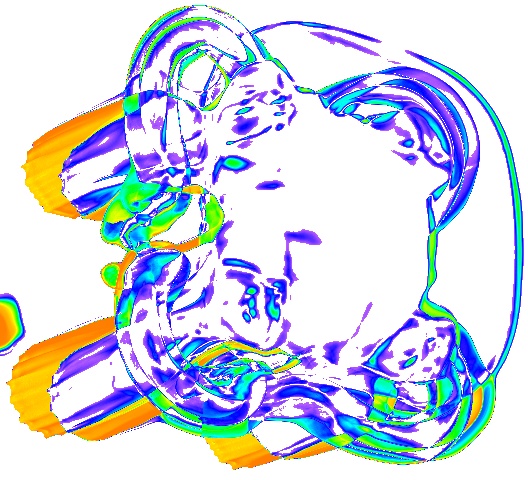}&
    \includegraphics[height=0.38in]{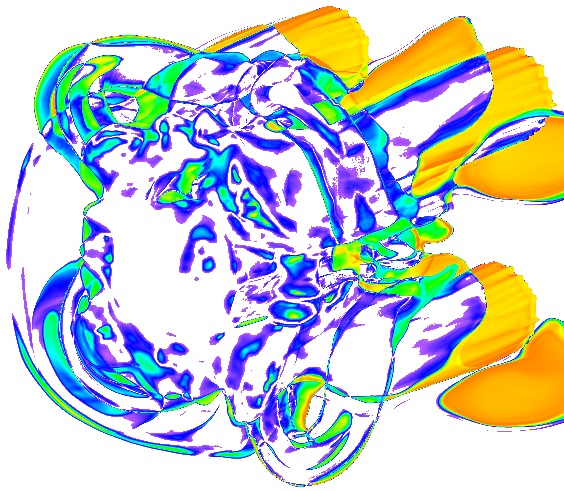}&
    \includegraphics[height=0.38in]{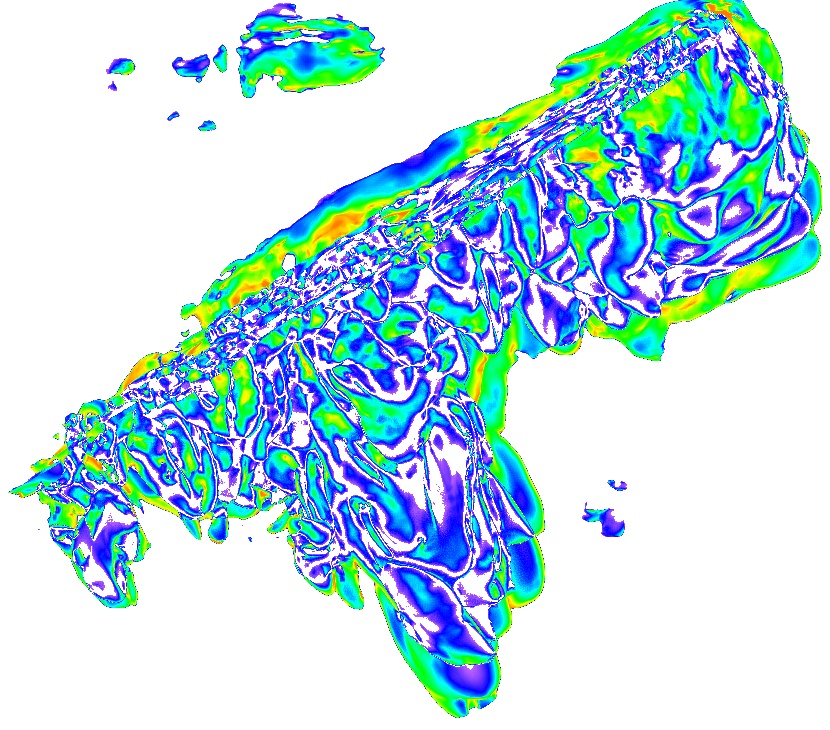}&
    \includegraphics[height=0.38in]{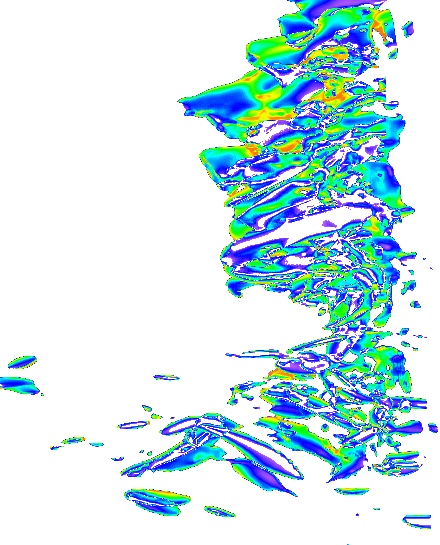}&
    \includegraphics[height=0.38in]{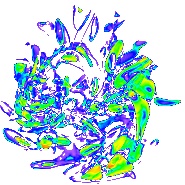}&
    \includegraphics[height=0.38in]{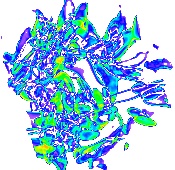}&
    \includegraphics[height=0.38in]{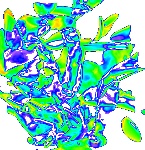}&
    \includegraphics[height=0.38in]{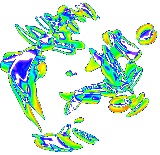}\\
    \raisebox{0.09in}{\rotatebox{90}{\tiny EG3D}}&
    \includegraphics[height=0.38in]{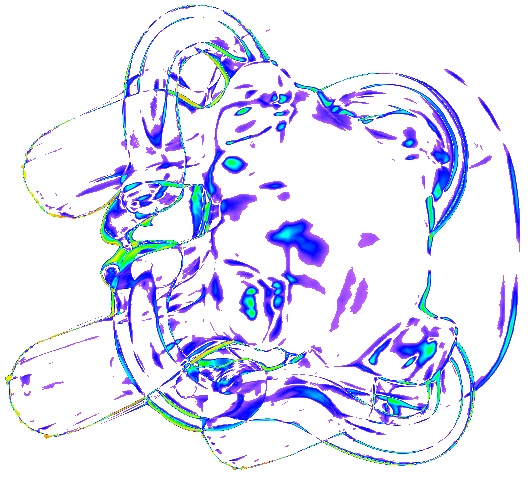}&
    \includegraphics[height=0.38in]{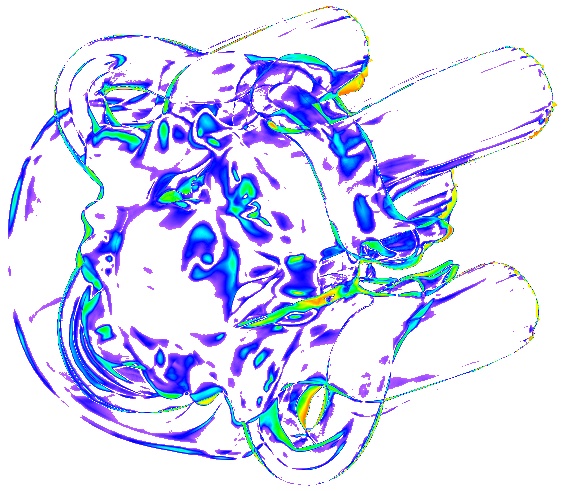}&
    \includegraphics[height=0.38in]{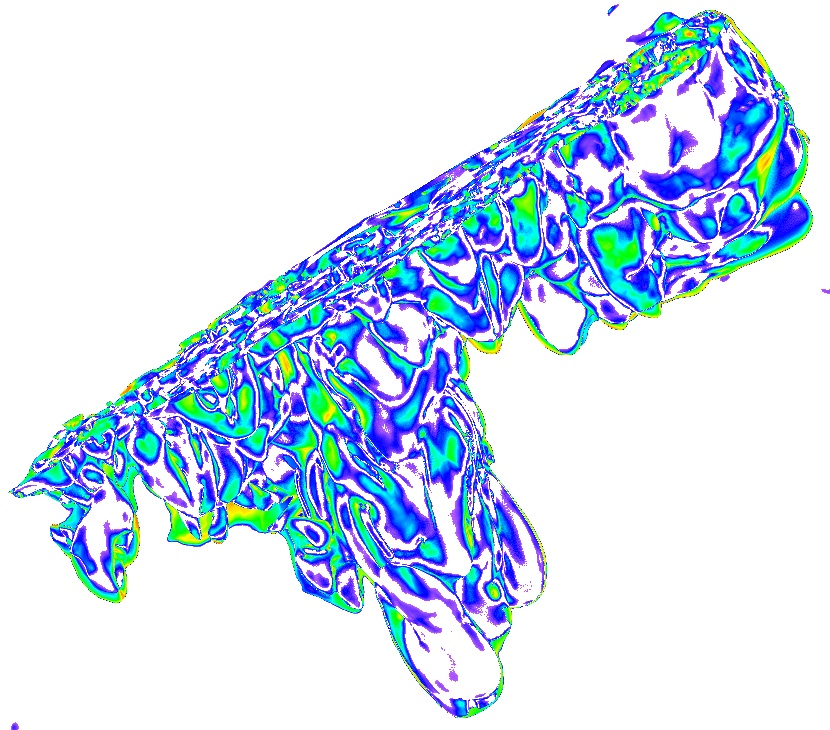}&
    \includegraphics[height=0.38in]{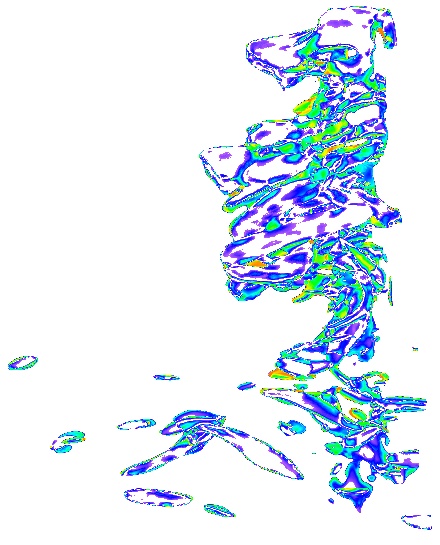}&
    \includegraphics[height=0.38in]{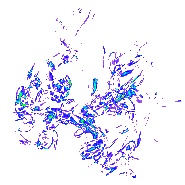}&
    \includegraphics[height=0.38in]{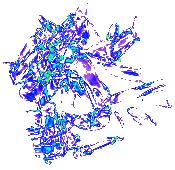}&
    \includegraphics[height=0.38in]{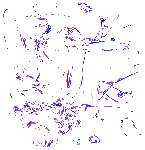}&
    \includegraphics[height=0.38in]{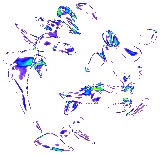}\\
    \raisebox{0.03in}{\rotatebox{90}{\tiny ViSNeRF}}&
    \includegraphics[height=0.38in]{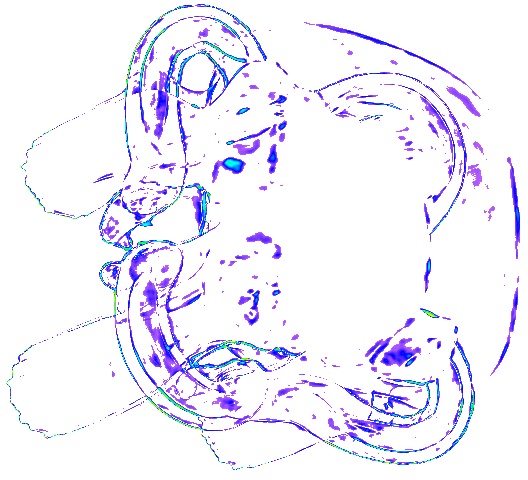}&
    \includegraphics[height=0.38in]{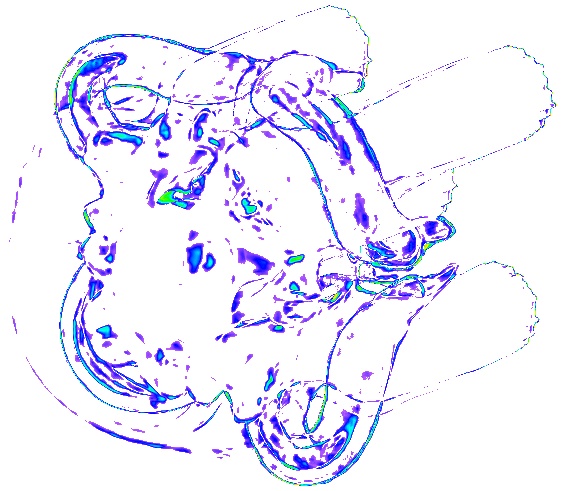}&
    \includegraphics[height=0.38in]{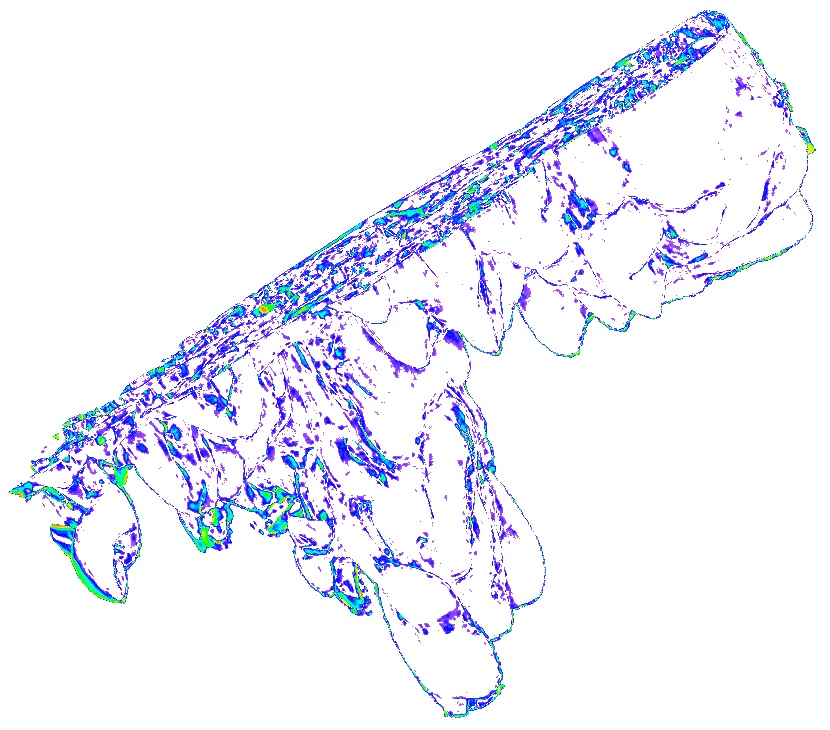}&
    \includegraphics[height=0.38in]{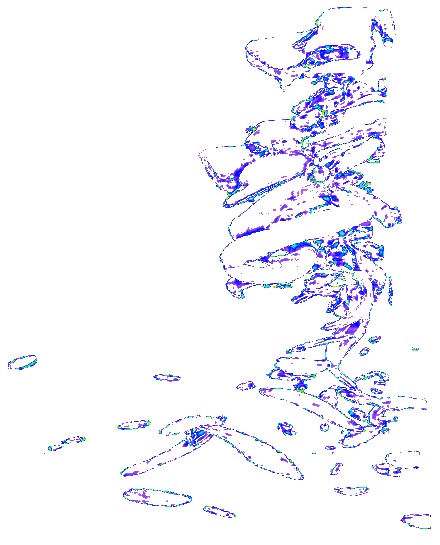}&
    \includegraphics[height=0.38in]{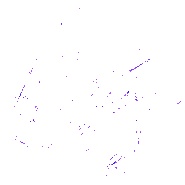}&
    \includegraphics[height=0.38in]{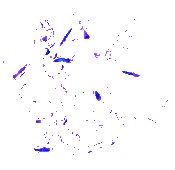}&
    \includegraphics[height=0.38in]{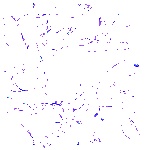}&
    \includegraphics[height=0.38in]{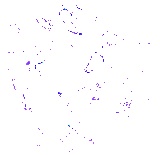}\\
    \raisebox{0.12in}{\rotatebox{90}{\tiny GT}}&
    \includegraphics[height=0.38in]{images/baselines/fivejets_time/GT-070-crop.png}&
    \includegraphics[height=0.38in]{images/baselines/fivejets_time/GT-112-crop.png}&
    \includegraphics[height=0.38in]{images/baselines/tangaroa_isoval/GT-016-crop.png}&
    \includegraphics[height=0.38in]{images/baselines/tangaroa_isoval/GT-134-crop.png}&
    \includegraphics[height=0.38in]{images/baselines/vortex_tf1/GT-030-crop.png}&
    \includegraphics[height=0.38in]{images/baselines/vortex_tf1/GT-170-crop.png}&
    \includegraphics[height=0.38in]{images/baselines/vortex_tf2/GT-000-crop.png}&
    \includegraphics[height=0.38in]{images/baselines/vortex_tf2/GT-113-crop.png}\\    
    &\mbox{\footnotesize (a)} & \mbox{\footnotesize (b)} & \mbox{\footnotesize (c)} & \mbox{\footnotesize (d)} & 
    \mbox{\footnotesize (e)} & \mbox{\footnotesize (f)} & \mbox{\footnotesize (g)} & \mbox{\footnotesize (h)}
  \end{array}$
 \end{center}
\vspace{-.25in} 
 \caption{Difference images for Figure~\ref{fig:comp-pse} of inferred images under different views. 
 (a) and (b) DVR images with interpolation over timesteps using the five jets dataset. 
 (c) and (d) IR images with interpolation over isovalues using the Tangaroa dataset.
 (e) to (h) DVR images with interpolation over TFs using the vortex dataset (TF-1: (e) and (f); TF-2: (g) and (h)).
 } 
 \label{fig:comp-pse-diff}
\end{figure}

\begin{figure}[htb]
  \begin{center}
  $\begin{array}{c@{\hspace{0.005in}}c@{\hspace{0.005in}}c@{\hspace{0.005in}}c@{\hspace{0.005in}}c@{\hspace{0.005in}}c@{\hspace{0.005in}}c@{\hspace{0.005in}}c@{\hspace{0.005in}}c}
    \raisebox{0.06in}{\rotatebox{90}{\tiny InSituNet}}&
    \includegraphics[height=0.38in]{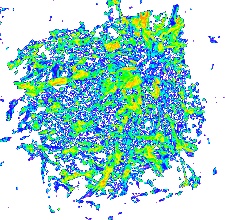}&
    \includegraphics[height=0.38in]{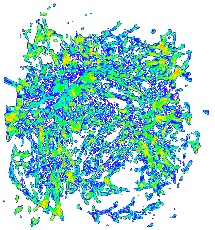}&
    \includegraphics[height=0.38in]{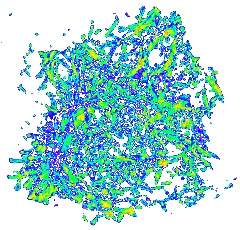}&
    \includegraphics[height=0.38in]{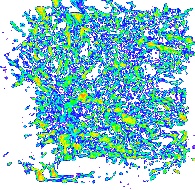}&
    \includegraphics[height=0.38in]{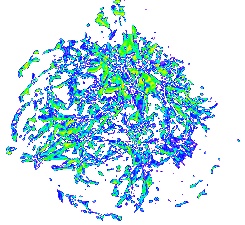}&
    \includegraphics[height=0.38in]{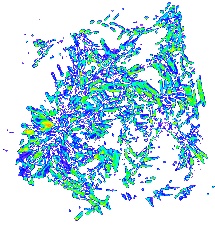}&
    \includegraphics[height=0.38in]{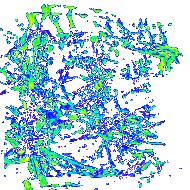}&
    \includegraphics[height=0.38in]{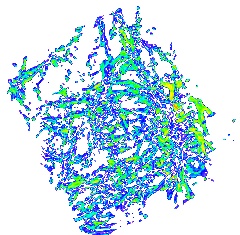}\\
    \raisebox{0.06in}{\rotatebox{90}{\tiny CoordNet}}&
    \includegraphics[height=0.38in]{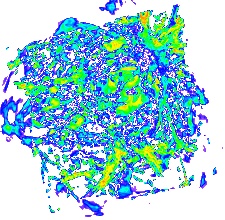}&
    \includegraphics[height=0.38in]{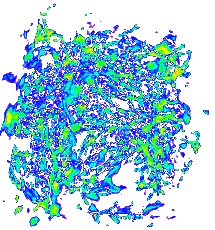}&
    \includegraphics[height=0.38in]{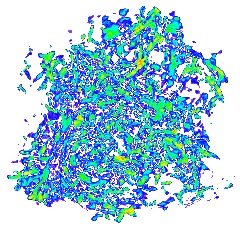}&
    \includegraphics[height=0.38in]{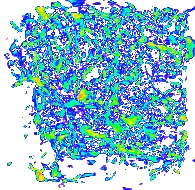}&    
    \includegraphics[height=0.38in]{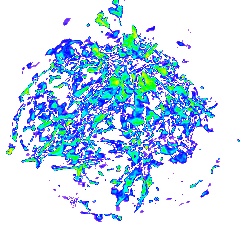}&
    \includegraphics[height=0.38in]{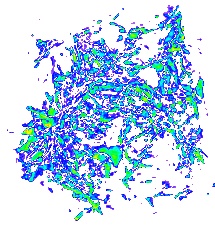}&
    \includegraphics[height=0.38in]{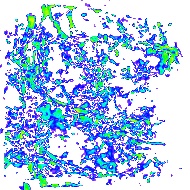}&
    \includegraphics[height=0.38in]{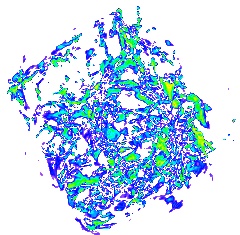}\\
    \raisebox{0.02in}{\rotatebox{90}{\tiny StyleGAN2}}&
    \includegraphics[height=0.38in]{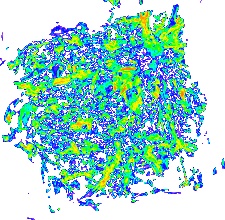}&
    \includegraphics[height=0.38in]{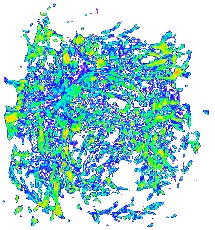}&
    \includegraphics[height=0.38in]{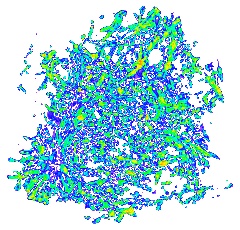}&
    \includegraphics[height=0.38in]{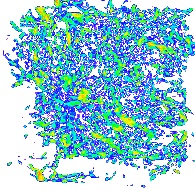}&
    \includegraphics[height=0.38in]{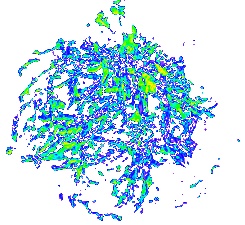}&
    \includegraphics[height=0.38in]{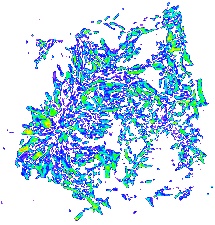}&
    \includegraphics[height=0.38in]{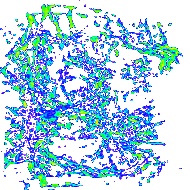}&
    \includegraphics[height=0.38in]{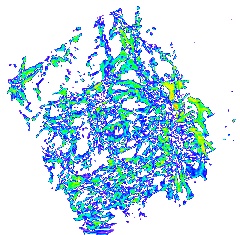}\\
    \raisebox{0.09in}{\rotatebox{90}{\tiny EG3D}}&
    \includegraphics[height=0.38in]{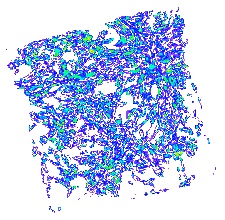}&
    \includegraphics[height=0.38in]{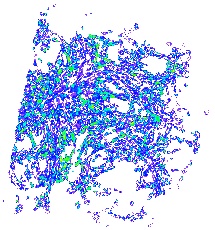}&
    \includegraphics[height=0.38in]{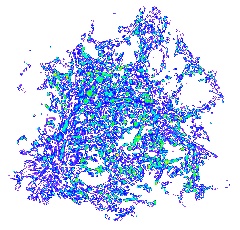}&
    \includegraphics[height=0.38in]{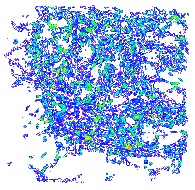}&
    \includegraphics[height=0.38in]{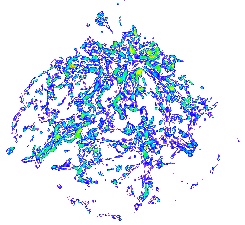}&
    \includegraphics[height=0.38in]{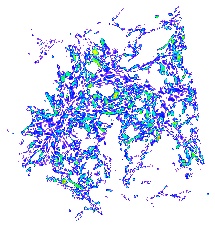}&
    \includegraphics[height=0.38in]{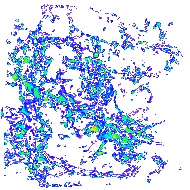}&
    \includegraphics[height=0.38in]{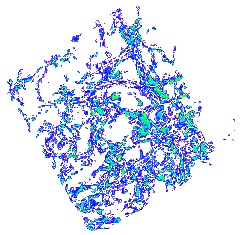}\\
    \raisebox{0.03in}{\rotatebox{90}{\tiny ViSNeRF}}&
    \includegraphics[height=0.38in]{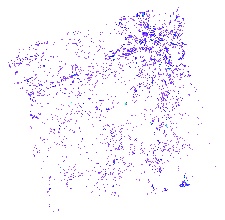}&
    \includegraphics[height=0.38in]{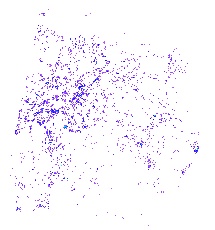}&
    \includegraphics[height=0.38in]{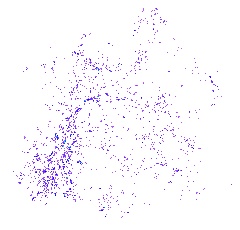}&
    \includegraphics[height=0.38in]{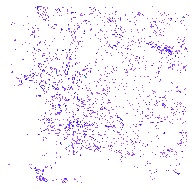}&
    \includegraphics[height=0.38in]{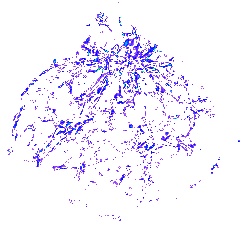}&
    \includegraphics[height=0.38in]{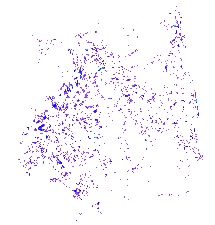}&
    \includegraphics[height=0.38in]{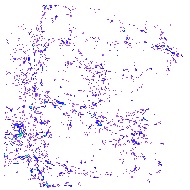}&
    \includegraphics[height=0.38in]{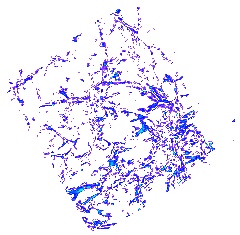}\\
    \raisebox{0.12in}{\rotatebox{90}{\tiny GT}}&
    \includegraphics[height=0.38in]{images/baselines/nyx_vr/GT-007-crop.png}&
    \includegraphics[height=0.38in]{images/baselines/nyx_vr/GT-038-crop.png}&
    \includegraphics[height=0.38in]{images/baselines/nyx_vr/GT-064-crop.png}&
    \includegraphics[height=0.38in]{images/baselines/nyx_vr/GT-096-crop.png}&
    \includegraphics[height=0.38in]{images/baselines/nyx_ir/GT-022-crop.png}&
    \includegraphics[height=0.38in]{images/baselines/nyx_ir/GT-052-crop.png}&
    \includegraphics[height=0.38in]{images/baselines/nyx_ir/GT-085-crop.png}&
    \includegraphics[height=0.38in]{images/baselines/nyx_ir/GT-164-crop.png}\\      
    & \mbox{\footnotesize (a)} & \mbox{\footnotesize (b)} & \mbox{\footnotesize (c)} & \mbox{\footnotesize (d)} & 
    \mbox{\footnotesize (e)} & \mbox{\footnotesize (f)} & \mbox{\footnotesize (g)} & \mbox{\footnotesize (h)}
  \end{array}$
 \end{center}
\vspace{-.25in} 
 \caption{
 Difference images for Figure~\ref{fig:comp-nyx} of inferred DVR (a to d) and IR (e to h) images with interpolation over simulation parameters using the Nyx dataset. 
 (a), (b), (c), (e), (f), and (g) are single-parameter variation results. 
 (d) and (h) are multiple-parameter variation results. 
 } 
 \label{fig:comp-nyx-diff}
\end{figure}

\begin{table}[htb]
  \caption{\prevhot{Simulation parameter values for the Nyx dataset shown in Figure~\ref{fig:comp-nyx}.}}
  \vspace{-0.1in}
  \centering
  \prevhot{
  {\scriptsize
  \begin{tabular}{c|ccc}

  subfigure & \hot{$OmM$} & \hot{$OmB$} & $h$ \\ \hline
  (a) & 0.128448 & 0.0215  & 0.55  \\
  (b) & 0.155 & 0.022052 & 0.55  \\
  (c) & 0.155 & 0.0235 & 0.591379 \\
  (d) & 0.150227 & 0.023227 & 0.809091 \\
  (e) & 0.146552 & 0.0215 & 0.55 \\
  (f) & 0.155 & 0.023017 & 0.55 \\
  (g) & 0.155 & 0.0235 & 0.808621 \\
  (h) & 0.142273 & 0.022773 & 0.740909
  \end{tabular}
  }}
  \label{tab:comp-nyx-params}
\end{table}

\hot{
\vspace{-0.15in}
\subsection{Network Training}
\label{subsec:nettrain}

We use PyTorch to implement ViSNeRF and train it using a single NVIDIA Tesla V100 GPU with 32 GB of video memory.
We set the number of low-rank components in the factorization of the 3D tensor $\mathcal{T}^3$ as $R_s=64$, where it is split to $R_\sigma=16$ for density and $R_c=48$ for color to represent the radiance field. 
Hence, the depths of density and appearance features are 16 and 48, respectively.
For the factorization of the $K$-D parameter space, $\mathcal{T}^K$, we set $R_p=4$. 
The length of each parameter vector corresponds to the number of distinct values available for each parameter in the training set.
The resolution of the spatial feature grid is initialized as $N \times N \times N$, where $N=128$.
For more complex datasets like Tangaroa, the grid resolution would increase at iterations 2,000, 4,000, 6,000, and 8,000, and the final grid has $300^3$ voxels. 
Note that the grid does not necessarily stay cubic, 
meaning the length, width, and height can differ depending on the shape of the bounding box. 

The number of cells in the parameter feature grid $M^K$ corresponds to the number of distinct values available for each parameter in the training set. Refer to Table~\ref{tab:pse-dataset}, for five jets (timestep) and vortex (TF-2), $K=1$ and $M^1=11$. 
For Tangaroa (isovalue), $K=1$ and $M^1=13$. 
For vortex (TF-1), $K=2$ and $M^2=36$ ($6 \times 6$). 
For Nyx-DVR/IR (simulation parameters), $K=3$ and $M^3=45$ ($3 \times 3 \times 5$).
The model size and training time should grow as the feature depth ($R_\sigma$, $R_c$, and $R_p$) or the feature grid size ($N^3$ and $M^K$) increases.

For all static scenes, we train ViSNeRF for 30,000 iterations, while the number of iterations is tripled for dynamic scenes in all parameter-space exploration experiments. 
At iteration 2,000, we create the mask volume for storing empty voxel information and shrink the bounding box according to the actual scene content.
At iteration 4,000, we update the mask volume and enable voxel skipping. 

The MLP decoders for density and appearance have one hidden layer, and each fully connected layer has 128 channels.
We set the batch size as 4096, considering the training efficiency and video memory consumption.
The number of sample points per ray starts at 192 and increases to 512 if the grid resolution increases to $300^3$ voxels.
The matrices and vectors and the two MLPs are optimized by an Adam optimizer with $\beta_1=0.9$ and $\beta_2=0.99$.
We set the initial learning rate of the matrices and vectors as 0.02 and that of the MLPs as 0.001. 
The weight of L1 loss ($\lambda_1$) is set as $10^{-4}$ initially and changed to $10^{-5}$ after 2,000 iterations. 
The weights of TV loss ($\lambda_2$) for density and appearance features are set as 1.0.
}

\vspace{-0.05in}
\subsection{Baselines}

For dynamic volumetric scenes, we compare ViSNeRF with methods that natively support the synthesis of visualizations, including InSituNet, CoordNet, StyleGAN2, and EG3D:
\begin{myitemize}
\vspace{-0.05in}
  \item InSituNet~\cite{He-InSituNet-TVCG20} is a GAN-based surrogate model supporting parameter-space exploration.
  We only use azimuth and elevation as the input and add extra upscaling residual blocks in the regressor to support output resolutions larger than 256$\times$256. 
  \item CoordNet~\cite{Han-CoordNet-TVCG} uses an INR architecture for visualization generation. We use the original implementation as it can handle any image resolution. 
  \item StyleGAN2~\cite{Karras-StyleGAN2-CVPR20} uses a style-based architecture with a discriminator to provide adversarial supervision.
  We use azimuth and elevation instead of a random latent vector to define the targets explicitly.
  \item EG3D~\cite{Chan-EG3D-CVPR22} can be seen as a style-based NeRF where a StyleGAN2 generator produces the triplane representation. 
\vspace{-0.05in}
\end{myitemize}

We use StyleGAN2 instead of StyleGAN3~\cite{Karras-StyleGAN3-NeurIPS21} due to two reasons: 
(1) In the context of visualization generation, where the emphasis is faithful reconstruction, the new features of StyleGAN3, such as translation and rotation equivariance, hardly improve the quality but significantly increase the training difficulty. 
(2) As we also incorporate EG3D with the StyleGAN2 backbone in our comparison, we aim to closely evaluate and compare the two approaches and illustrate the transition from a 2D approach to a 3D method. 


\vspace{-0.05in}
\subsection{\hot{Results}}

The quantitative results are displayed in Table~\ref{tab:psnr-lpips-pse}. 
To evaluate the quality of generated visualizations, we employ three metrics: {\em peak signal-to-noise ratio} (PSNR), {\em structural similarity index} (SSIM)~\cite{Wang-TIP04}, and {\em learned perceptual image patch similarity} (LPIPS)~\cite{Zhang-CVPR18}. 
\hot{Overall, ViSNeRF significantly outperforms the four baseline methods across all three metrics, EG3D comes a distant second, and the rest of the three methods are the worst.}

\hot{InSituNet, CoordNet, and StyleGAN2 perform poorly because they depend heavily on large training sets. 
With a small number of training images, 
as in our experiments, these methods struggle to produce desirable results. Moreover, they require far more scene frames to capture the dynamic scenes adequately.
For example, under identical Nyx simulation settings, He et al.\ \cite{He-InSituNet-TVCG20} used 100 images per scene frame and 400 frames, resulting in 40,000 training images for InSituNet. In contrast, ViSNeRF produces accurate results with 1,890 images, reducing the training data by 95\%. This highlights ViSNeRF's robustness and efficiency, offering a significant advantage in dynamic scene reconstruction, especially as the demand for training data increases with growing dynamic ranges.}

\hot{EG3D, despite being 3D-aware, produces less accurate results due to its reliance on convolutional layers to upscale low-resolution synthesized images. While this approach helps address GPU memory consumption issues, it fails to leverage the advantages of 3D-aware reconstruction fully and, like 2D-based methods, depends on a large number of training images to achieve satisfactory results.}

For model size, CoordNet wins as the fully implicit representation, ViSNeRF comes second as the hybrid neural representation, and the rest of the three methods incur much larger model sizes. 
ViSNeRF is the fastest for training time, followed by StyleGAN2 and InSituNet, and CoordNet and EG3D are the slowest.  
InSituNet and StyleGAN2 are the fastest for inference time, followed by EG3D and CoordNet, and ViSNeRF is the slowest. 
Nevertheless, for 256$\times$256 images, the frame rate is still at least 1.86 FPS for ViSNeRF. 
For 1024$\times$1024 images, the frame rate drops substantially. 


For qualitative results, the synthesized visualization images are presented in Figure~\ref{fig:comp-pse} and Figure~\ref{fig:comp-nyx}.
Figure~\ref{fig:comp-pse-diff} and Figure~\ref{fig:comp-nyx-diff} show the corresponding difference images.
\hot{In the following, we examine the qualitative results on a case-by-case basis.}

{\bf Interpolation over timesteps.}
We use 1,000 timesteps of the five jets dataset for temporal interpolation. 
We pick every 100th timestep for training, resulting in 11 samples (including timesteps 1 and 1,000). 
To test the performance of ViSNeRF, we infer 181 images evenly from timestep 1 to timestep 1,000 in a full 360-degree view. 
The inferred images are used as the test set for evaluation.
Figure~\ref{fig:comp-pse} (a) and (b) show that CoordNet, StyleGAN2, and EG3D generate blurry results. InSituNet produces results with better clarity, but the reconstructed viewpoints deviate from GT, leading to the worst quantitative performance (refer to Table~\ref{tab:psnr-lpips-pse}). The results of ViSNeRF are the best. 

{\bf Interpolation over isovalues.}
We select an isovalue range [0.01, 0.3] for the Tangaroa dataset, producing meaningful IR occupying the resulting images well. 
Within the isovalue range, we evenly pick 13 samples (including the two ends) for training and 181 samples in a full 360-degree view for inference.
Figure~\ref{fig:comp-pse} (c) and (d) show that the reconstructed viewpoints from InSituNet are completely wrong, and ViSNeRF produces results with the best detail and clarity compared with CoordNet, StyleGAN2, and EG3D. 

{\bf Interpolation over TFs.}
In this experiment, we provide two ways to interpolate TFs, denoted as TF-1 and TF-2, using the vortex dataset.
For TF-1, we interpolate the opacity for each visible value range while keeping the ranges intact. 
As shown in Figure~\ref{fig:tf-illustration} (a), we control the opacities of the red and blue value ranges. 
We sample 36 steps, including the cases where both red and blue are opaque, one is opaque and the other is transparent, and both are transparent. 
For TF-2, we interpolate from a visible value range (source) to another (target) using 11 steps. This is achieved by translating the source while updating the bound and opacity to match those of the target. 
As shown in Figure~\ref{fig:tf-illustration} (b), the blue range is shifted from low values to match the red range at high values. The interpolation will show intermediate ranges with interpolated colors (such as purple). 
In either case (TF-1 or TF-2), we infer 181 images evenly from the corresponding interpolation scheme. 
For TF-1, Figure~\ref{fig:comp-pse} (e) and (f) show that ViSNeRF is the best, followed by EG3D. CoordNet and StyleGAN2 produce rather blurry results, and InSituNet yields significant viewpoint deviations. Similar conclusions can be drawn for TF-2 from Figure~\ref{fig:comp-pse} (g) and (h). 

\prevhot{{\bf Interpolation over simulation parameters.}
For this study, we use an additional dataset, Nyx, derived from a cosmological simulation software developed by Almgren et al.\ \cite{Nyx}. 
Our investigation focuses on three critical parameters selected based on expert recommendations. 
These parameters and their respective ranges, inspired by the methodology used in InSituNet, include 
the total matter density $(OmM\in[0.12,0.155])$, 
the total density of baryons $(OmB\in[0.0215,0.235])$, and 
the Hubble constant $(h\in[0.55,0.85])$. 
For the training phase, we selectively use 45 parameter configurations derived from a combination of 
three values of \hot{$OmM$} (0.12, 0.1375, 0.155), 
three values of \hot{$OmB$} (0.0215, 0.0225, 0.0235), and 
five values of $h$ (0.55, 0.625, 0.70, 0.775, 0.85).
The number of intervals chosen for each parameter is determined by its relative impact on the simulation outcome.

During inference, we analyze 181 rendered images, categorized as follows:
(1) single-parameter variation: 90 images for isolated changes in each parameter (\hot{$OmM$, $OmB$}, or $h$), along with 30 different views per parameter. 
\hot{This kind of variation results in 90 simulation parameter configurations.} 
(2) multiple-parameter variation: 91 images with concurrent changes in all three parameters (\hot{$OmM$, $OmB$}, and $h$) \hot{back and forth}, each with a different view, testing complex interpolation conditions. 
\hot{This kind of variation results in 46 simulation parameter configurations. 
As five simulation parameter configurations are the same across these two kinds of variation, we end up with $90+46-5=131$ simulation parameter configurations for inference.}

Figure~\ref{fig:comp-nyx} shows the results obtained using different variations on the Nyx dataset, encompassing both DVR and IR. The comparison highlights that ViSNeRF produces the highest quality and most accurate results. In contrast, InSituNet fails to generate correct views, whereas CoordNet, StyleGAN2, and EG3D tend to produce blurry outputs. The superior performance of ViSNeRF in simulation parameter interpolation is further confirmed by quantitative results given in Table~\ref{tab:psnr-lpips-pse}.}

\vspace{-0.05in}
\subsection{Limitations and Future Work}

Our ViSNeRF framework has the following limitations, which we will improve in future work. 
First, we use an icosphere to evenly select sampled viewpoints and choose 42 samples for novel view synthesis. 
This is not a tight lower bound, as the subdivision level determines the number of samples, which increases significantly with each subdivision level. 
An alternative solution, such as Poisson disk sampling, provides a good balance between flexibility and control over the density of camera viewpoints on a sphere's surface, even though the random sampling process may not guarantee complete uniformity. 

Second, ViSNeRF incorporates both DVR and IR from volume visualization. 
For IR, there are better ways to synthesize rendering images for isosurfaces than the underlying neural volumetric representation. 
A more accurate representation should consider implicit surfaces such as signed or unsigned distance functions~\cite{Yao-VISSP24}. 

Third, as indicated in Table~\ref{tab:psnr-lpips-pse}, ViSNeRF renders images much slower than InSituNet, which poses a challenge for real-time parameter-space exploration. 
A potential solution 
is to adopt 3DGS in the NeRF representation to improve rendering speed.
\hot{Unfortunately, state-of-the-art Gaussian splatting methods for dynamic scenes~\cite{Wu-CVPR24, Huang-CVPR24, Yang-CVPR24, Huang-MICCAI24} are based on deforming a fixed set of 3D Gaussians. This works well for real-world scenarios where motion, like human movement, is the main source of dynamics. However, it creates major problems in cases where 
different parts of a scene could move, merge, split, appear, or disappear simultaneously, 
which is common in scientific visualization.
When parts appear or disappear, these methods move Gaussians from nearby components with similar colors instead of adding or removing them locally. This leads to visible artifacts and reduces the reconstruction quality of interpolated scene frames. Adding more Gaussians can help by making them smaller and less noticeable when moving, but such a practice requires much more GPU memory. Thus, in the context of scientific visualization, there is a clear need for more efficient and effective methods to handle Gaussians in dynamic scenes to achieve real-time parameter-space exploration.}

\hot{Finally, fully implementing ViSNeRF within the CUDA framework could speed up rendering.
However, it may present a barrier for researchers aiming to extend ViSNeRF to other scientific visualization tasks, such as segmentation or language embedding.}

\vspace{-0.05in}
\section{Conclusions}
 
We have presented ViSNeRF, a novel solution for \prevhot{dynamic} visualization synthesis using NeRF. 
Compared with representative NeRF methods, ViSNeRF achieves better quality across PSNR, SSIM, and LPIPS using the same number of training views.
Furthermore, ViSNeRF excels in static \hot{scene} reconstruction and showcases its scalability and efficiency in handling dynamic \hot{scenes}, thanks to its novel factorization strategy.
For parameter-space exploration tasks, ViSNeRF also outperforms other deep learning-based methods, including InSituNet, CoordNet, StyleGAN2, and EG3D, in terms of visualization generation quality. 
Its superior performance suggests that ViSNeRF is a well-designed framework for the visualization synthesis of dynamic volumetric scenes.

\vspace{-0.05in}
\acknowledgments{
This research was supported in part by the U.S.\ National Science Foundation through grants IIS-1955395, IIS-2101696, OAC-2104158, and IIS-2401144, and the U.S.\ Department of Energy through grant DESC0023145. The authors would like to thank the anonymous reviewers for their insightful comments.
}


\setcounter{section}{0}
\setcounter{figure}{0}
\setcounter{table}{0}

\vspace{-0.05in}
\section*{Appendix}
\label{sec:appendix}

\section{Optimization Schemes}
\label{subsec:rayopt}

{\bf Coarse-to-fine feature grid.}
To reduce the training cost at the initial stage and regularize the matrices and vectors, like TensoRF and HexPlane, we apply a coarse-to-fine scheme to the 3D feature grid. 
The grid is initialized with a coarse resolution. 
After the initial stage, 
the matrices and vectors are bi-linearly and linearly upscaled to higher resolution multiple times in subsequent iterations. 
The final fine resolution 
depends on the complexity of the visual content in the scene (i.e., input DVR or IR images).

\prevhot{
{\bf Two-phase progressive sampling and importance sampling.}
The training is divided into two phases to optimize efficiency: {\em warm-up} and {\em finetuning}. During the warm-up phase, the sample points per ray begin at a small number, such as 64, and increase steadily until they reach the predetermined target number at the phase's conclusion. For the finetuning phase, the number of samples per ray matches the final count of the warm-up phase, plus an additional small number of sample points, which is 64 in our case. These points are sampled using an importance sampling technique akin to NeRF. This two-phase method not only accelerates convergence at the outset but also minimizes early-stage overfitting artifacts. Additionally, it aids the network in learning more intricate details by employing denser sampling in higher complexity areas.}

{\bf Ray skipping.}
If a ray does not pass through the scene with any actual content, it is invalid, and we can skip it completely for efficiency. 
To achieve this, we define a bounding box, which is initialized to the full extent and shrunk to only capture the scene's actual content when the initial training stage completes. 
Such a tight bounding box contains fewer empty voxels, making the ray-skipping operation more effective. 

{\bf Voxel skipping.}
Although a compact bounding box eliminates invalid rays, we could still waste time sampling at empty voxels, forming substantial gaps along a valid ray. 
After the initial training stage, we create a mask volume (i.e., a binary volume) recording whether each voxel is empty so that we can skip empty voxels to improve subsequent training efficiency. 
When ViSNeRF is trained for a dynamic scene, we update the mask volume based on the empty voxels determined from multiple static scenes.

\vspace{-0.05in}
\section{Novel View Synthesis of Static Scenes}

\subsection{Datasets}
We evaluate ViSNeRF using four simulation datasets, each containing both a DVR scene and an IR scene.
As listed in Table~\ref{tab:nvs-dataset}, in addition to the vortex, five jets, and Tangaroa datasets, we include the more complex supernova dataset to test ViSNeRF's performance on more challenging scenarios.
Image resolutions are adjusted according to the content of each dataset.
For every scene, we use 42 views for training and 181 views for inference, while the parameter settings remain constant throughout.

\vspace{-0.05in}
\subsection{Baselines}

In addition to the methods InSituNet, CoordNet, StyleGAN2, and EG3D we use for dynamic scenes, for static scenes, we also include four representative NeRF methods (i.e., NeRF, 3DGS, Instant-NGP, and TensoRF):
\begin{myitemize}
\vspace{-0.05in}
\item NeRF~\cite{Mildenhall-NeRF-ECCV20} is a 3D-aware novel view synthesis method that uses neural radiance fields implemented with an MLP. 
\item 3DGS~\cite{Kerbl-TOG23} uses Gaussian ellipsoids to represent radiance fields, enabling efficient optimization and real-time rendering.
\item Instant-NGP~\cite{Thomas-InstantNGP} utilizes multiresolution hash table encoding to achieve rapid NeRF training and fast rendering while maintaining a compact model size even for large and complex scenes.
\item TensoRF~\cite{Chen-TensoRF-ECCV22} employs tensor factorization in the radiance field to speed up NeRF training and rendering while improving the quality of reconstructed scenes. 
\vspace{-0.05in}
\end{myitemize}

\begin{table}[t!]
\caption{Novel view synthesis of static scenes: numbers of training images sampled and their resolutions.}
\vspace{-0.1in}
\centering
{\fontsize{6}{7.2}\selectfont
\begin{tabular}{c|cccc}
        & \prevhot{volume} & \# DVR & \# IR & image \\ 
dataset & \prevhot{resolution} & images & images & resolution\\ \hline
vortex & \prevhot{128$\times$128$\times$128} & 42 & 42 & 256$\times$256 \\
five jets & \prevhot{128$\times$128$\times$128} & 42 & 42 & 256$\times$256 \\
Tangaroa & \prevhot{300$\times$180$\times$120} & 42 & 42 & 1024$\times$1024 \\
supernova & \prevhot{432$\times$432$\times$432} & 42 & 42 & 1024$\times$1024 \\
\end{tabular}
}
\label{tab:nvs-dataset}
\end{table}

\begin{figure*}[t!]
  \begin{center}
  $\begin{array}{c@{\hspace{0.005in}}c@{\hspace{0.005in}}c@{\hspace{0.005in}}c@{\hspace{0.005in}}c@{\hspace{0.005in}}c@{\hspace{0.005in}}c@{\hspace{0.005in}}c@{\hspace{0.005in}}c@{\hspace{0.005in}}c}
    \includegraphics[width=0.097\linewidth]{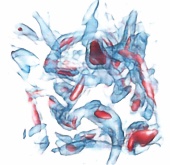}&
    \includegraphics[width=0.097\linewidth]{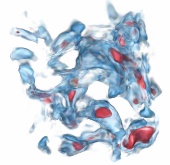}&
    \includegraphics[width=0.097\linewidth]{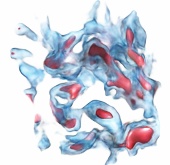}&
    \includegraphics[width=0.097\linewidth]{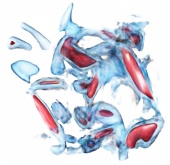}&
    \includegraphics[width=0.097\linewidth]{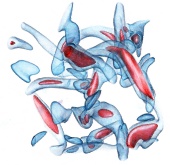}&
    \includegraphics[width=0.097\linewidth]{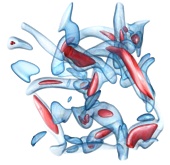}&
    \includegraphics[width=0.097\linewidth]{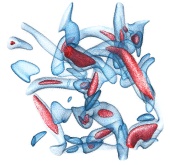}&
    \includegraphics[width=0.097\linewidth]{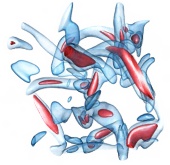}&
    \includegraphics[width=0.097\linewidth]{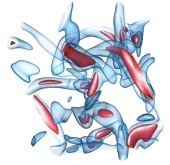}&
    \includegraphics[width=0.097\linewidth]{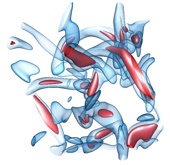}\\
    \includegraphics[width=0.097\linewidth]{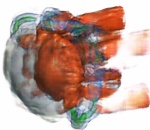}&
    \includegraphics[width=0.097\linewidth]{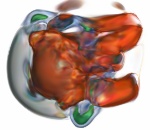}&
    \includegraphics[width=0.097\linewidth]{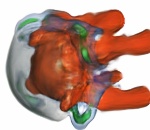}&
    \includegraphics[width=0.097\linewidth]{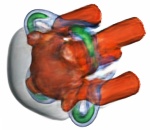}&
    \includegraphics[width=0.097\linewidth]{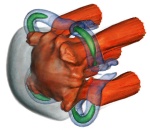}&
    \includegraphics[width=0.097\linewidth]{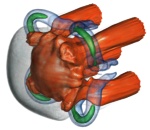}&
    \includegraphics[width=0.097\linewidth]{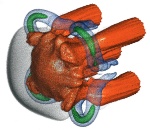}&
    \includegraphics[width=0.097\linewidth]{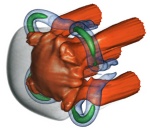}&
    \includegraphics[width=0.097\linewidth]{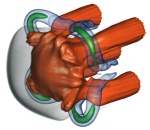}&
    \includegraphics[width=0.097\linewidth]{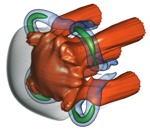}\\
    \includegraphics[width=0.097\linewidth]{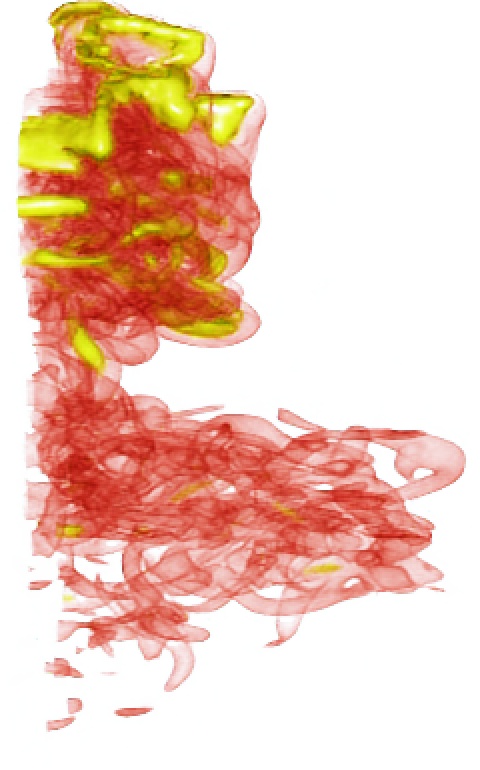}&
    \includegraphics[width=0.097\linewidth]{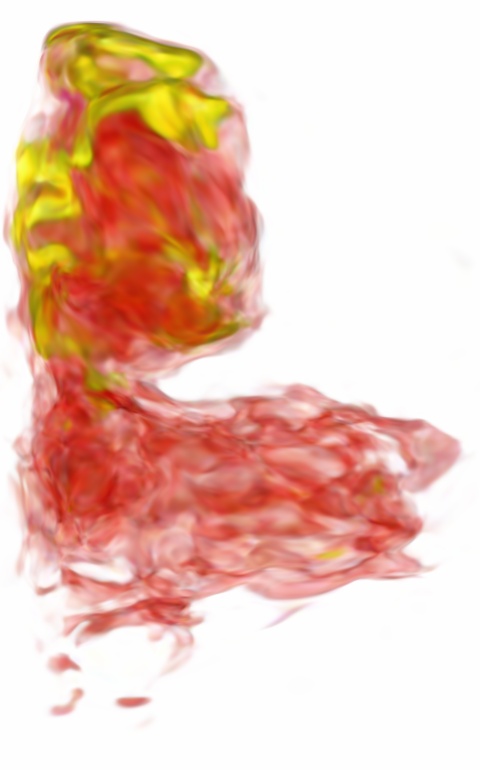}&
    \includegraphics[width=0.097\linewidth]{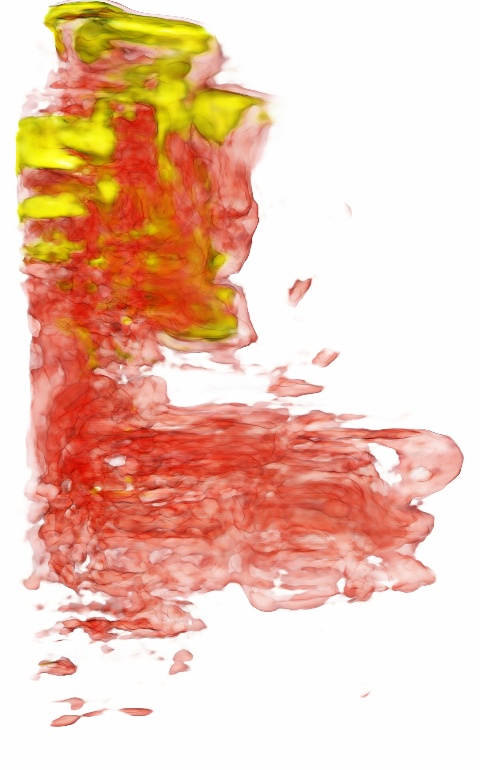}&
    \includegraphics[width=0.097\linewidth]{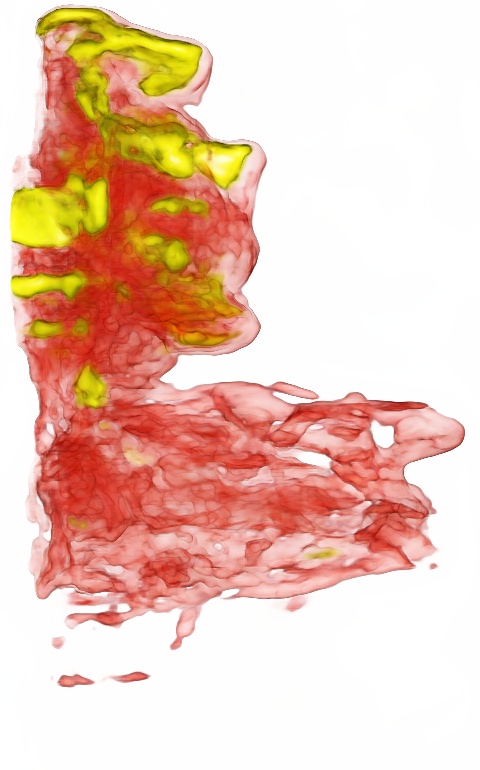}&
    \includegraphics[width=0.097\linewidth]{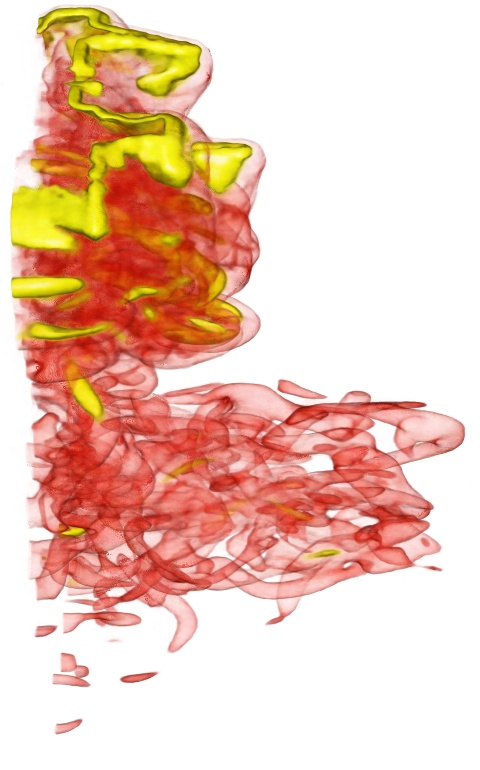}&
    \includegraphics[width=0.097\linewidth]{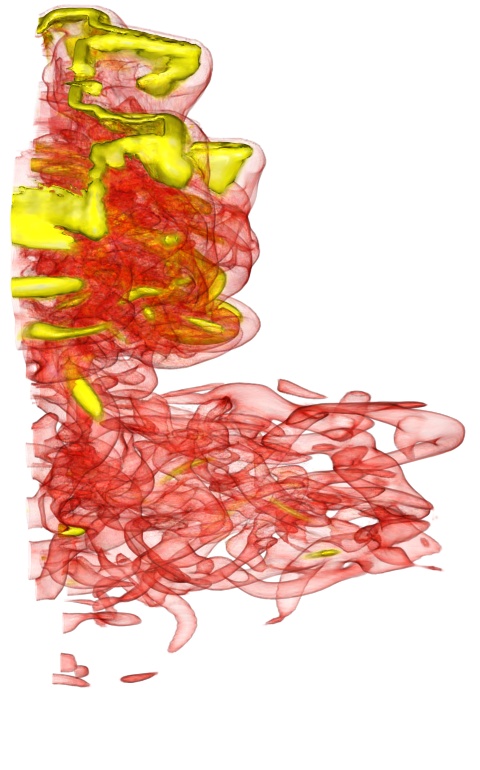}&
    \includegraphics[width=0.097\linewidth]{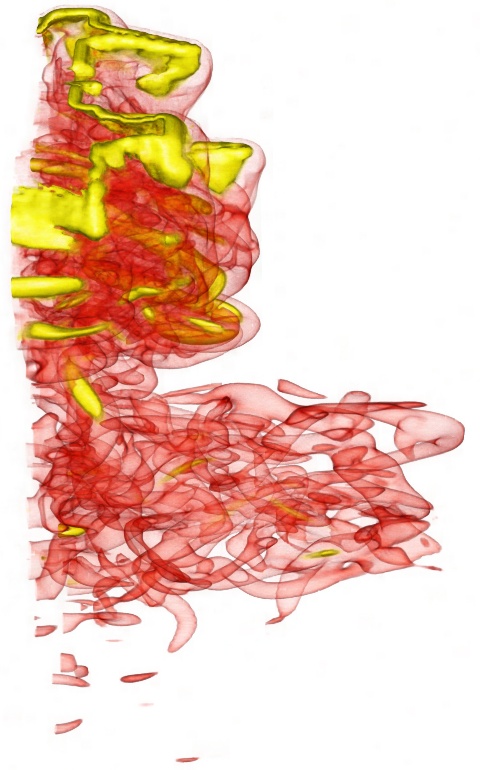}&
    \includegraphics[width=0.097\linewidth]{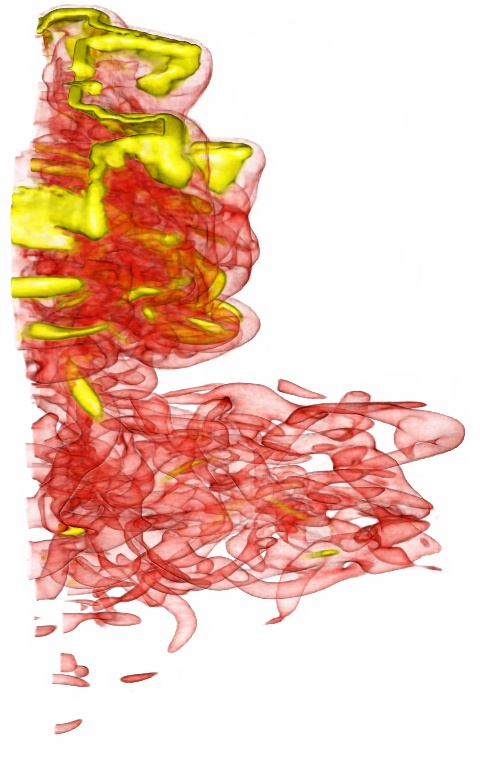}&
    \includegraphics[width=0.097\linewidth]{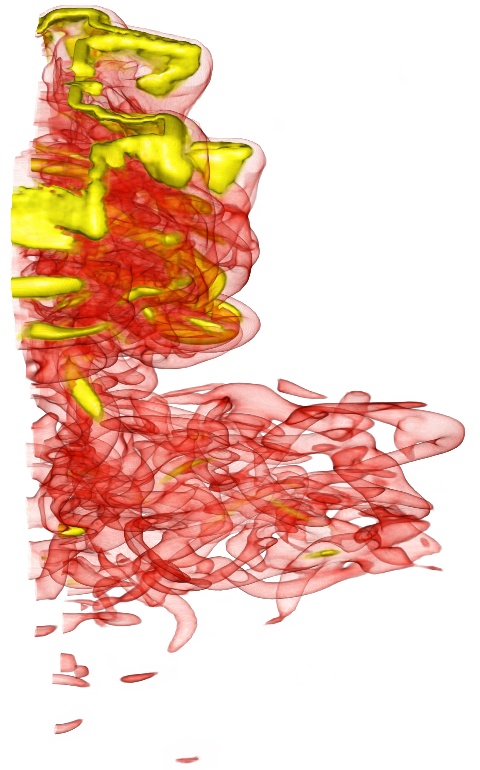}&
    \includegraphics[width=0.097\linewidth]{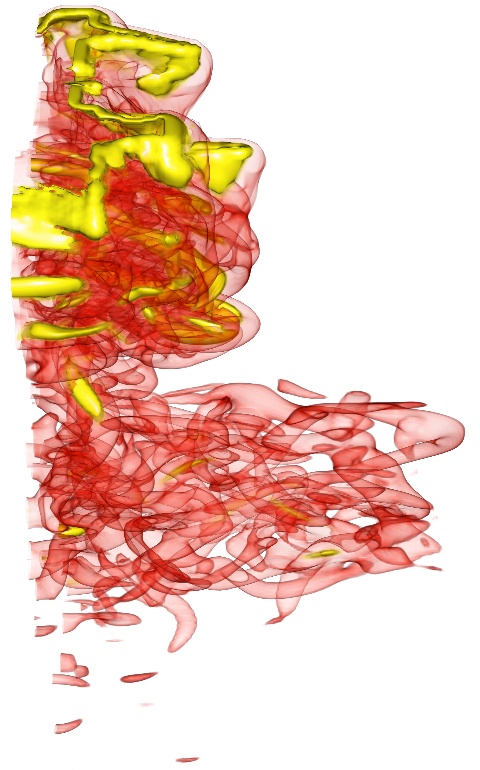}\\
    \includegraphics[width=0.097\linewidth]{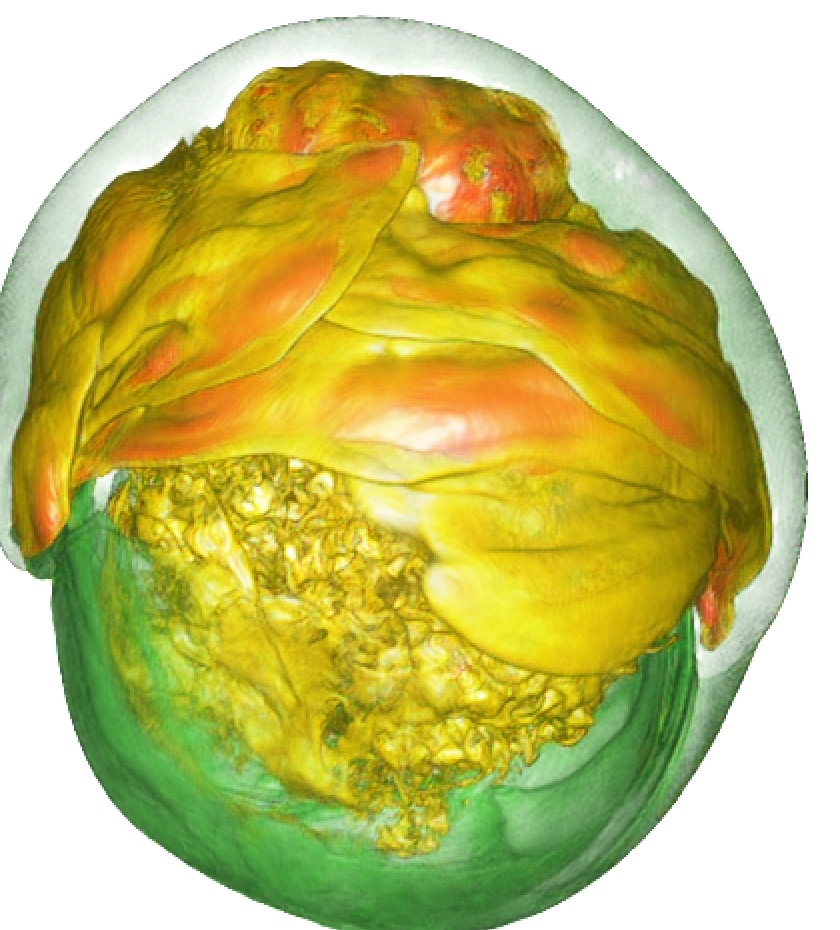}&
    \includegraphics[width=0.097\linewidth]{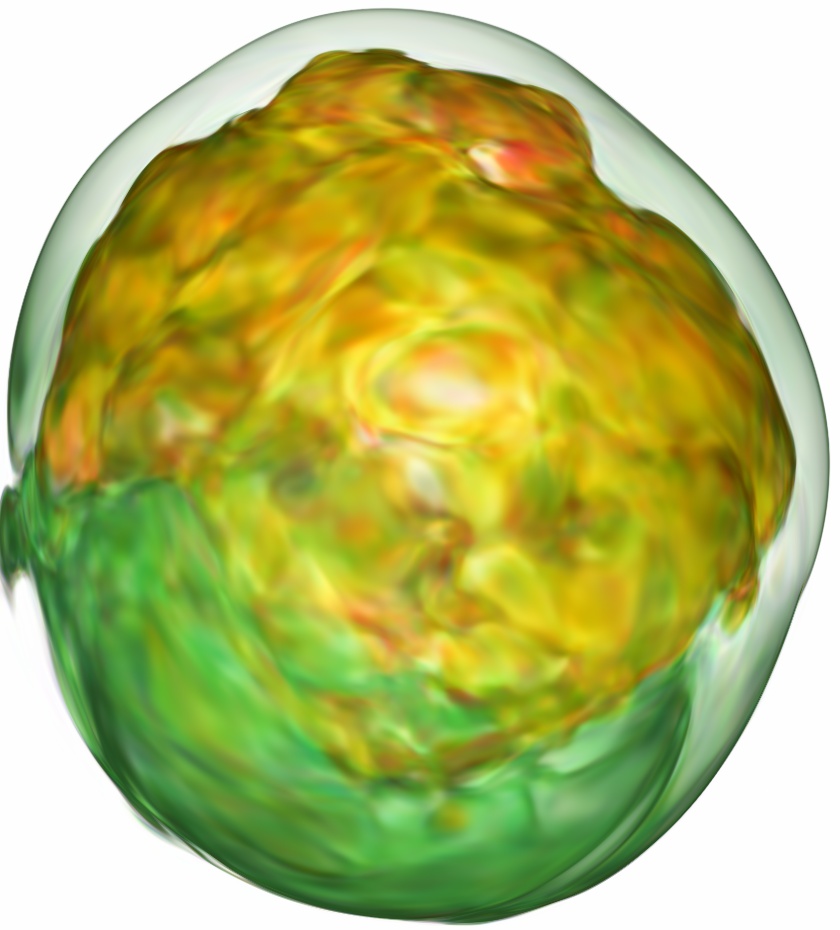}&
    \includegraphics[width=0.097\linewidth]{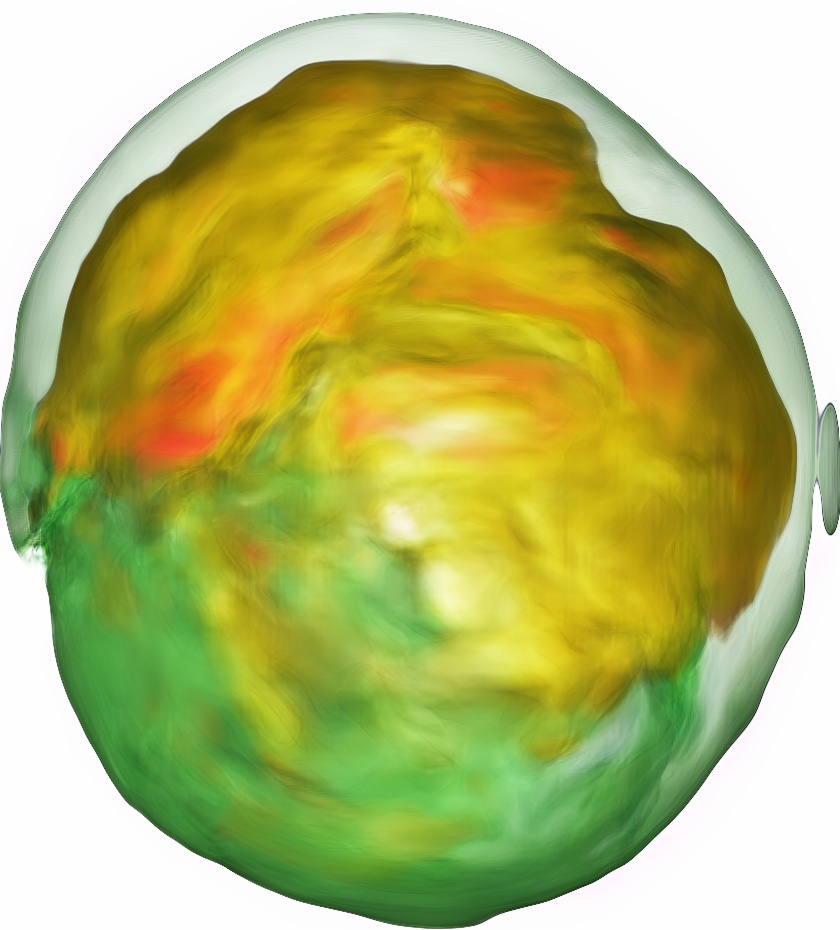}&
    \includegraphics[width=0.097\linewidth]{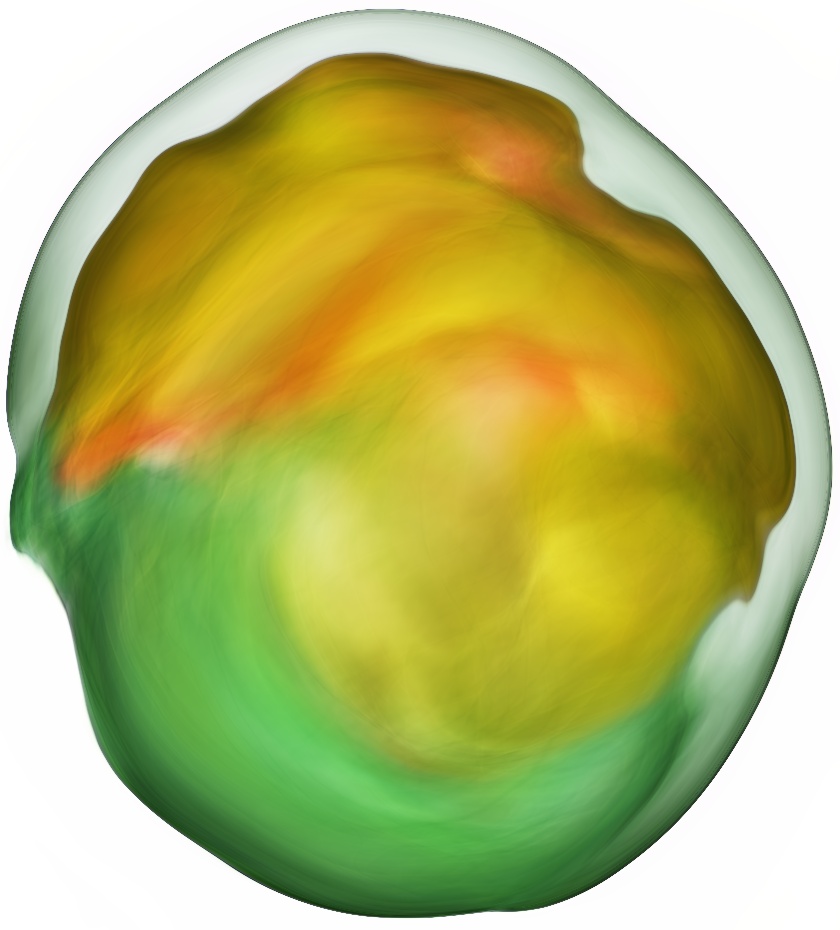}&
    \includegraphics[width=0.097\linewidth]{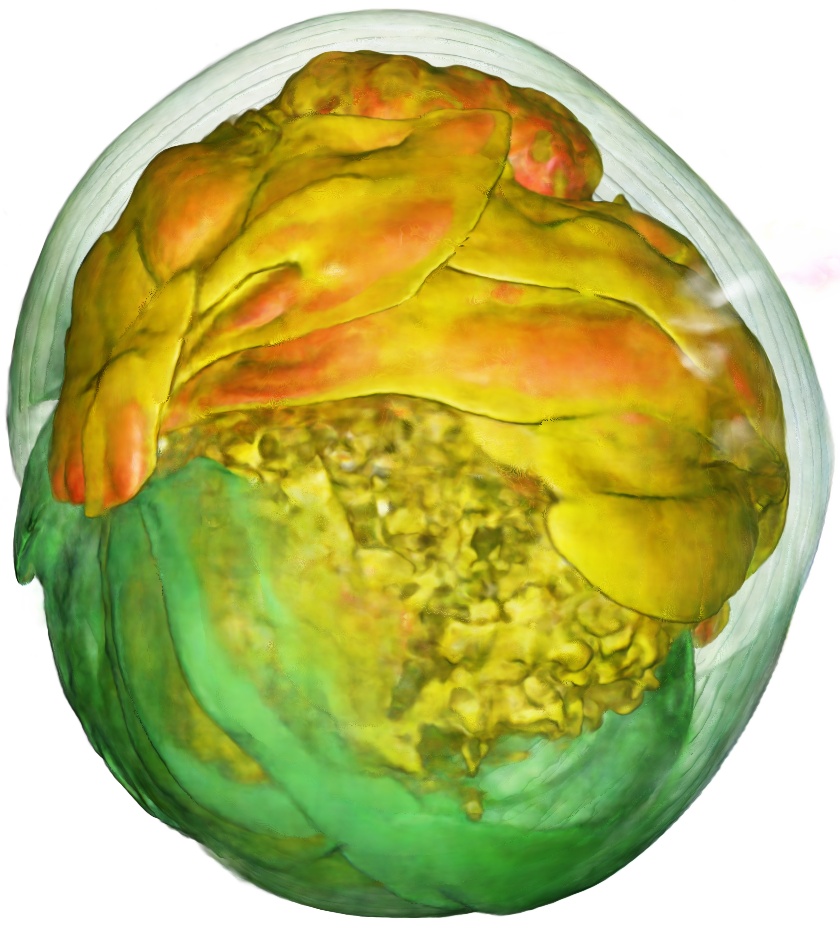}&
    \includegraphics[width=0.097\linewidth]{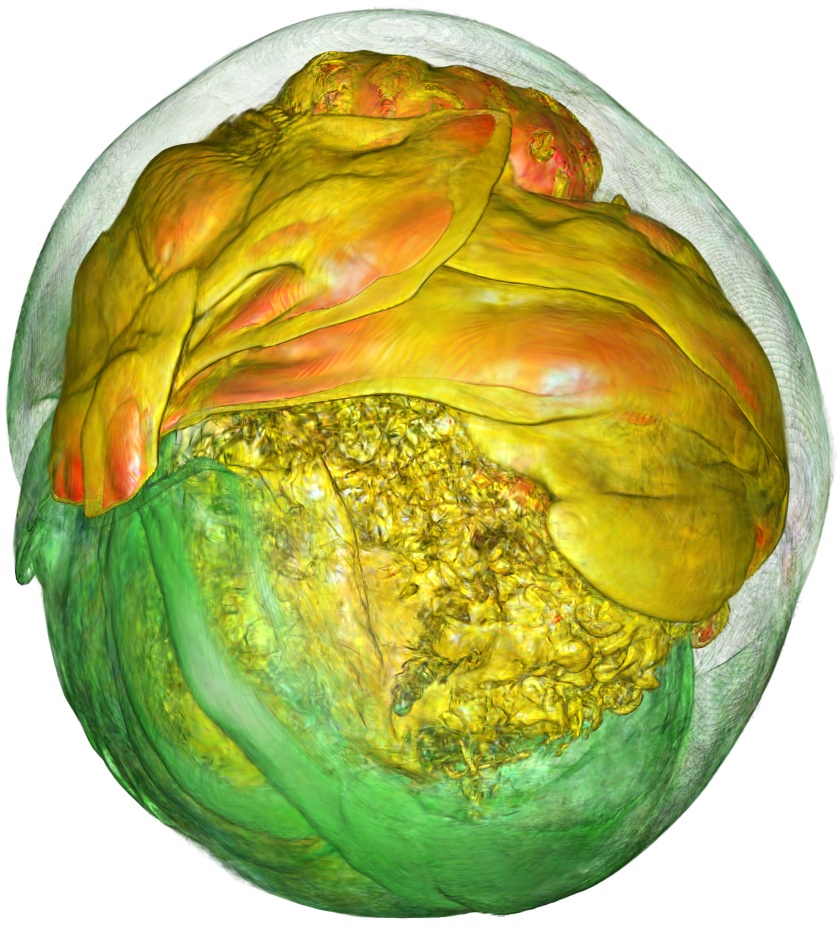}&
    \includegraphics[width=0.097\linewidth]{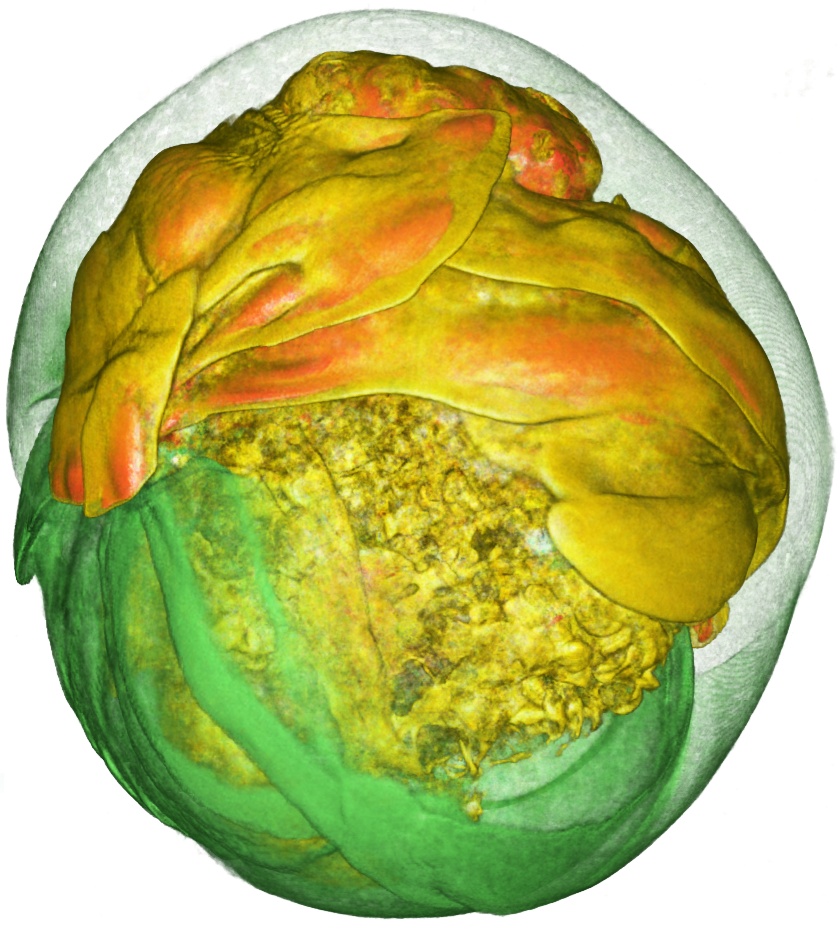}&
    \includegraphics[width=0.097\linewidth]{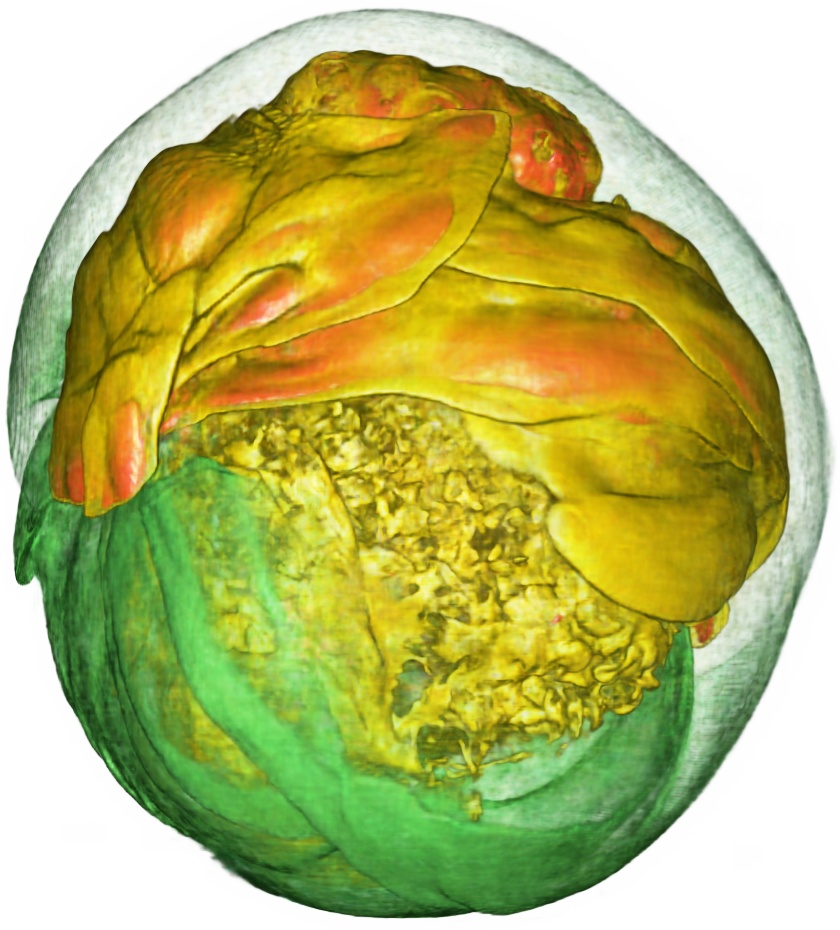}&
    \includegraphics[width=0.097\linewidth]{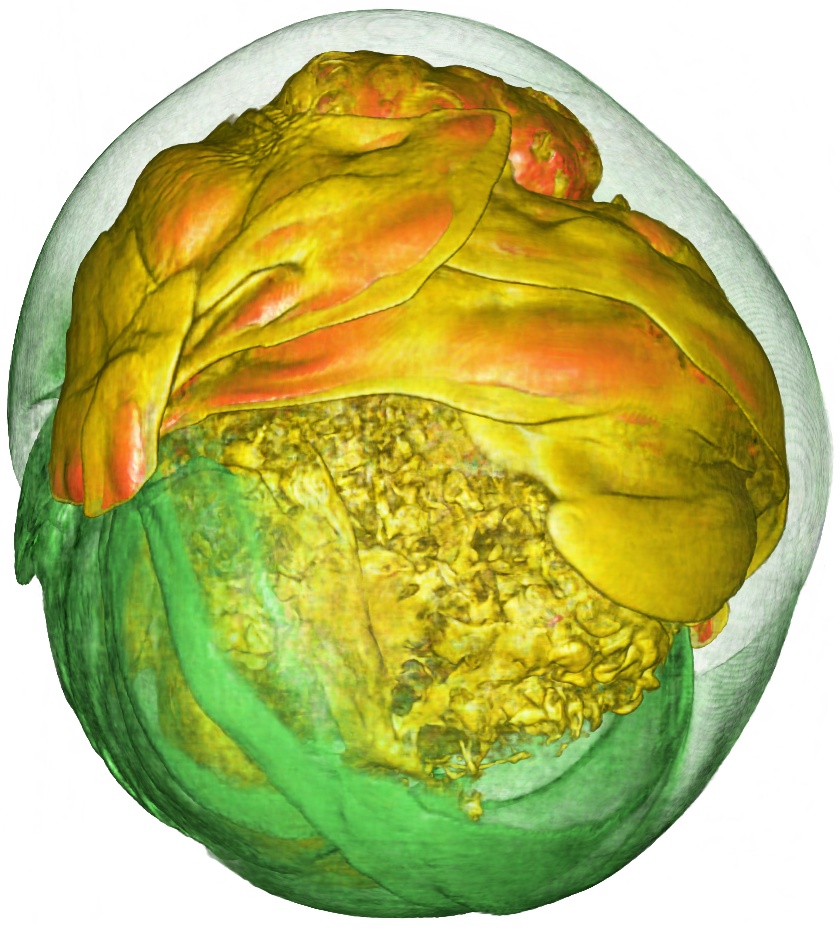}&
    \includegraphics[width=0.097\linewidth]{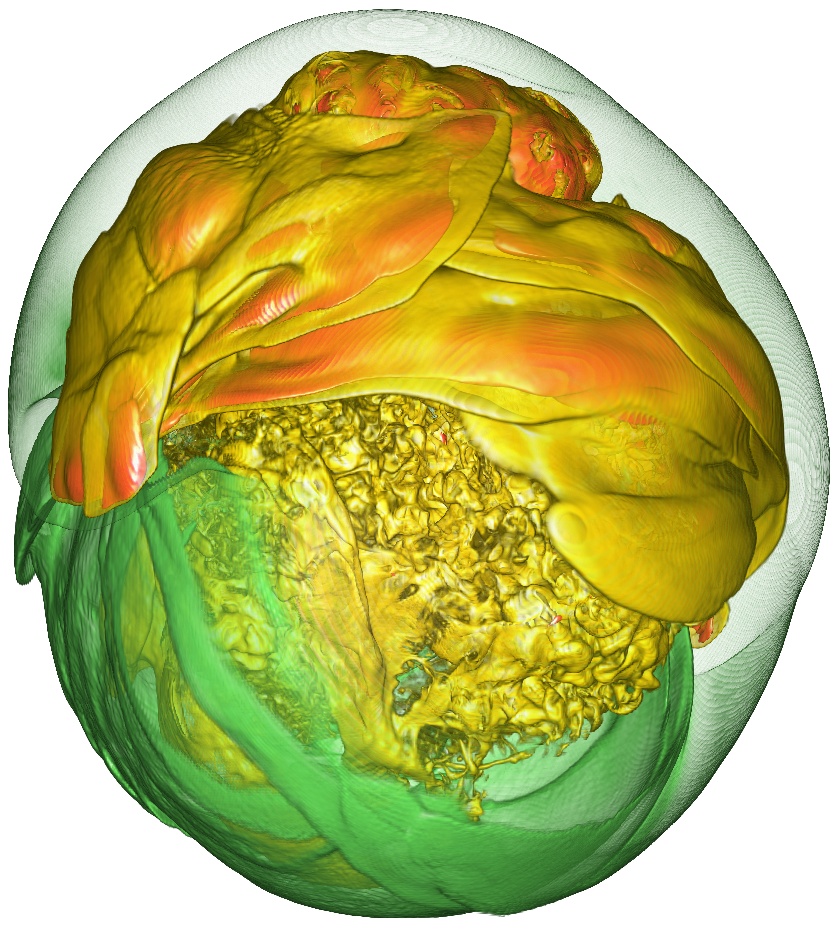}\\
    \mbox{\footnotesize (a)} & \mbox{\footnotesize (b)} & \mbox{\footnotesize (c)} & \mbox{\footnotesize (d)} & \mbox{\footnotesize (e)} & \mbox{\footnotesize \prevhot{(f)}} & \mbox{\footnotesize \prevhot{(g)}} & \mbox{\footnotesize (h)} & \mbox{\footnotesize \prevhot{(i)}} & \mbox{\footnotesize (j)}
  \end{array}$
 \end{center}
 \vspace{-.25in} 
 \caption{Novel view synthesis of DVR images for static scenes. 
 (a) to (h): InSituNet, CoordNet, StyleGAN2, EG3D, NeRF, 3DGS, Instant-NGP, TensoRF, ViSNeRF, and GT. 
 Top to bottom: vortex, five jets, Tangaroa, and supernova.} 
 \label{fig:comp-nvs-dvr}
\end{figure*}

\begin{figure*}[t!]
  \begin{center}
  $\begin{array}{c@{\hspace{0.005in}}c@{\hspace{0.005in}}c@{\hspace{0.005in}}c@{\hspace{0.005in}}c@{\hspace{0.005in}}c@{\hspace{0.005in}}c@{\hspace{0.005in}}c@{\hspace{0.005in}}c@{\hspace{0.005in}}c}
    \includegraphics[width=0.097\linewidth]{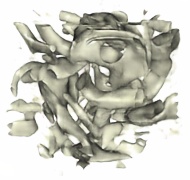}&
    \includegraphics[width=0.097\linewidth]{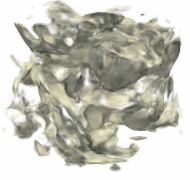}&
    \includegraphics[width=0.097\linewidth]{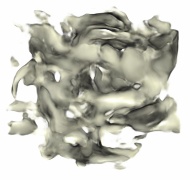}&
    \includegraphics[width=0.097\linewidth]{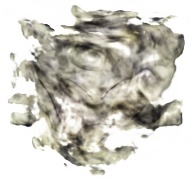}&
    \includegraphics[width=0.097\linewidth]{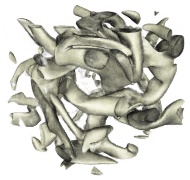}&
    \includegraphics[width=0.097\linewidth]{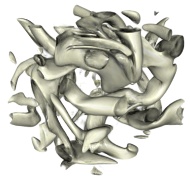}&
    \includegraphics[width=0.097\linewidth]{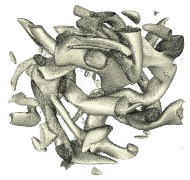}&
    \includegraphics[width=0.097\linewidth]{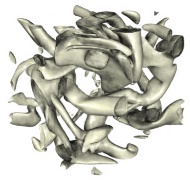}&
    \includegraphics[width=0.097\linewidth]{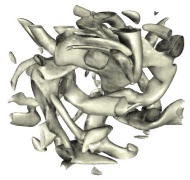}&
    \includegraphics[width=0.097\linewidth]{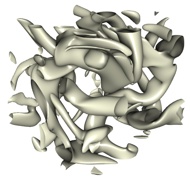}\\
    \includegraphics[width=0.097\linewidth]{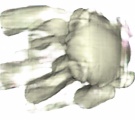}&
    \includegraphics[width=0.097\linewidth]{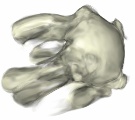}&
    \includegraphics[width=0.097\linewidth]{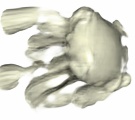}&
    \includegraphics[width=0.097\linewidth]{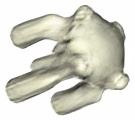}&
    \includegraphics[width=0.097\linewidth]{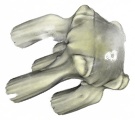}&
    \includegraphics[width=0.097\linewidth]{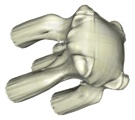}&
    \includegraphics[width=0.097\linewidth]{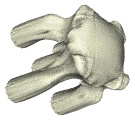}&
    \includegraphics[width=0.097\linewidth]{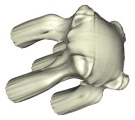}&
    \includegraphics[width=0.097\linewidth]{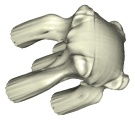}&
    \includegraphics[width=0.097\linewidth]{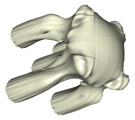}\\
    \includegraphics[width=0.097\linewidth]{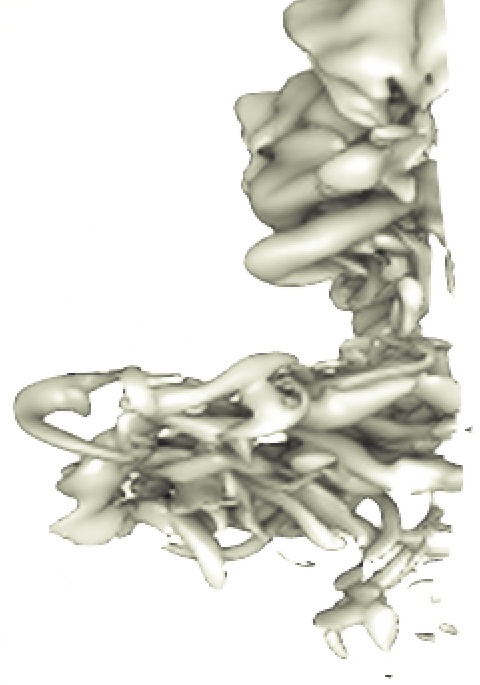}&
    \includegraphics[width=0.097\linewidth]{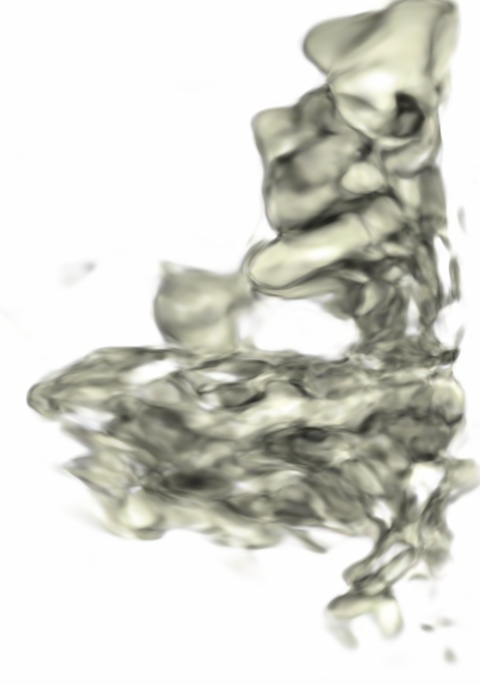}&
    \includegraphics[width=0.097\linewidth]{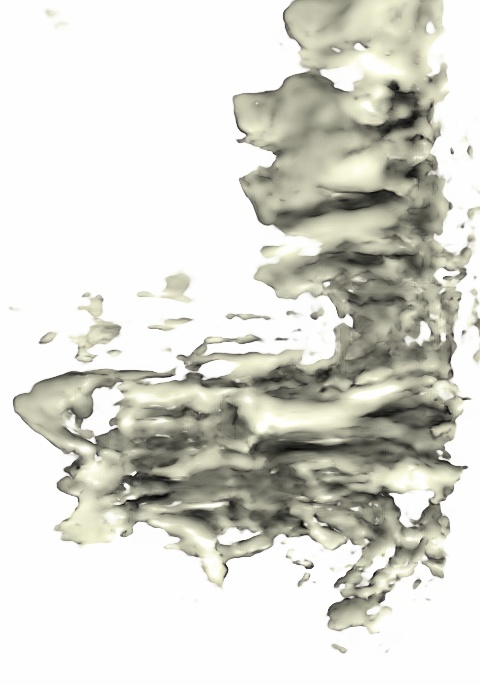}&
    \includegraphics[width=0.097\linewidth]{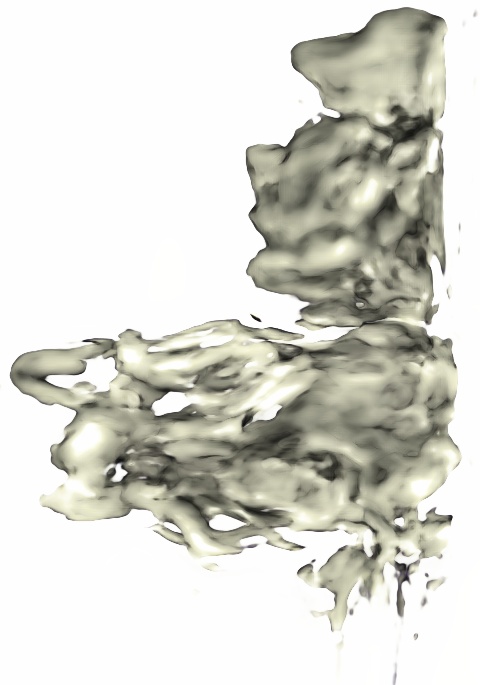}&
    \includegraphics[width=0.097\linewidth]{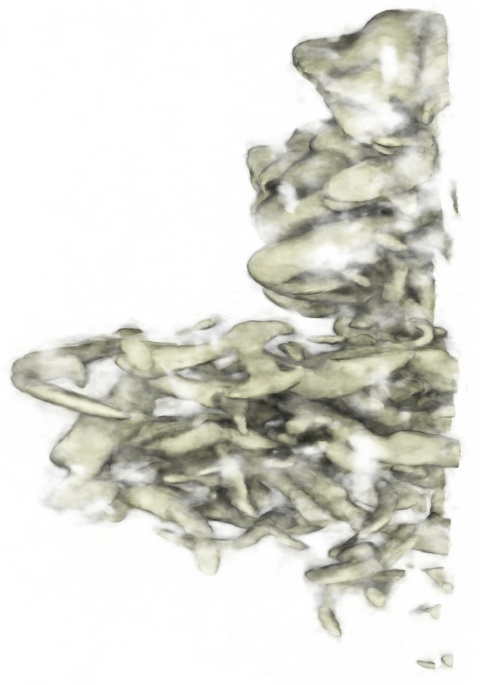}&
    \includegraphics[width=0.097\linewidth]{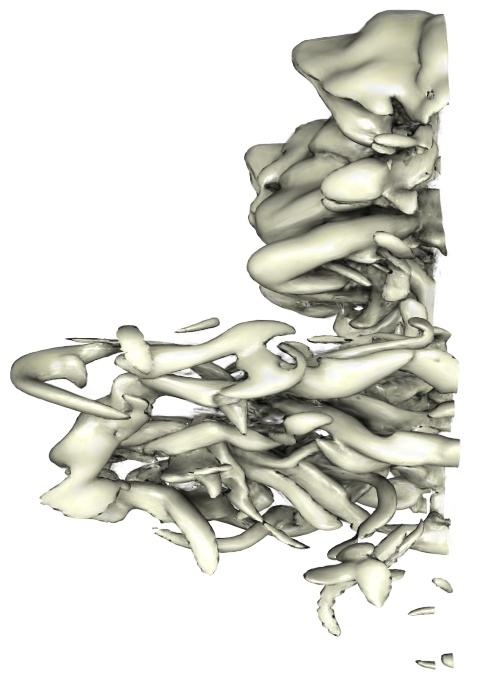}&
    \includegraphics[width=0.097\linewidth]{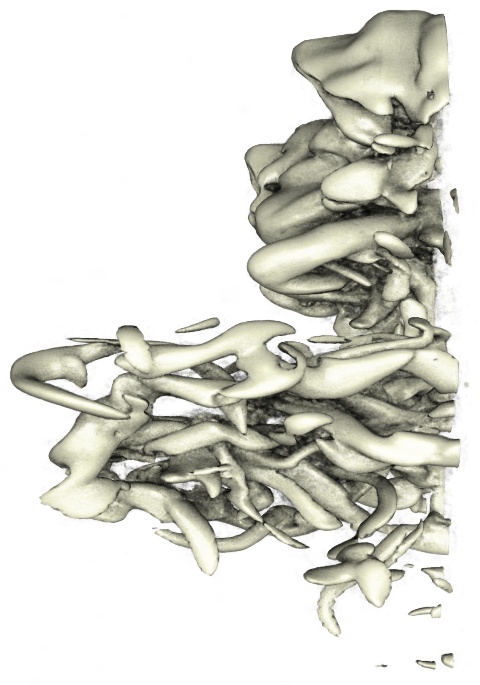}&
    \includegraphics[width=0.097\linewidth]{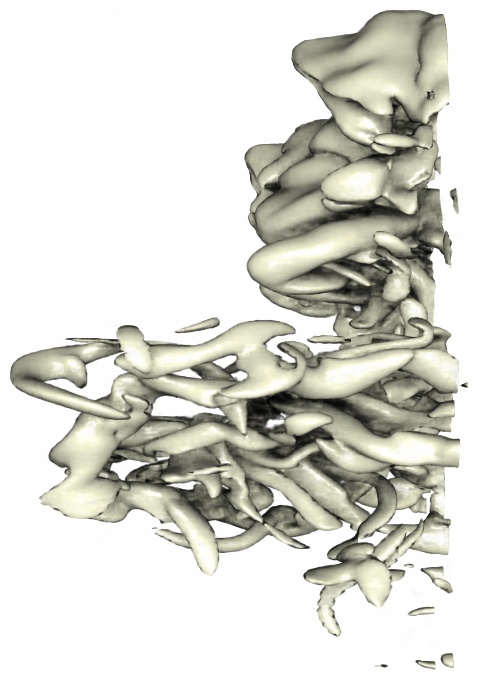}&
    \includegraphics[width=0.097\linewidth]{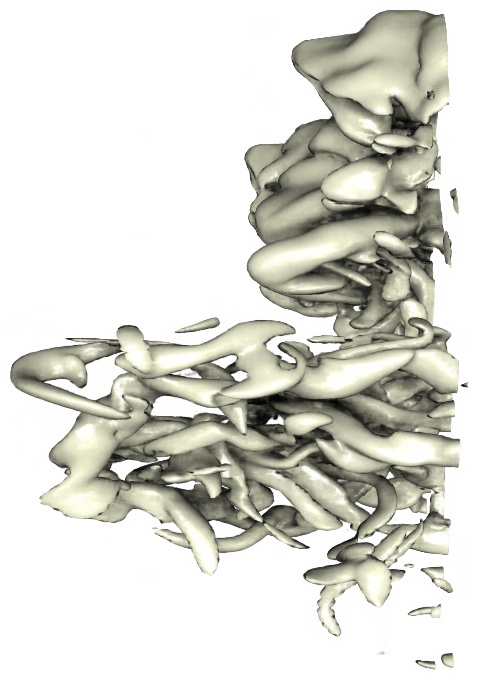}&
    \includegraphics[width=0.097\linewidth]{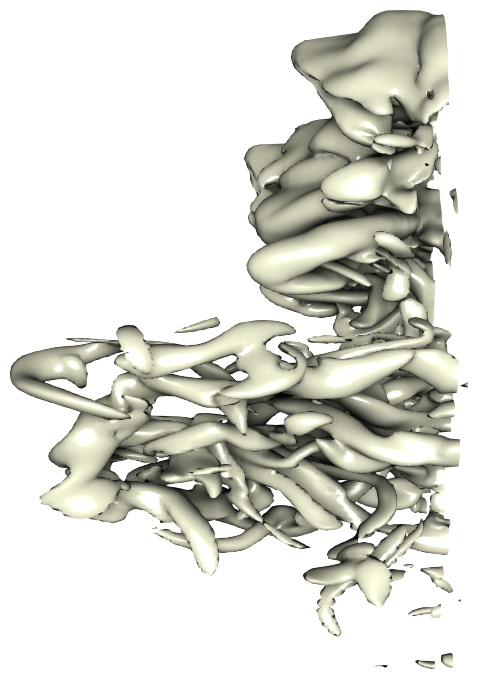}\\
    \includegraphics[width=0.097\linewidth]{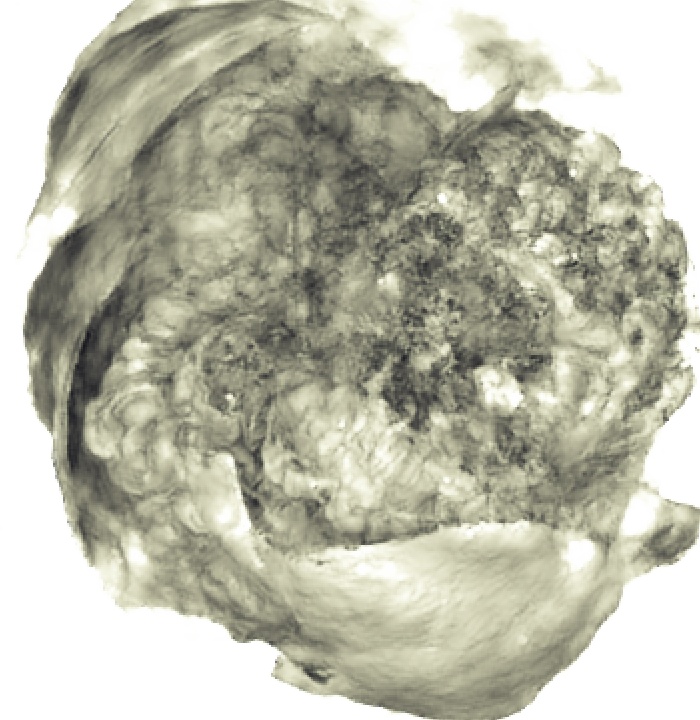}&
    \includegraphics[width=0.097\linewidth]{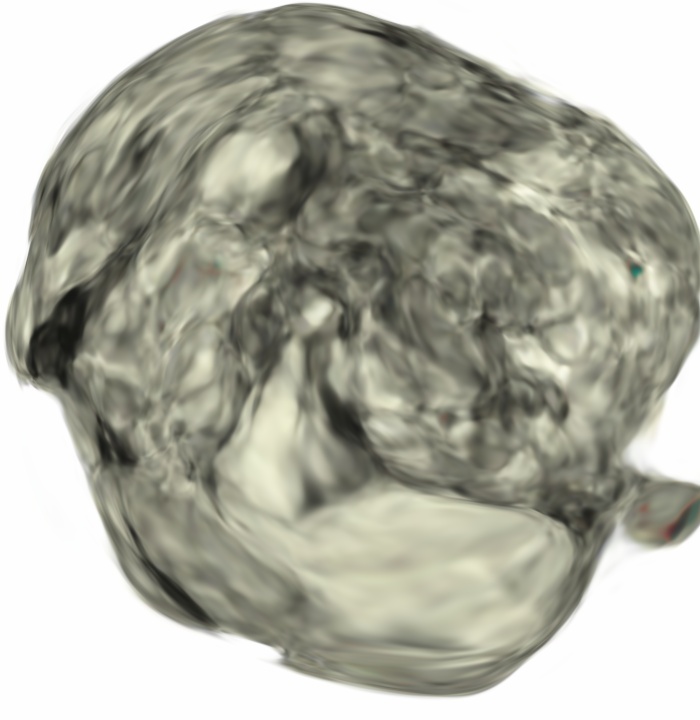}&
    \includegraphics[width=0.097\linewidth]{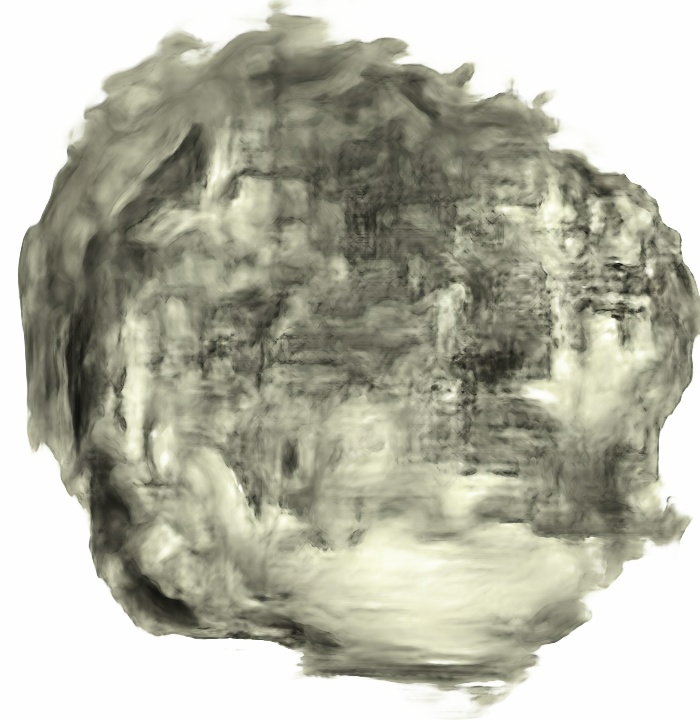}&
    \includegraphics[width=0.097\linewidth]{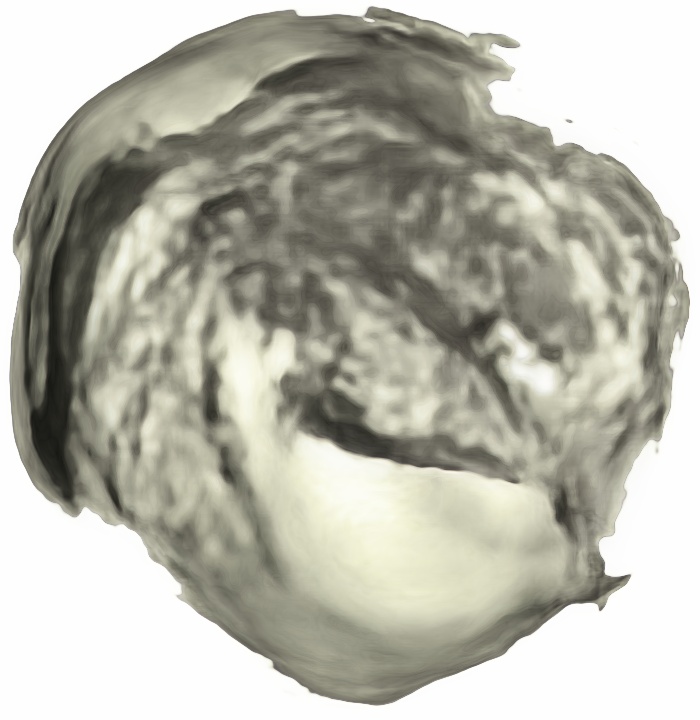}&
    \includegraphics[width=0.097\linewidth]{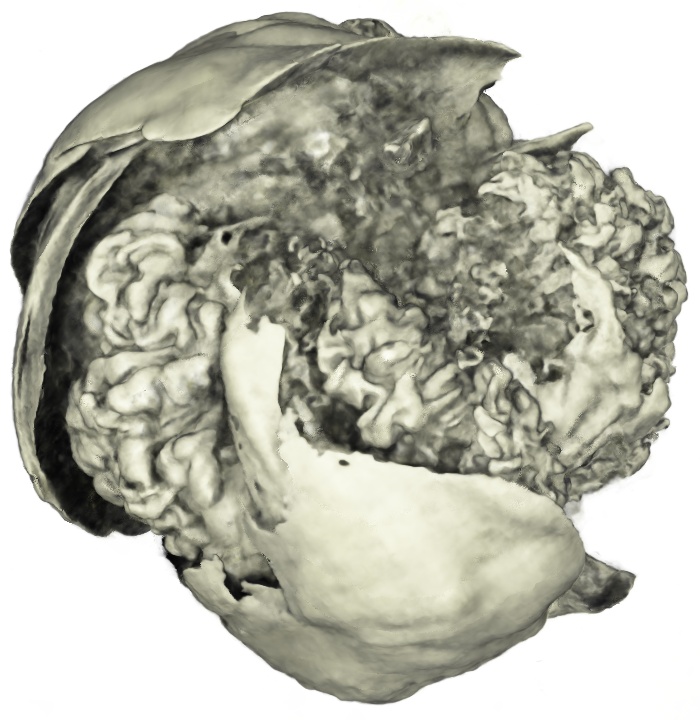}&
    \includegraphics[width=0.097\linewidth]{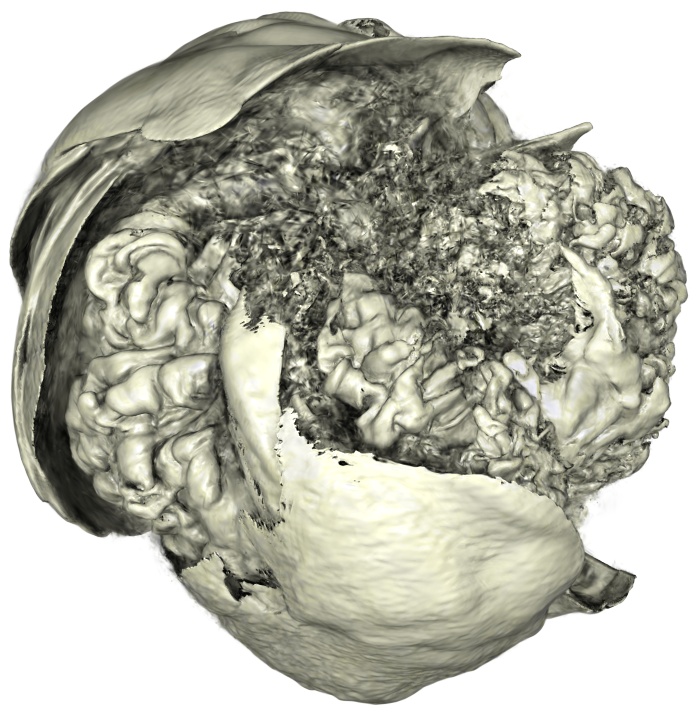}&
    \includegraphics[width=0.097\linewidth]{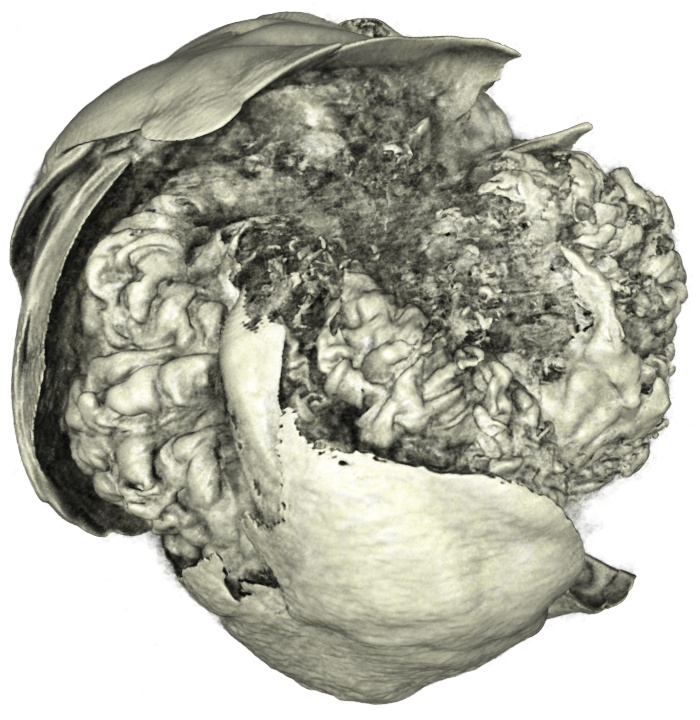}&
    \includegraphics[width=0.097\linewidth]{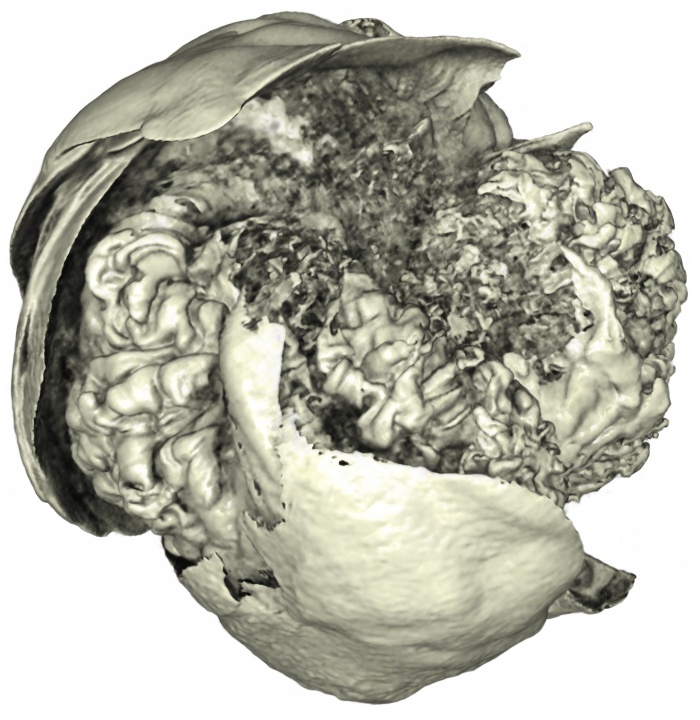}&
    \includegraphics[width=0.097\linewidth]{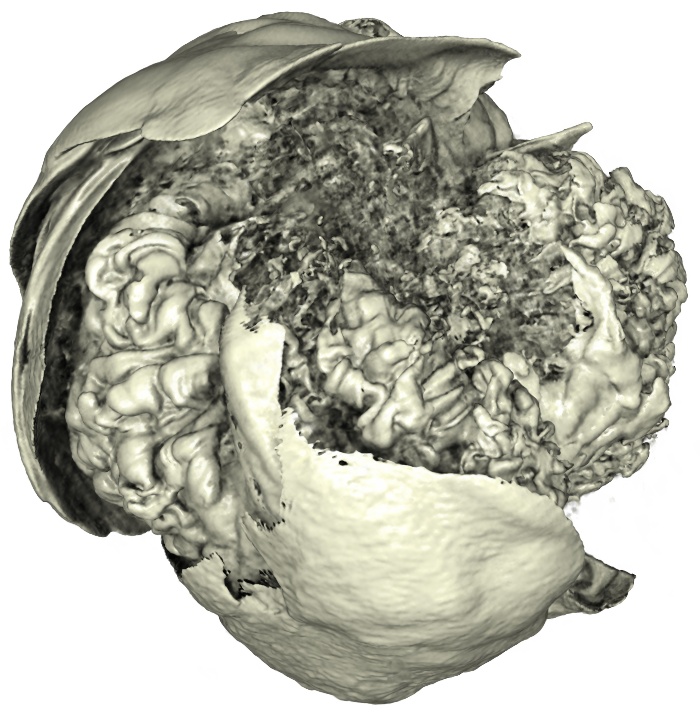}&
    \includegraphics[width=0.097\linewidth]{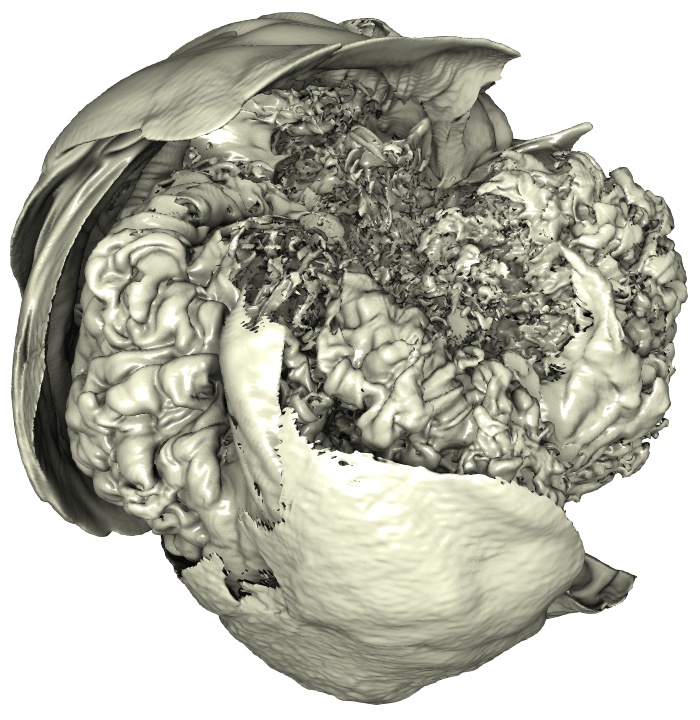}\\
    \mbox{\footnotesize (a)} & \mbox{\footnotesize (b)} & \mbox{\footnotesize (c)} & \mbox{\footnotesize (d)} & \mbox{\footnotesize (e)} & \mbox{\footnotesize \prevhot{(f)}} & \mbox{\footnotesize \prevhot{(g)}} & \mbox{\footnotesize (h)} & \mbox{\footnotesize \prevhot{(i)}} & \mbox{\footnotesize (j)}     
  \end{array}$
 \end{center}
\vspace{-.25in} 
 \caption{Novel view synthesis of IR images for static scenes. 
 (a) to (h): InSituNet, CoordNet, StyleGAN2, EG3D, NeRF, 3DGS, Instant-NGP, \prevhot{TensoRF}, \prevhot{ViSNeRF}, and GT.  
 Top to bottom: vortex, five jets, Tangaroa, and supernova.} 
 \label{fig:comp-nvs-ir}
\end{figure*}

\begin{figure}[t!]
  \begin{center}
  $\begin{array}{c@{\hspace{0.005in}}c@{\hspace{0.005in}}c@{\hspace{0.005in}}c@{\hspace{0.005in}}c@{\hspace{0.005in}}c@{\hspace{0.005in}}c@{\hspace{0.005in}}c@{\hspace{0.005in}}c@{\hspace{0.005in}}c}
    \includegraphics[width=0.095\linewidth]{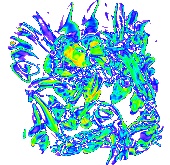}&
    \includegraphics[width=0.095\linewidth]{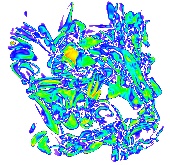}&
    \includegraphics[width=0.095\linewidth]{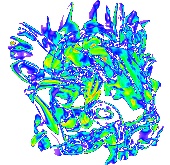}&
    \includegraphics[width=0.095\linewidth]{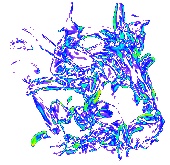}&
    \includegraphics[width=0.095\linewidth]{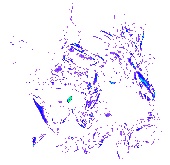}&
    \includegraphics[width=0.095\linewidth]{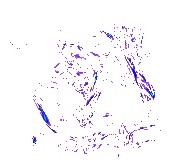}&
    \includegraphics[width=0.095\linewidth]{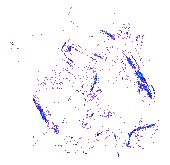}&
    \includegraphics[width=0.095\linewidth]{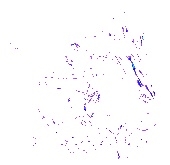}&
    \includegraphics[width=0.095\linewidth]{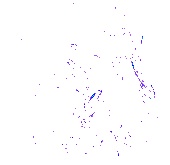}&
    \includegraphics[width=0.095\linewidth]{images/baselines/vortex_vr/GT-065-crop.png}\\
    \includegraphics[width=0.095\linewidth]{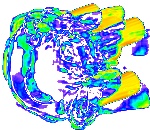}&
    \includegraphics[width=0.095\linewidth]{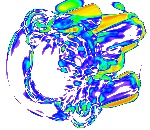}&
    \includegraphics[width=0.095\linewidth]{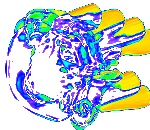}&
    \includegraphics[width=0.095\linewidth]{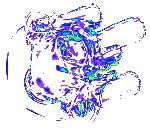}&
    \includegraphics[width=0.095\linewidth]{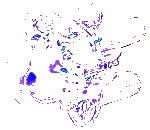}&
    \includegraphics[width=0.095\linewidth]{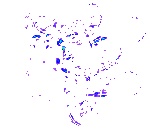}&
    \includegraphics[width=0.095\linewidth]{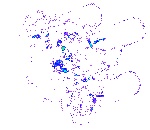}&
    \includegraphics[width=0.095\linewidth]{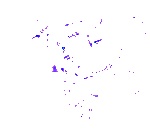}&
    \includegraphics[width=0.095\linewidth]{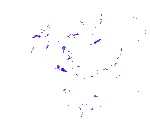}&
    \includegraphics[width=0.095\linewidth]{images/baselines/fivejets_vr/GT-115-crop.png}\\
    \includegraphics[width=0.095\linewidth]{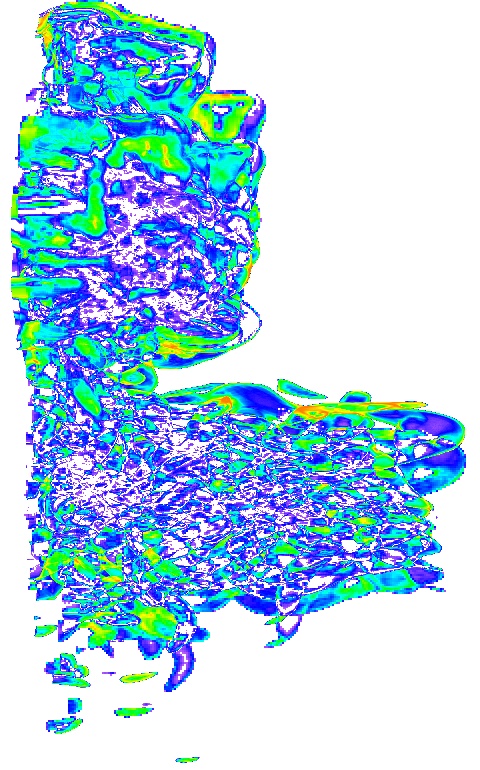}&
    \includegraphics[width=0.095\linewidth]{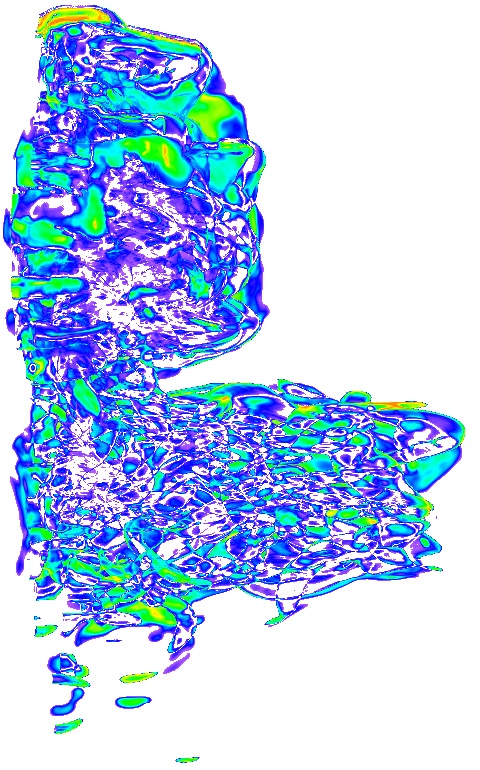}&
    \includegraphics[width=0.095\linewidth]{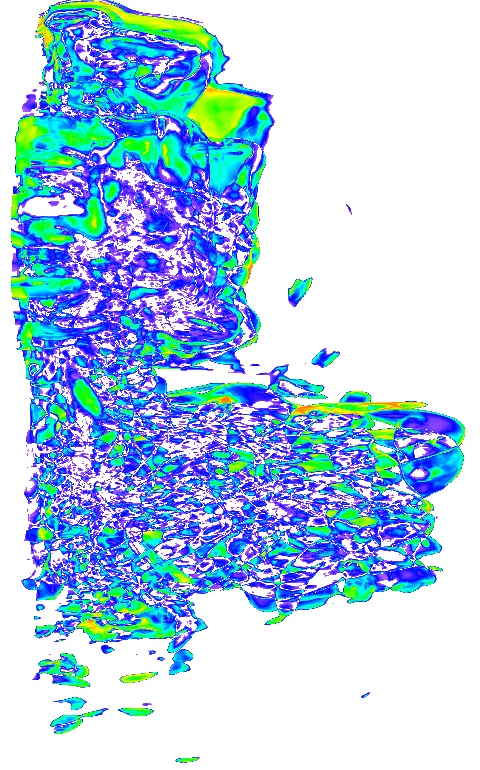}&
    \includegraphics[width=0.095\linewidth]{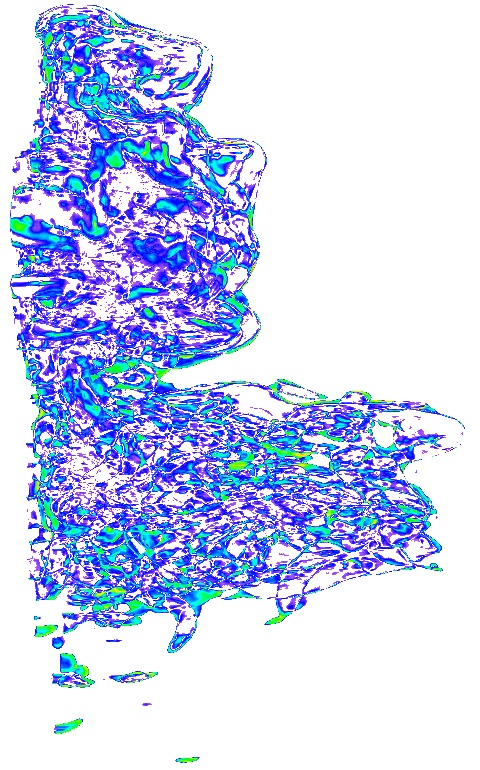}&
    \includegraphics[width=0.095\linewidth]{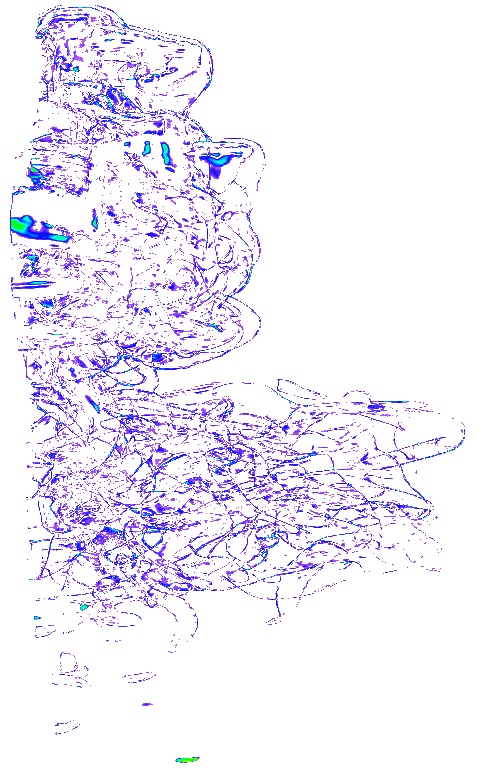}&
    \includegraphics[width=0.095\linewidth]{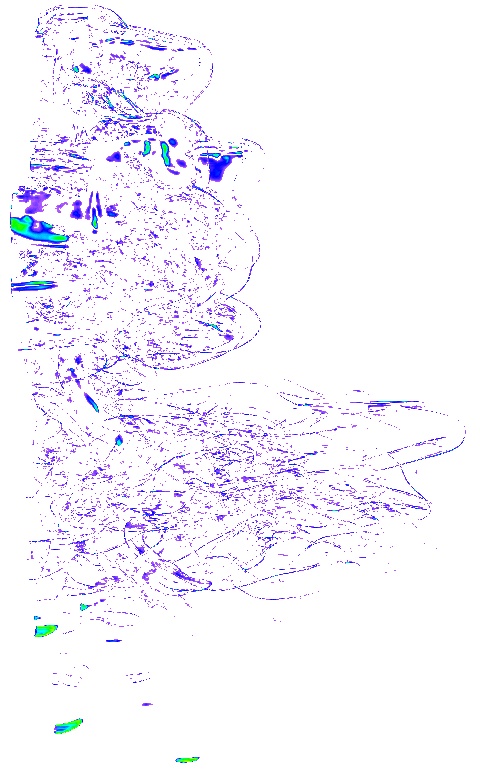}&
    \includegraphics[width=0.095\linewidth]{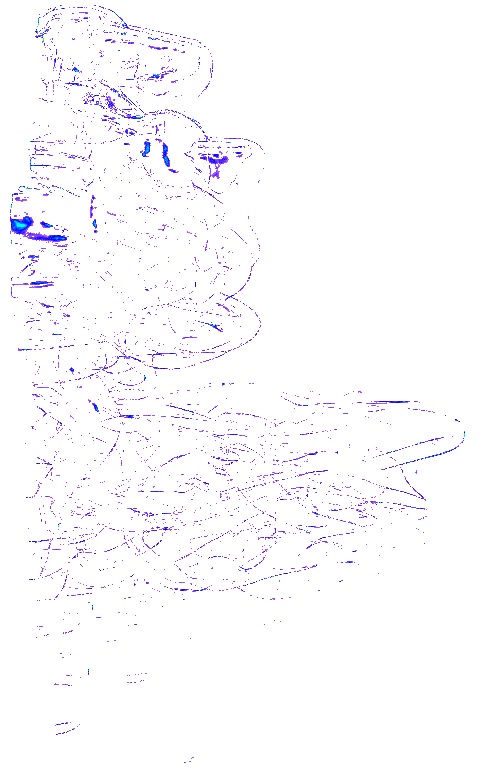}&
    \includegraphics[width=0.095\linewidth]{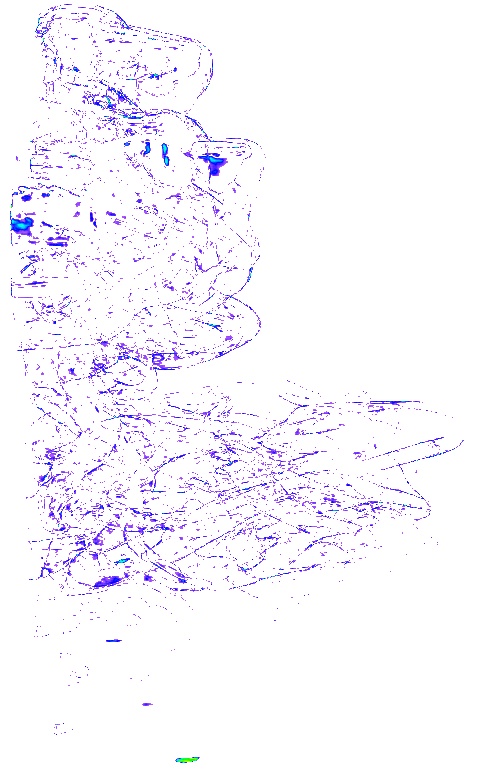}&
    \includegraphics[width=0.095\linewidth]{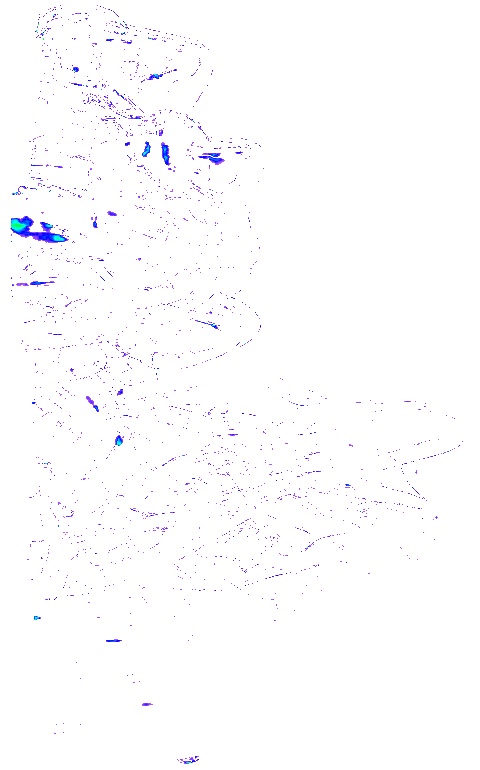}&
    \includegraphics[width=0.095\linewidth]{images/baselines/tangaroa_vr/GT-045-crop.png}\\
    \includegraphics[width=0.095\linewidth]{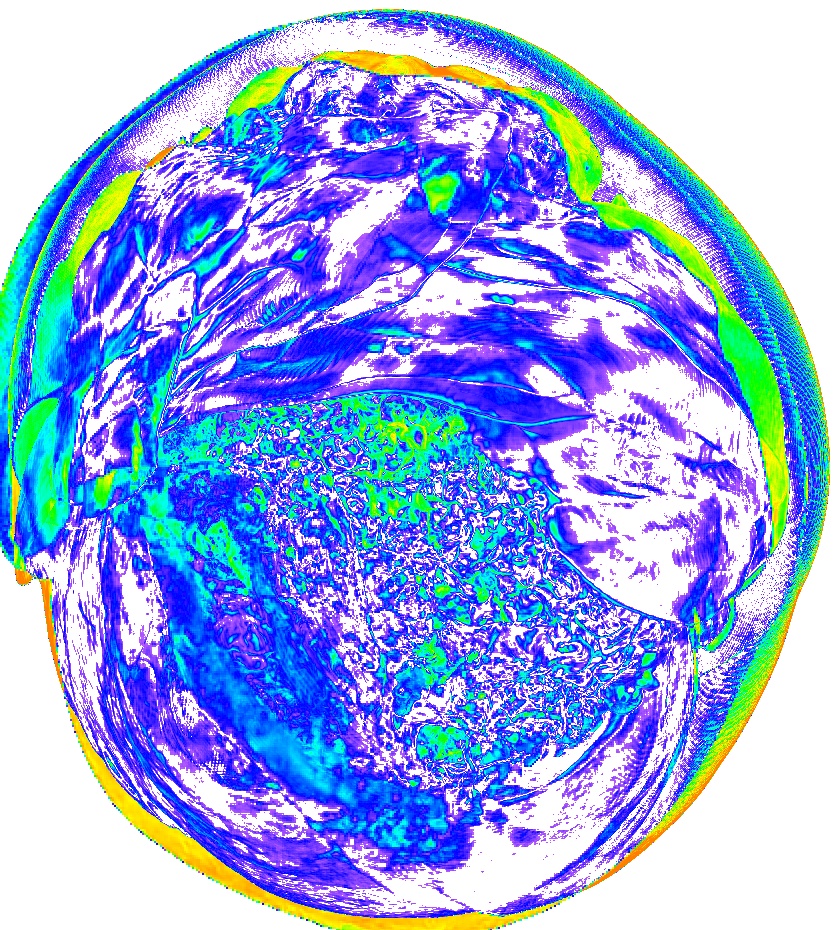}&
    \includegraphics[width=0.095\linewidth]{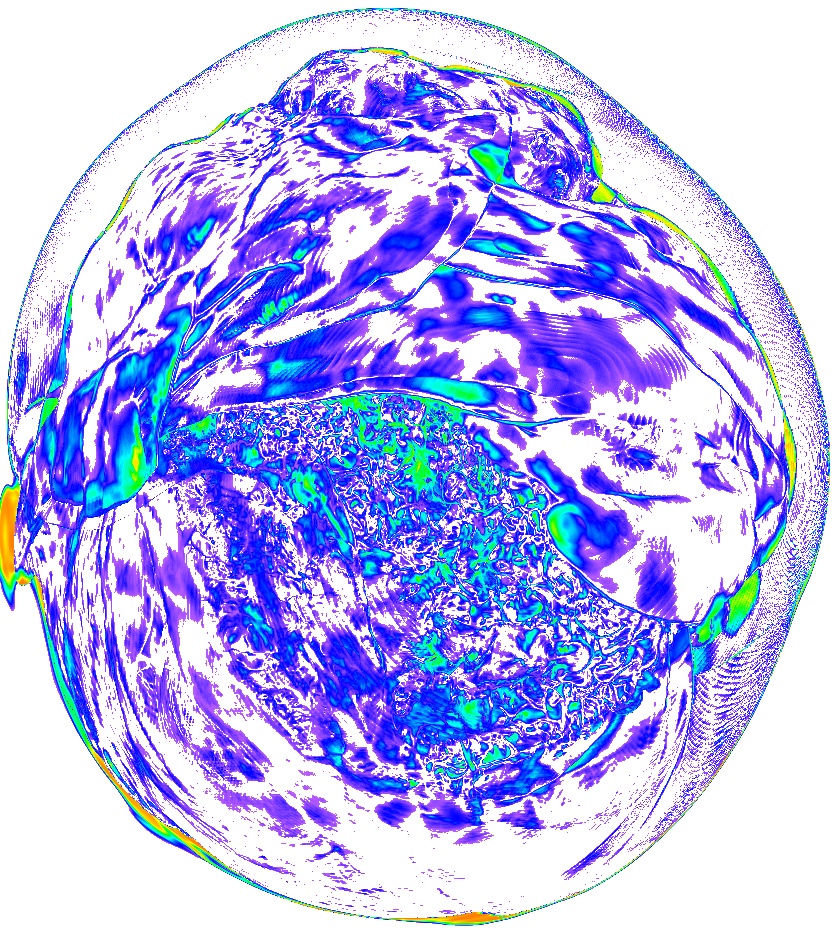}&
    \includegraphics[width=0.095\linewidth]{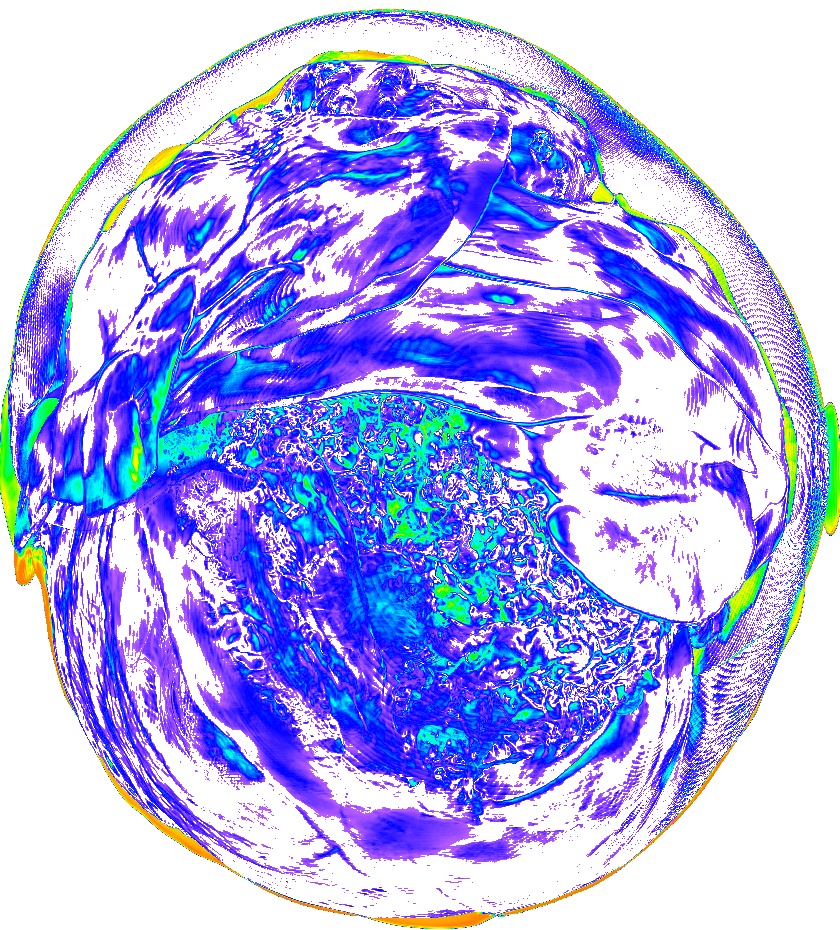}&
    \includegraphics[width=0.095\linewidth]{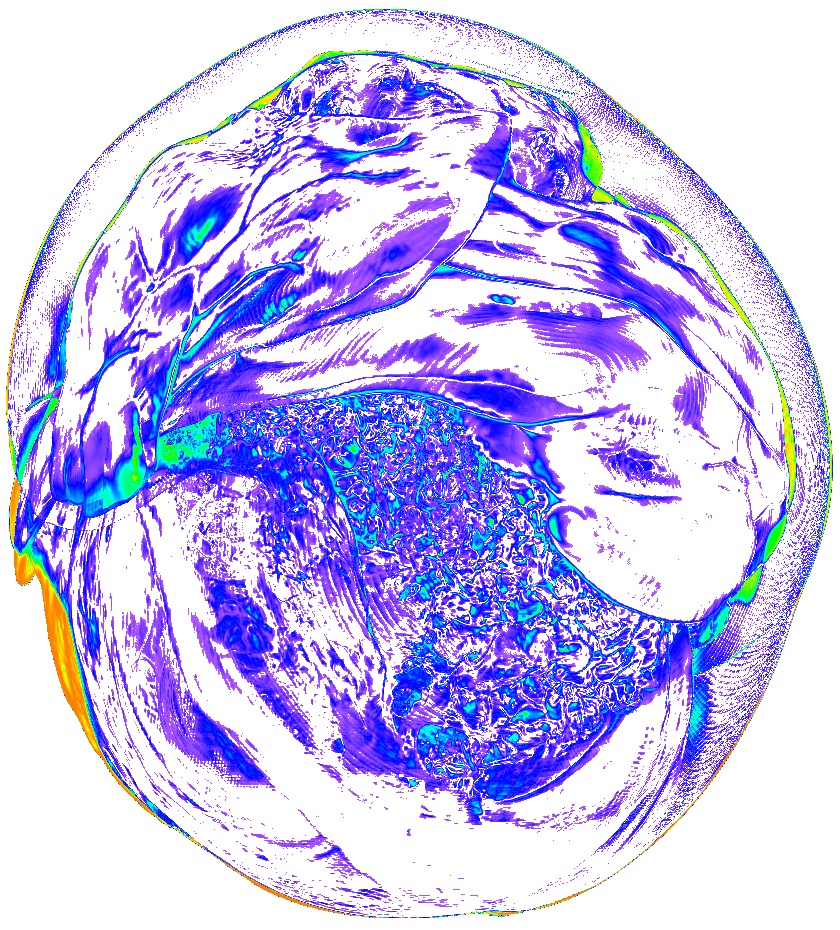}&
    \includegraphics[width=0.095\linewidth]{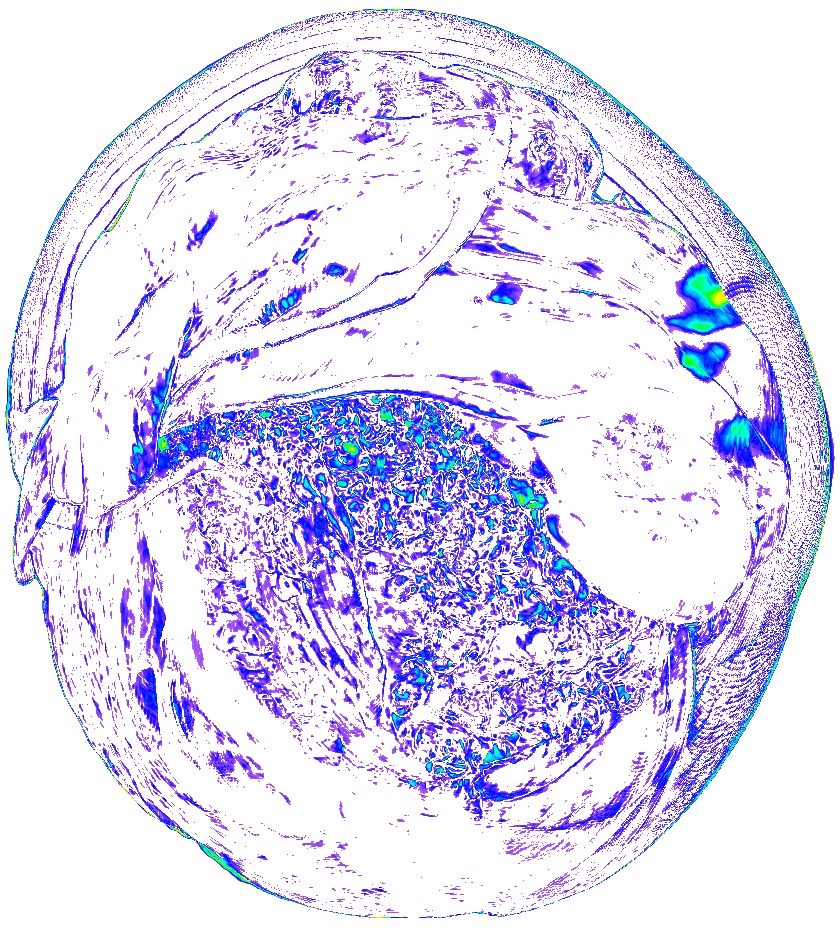}&
    \includegraphics[width=0.095\linewidth]{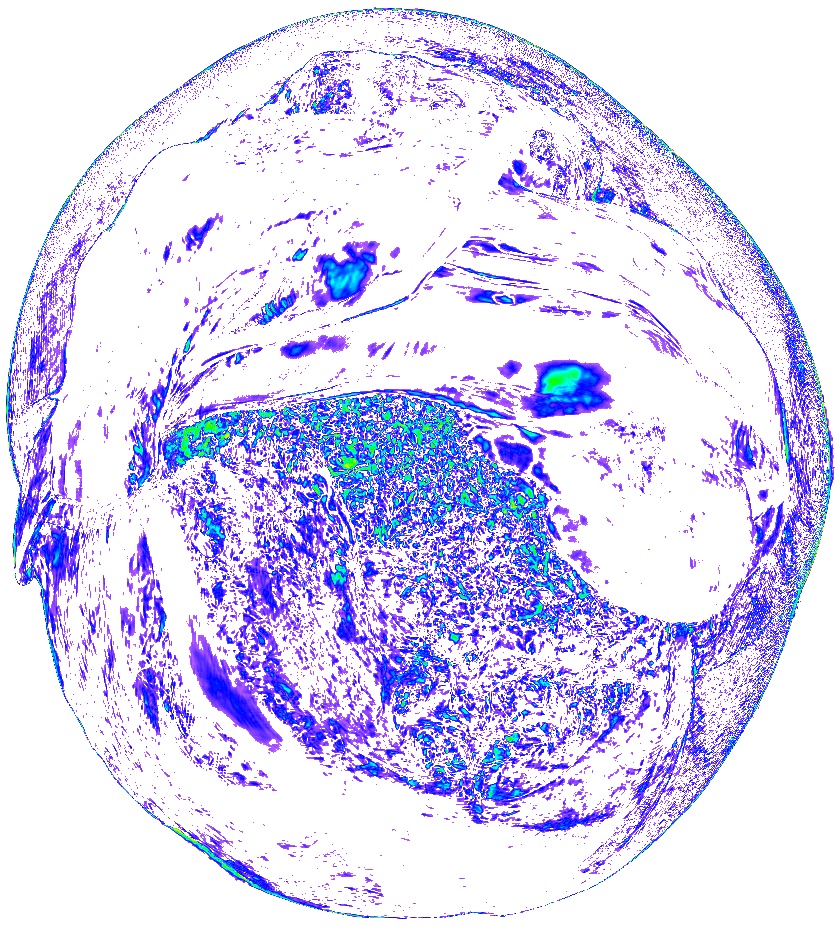}&
    \includegraphics[width=0.095\linewidth]{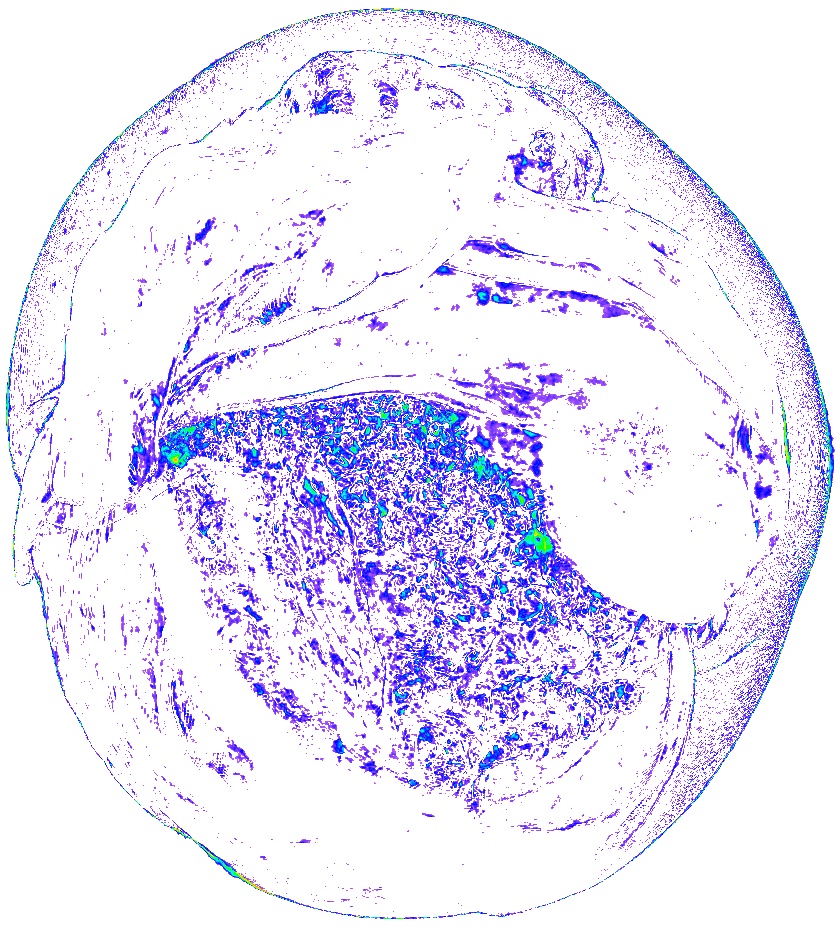}&
    \includegraphics[width=0.095\linewidth]{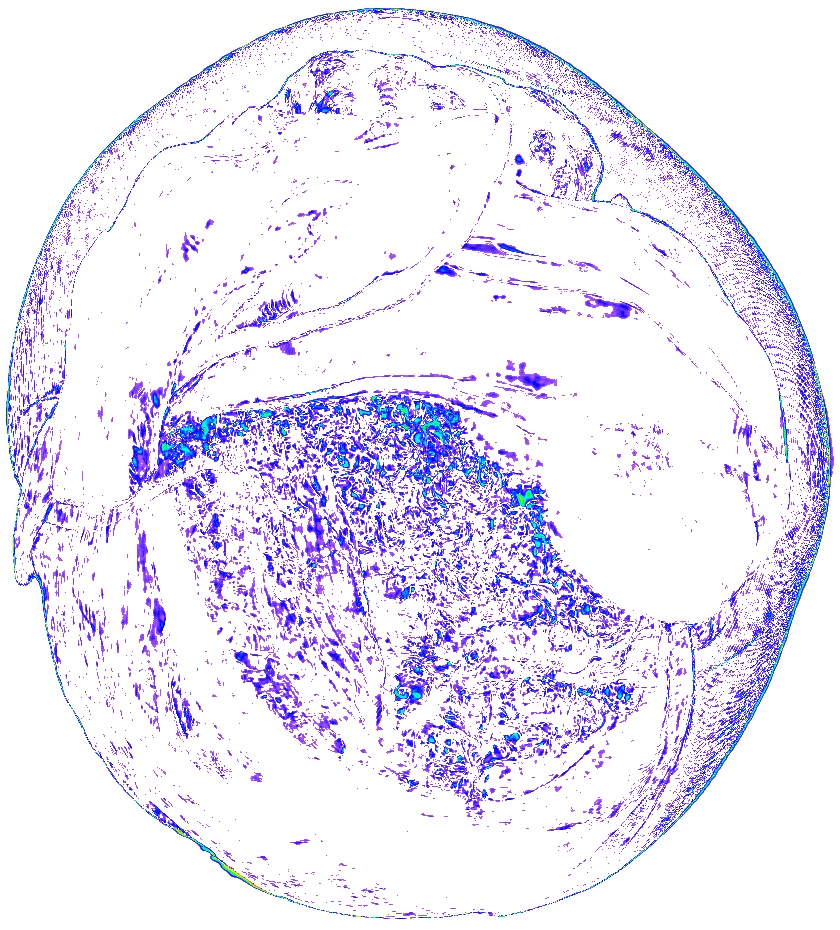}&
    \includegraphics[width=0.095\linewidth]{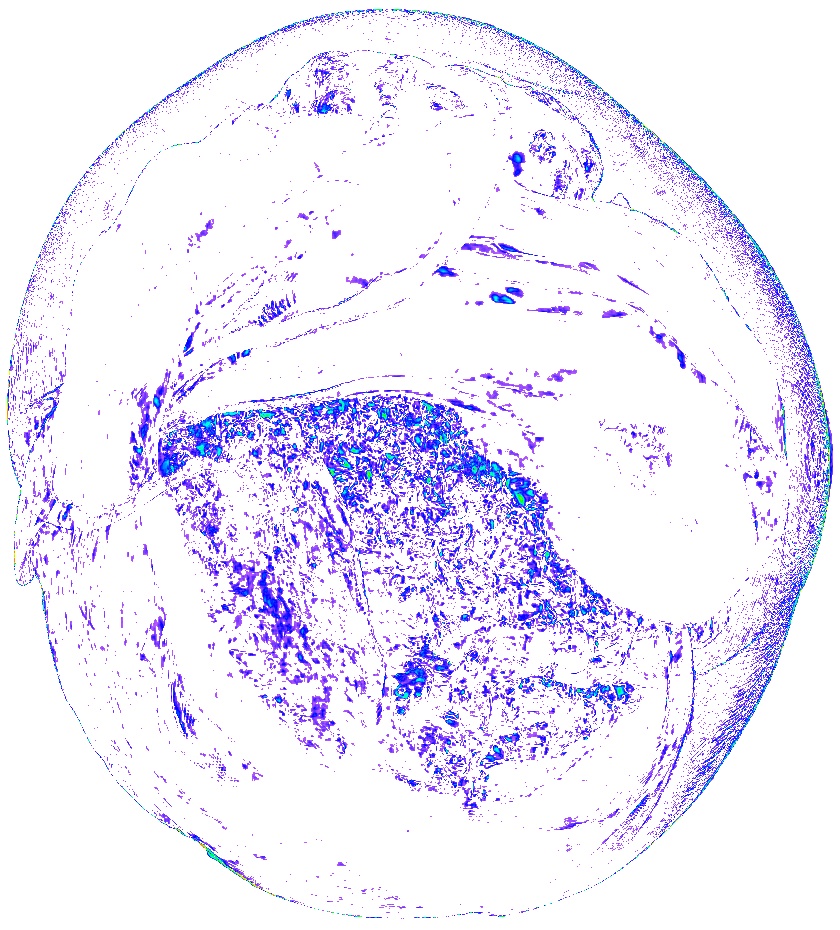}&
    \includegraphics[width=0.095\linewidth]{images/baselines/supernova_vr/GT-083-crop.png}\\
    \mbox{\footnotesize (a)} & \mbox{\footnotesize (b)} & \mbox{\footnotesize (c)} & \mbox{\footnotesize (d)} & \mbox{\footnotesize (e)} & \mbox{\footnotesize \prevhot{(f)}} & \mbox{\footnotesize \prevhot{(g)}} & \mbox{\footnotesize (h)} & \mbox{\footnotesize \prevhot{(i)}} & \mbox{\footnotesize (j)}
  \end{array}$
 \end{center}
\vspace{-.25in} 
 \caption{Difference images of novel view synthesis of DVR images for static scenes. 
 (a) to (h): InSituNet, CoordNet, StyleGAN2, EG3D, NeRF, 3DGS, Instant-NGP, \prevhot{TensoRF}, \prevhot{ViSNeRF}, and GT. 
 Top to bottom: vortex, five jets, Tangaroa, and supernova.} 
 \label{fig:comp-nvs-dvr-diff}
\end{figure}

\begin{figure}[t!]
  \begin{center}
  $\begin{array}{c@{\hspace{0.005in}}c@{\hspace{0.005in}}c@{\hspace{0.005in}}c@{\hspace{0.005in}}c@{\hspace{0.005in}}c@{\hspace{0.005in}}c@{\hspace{0.005in}}c@{\hspace{0.005in}}c@{\hspace{0.005in}}c}
    \includegraphics[width=0.095\linewidth]{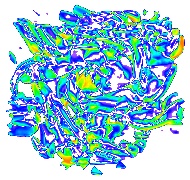}&
    \includegraphics[width=0.095\linewidth]{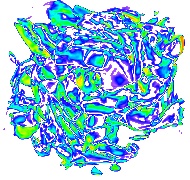}&
    \includegraphics[width=0.095\linewidth]{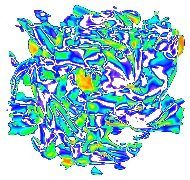}&
    \includegraphics[width=0.095\linewidth]{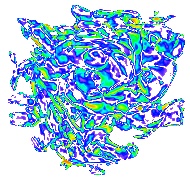}&
    \includegraphics[width=0.095\linewidth]{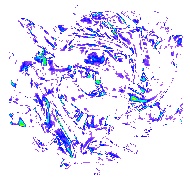}&
    \includegraphics[width=0.095\linewidth]{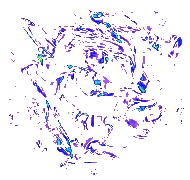}&
    \includegraphics[width=0.095\linewidth]{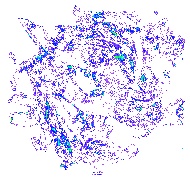}&
    \includegraphics[width=0.095\linewidth]{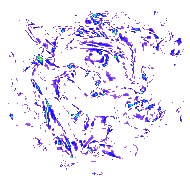}&
    \includegraphics[width=0.095\linewidth]{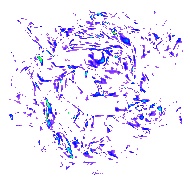}&
    \includegraphics[width=0.095\linewidth]{images/baselines/vortex_ir/GT-113-crop.png}\\
    \includegraphics[width=0.095\linewidth]{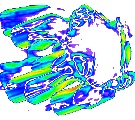}&
    \includegraphics[width=0.095\linewidth]{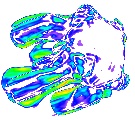}&
    \includegraphics[width=0.095\linewidth]{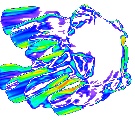}&
    \includegraphics[width=0.095\linewidth]{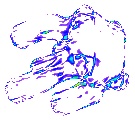}&
    \includegraphics[width=0.095\linewidth]{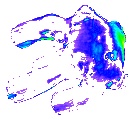}&
    \includegraphics[width=0.095\linewidth]{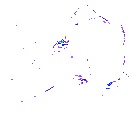}&
    \includegraphics[width=0.095\linewidth]{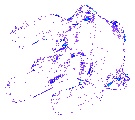}&
    \includegraphics[width=0.095\linewidth]{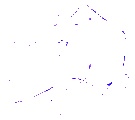}&
    \includegraphics[width=0.095\linewidth]{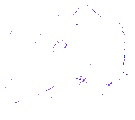}&
    \includegraphics[width=0.095\linewidth]{images/baselines/fivejets_ir/GT-066-crop.png}\\
    \includegraphics[width=0.095\linewidth]{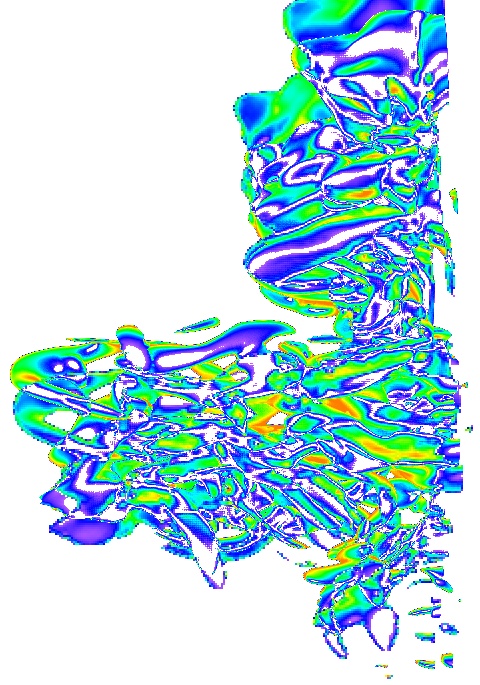}&
    \includegraphics[width=0.095\linewidth]{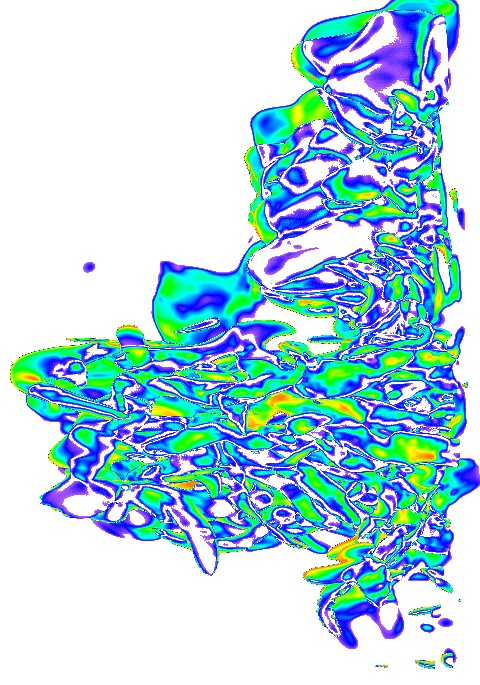}&
    \includegraphics[width=0.095\linewidth]{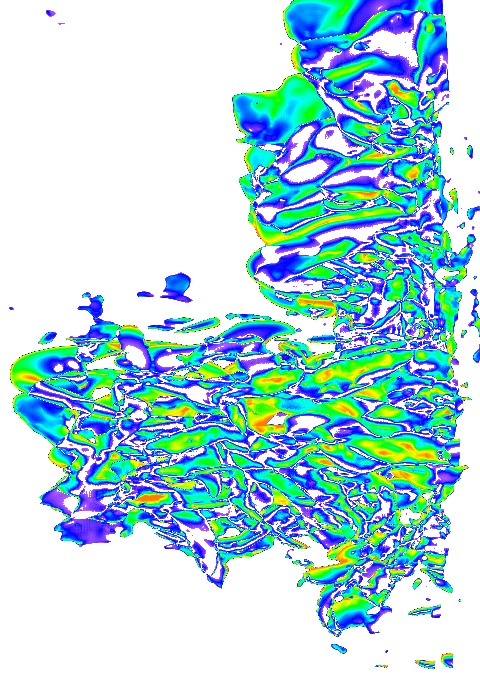}&
    \includegraphics[width=0.095\linewidth]{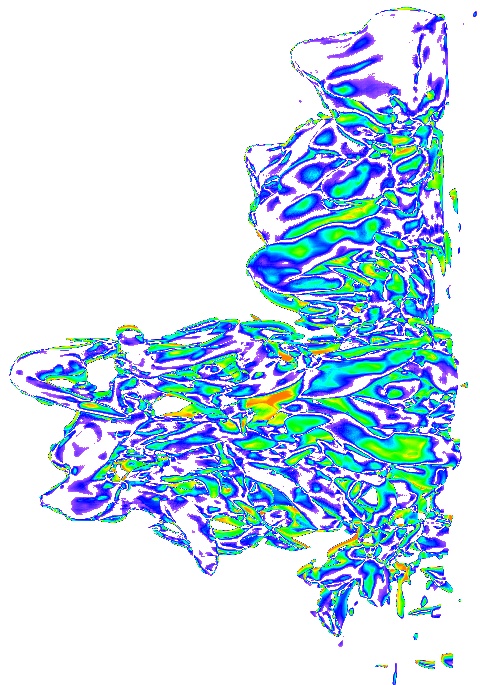}&
    \includegraphics[width=0.095\linewidth]{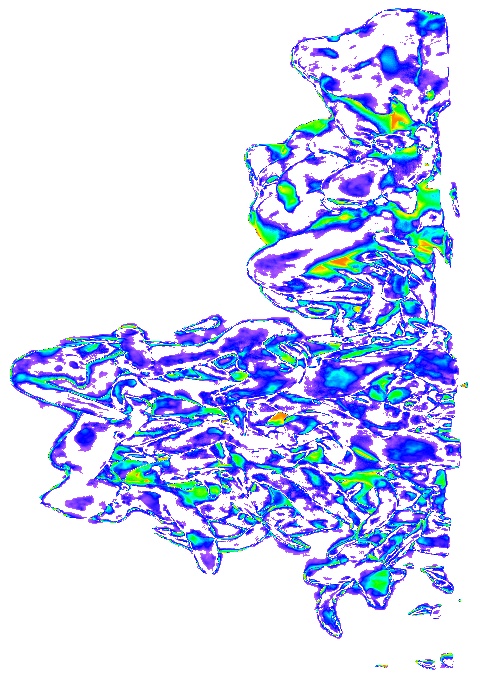}&
    \includegraphics[width=0.095\linewidth]{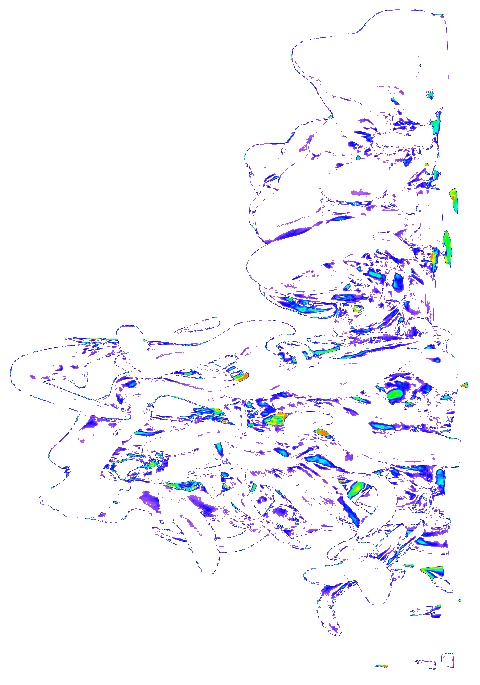}&
    \includegraphics[width=0.095\linewidth]{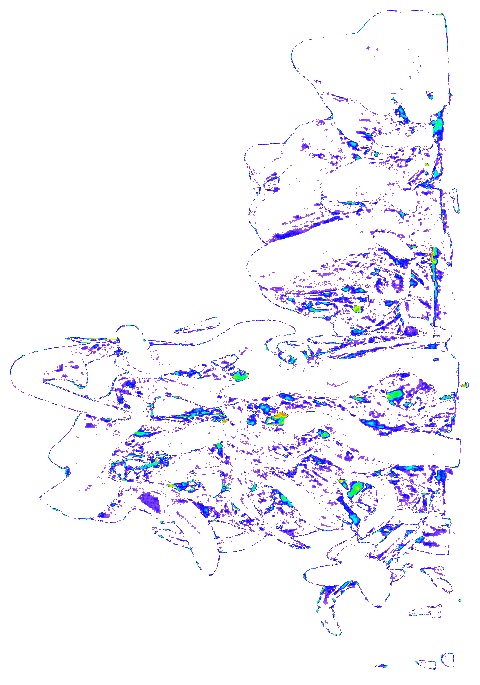}&
    \includegraphics[width=0.095\linewidth]{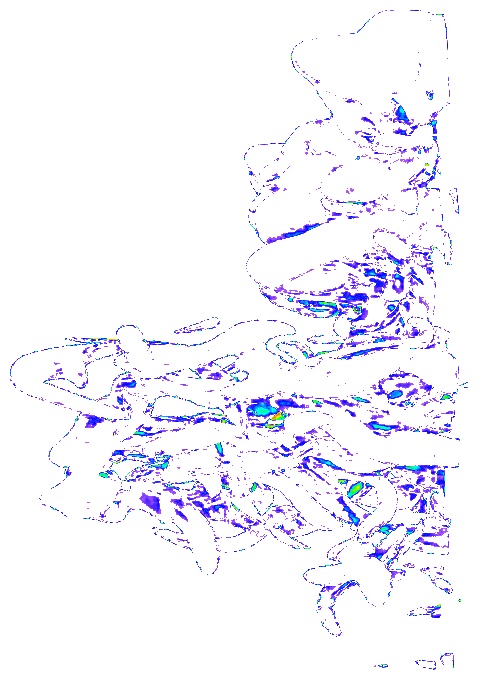}&
    \includegraphics[width=0.095\linewidth]{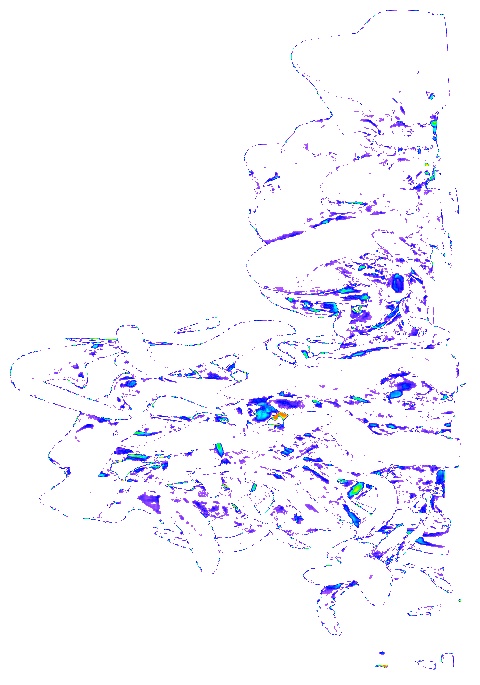}&
    \includegraphics[width=0.095\linewidth]{images/baselines/tangaroa_ir/GT-136-crop.png}\\
    \includegraphics[width=0.095\linewidth]{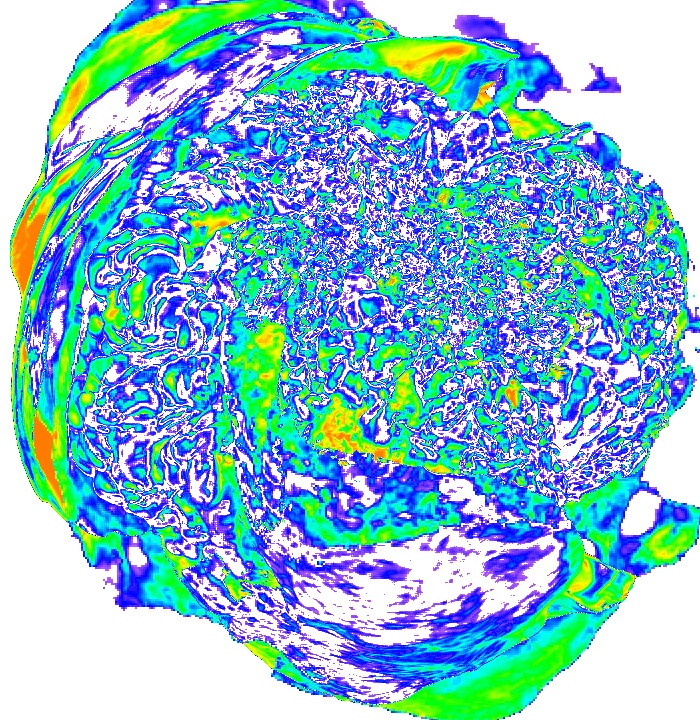}&
    \includegraphics[width=0.095\linewidth]{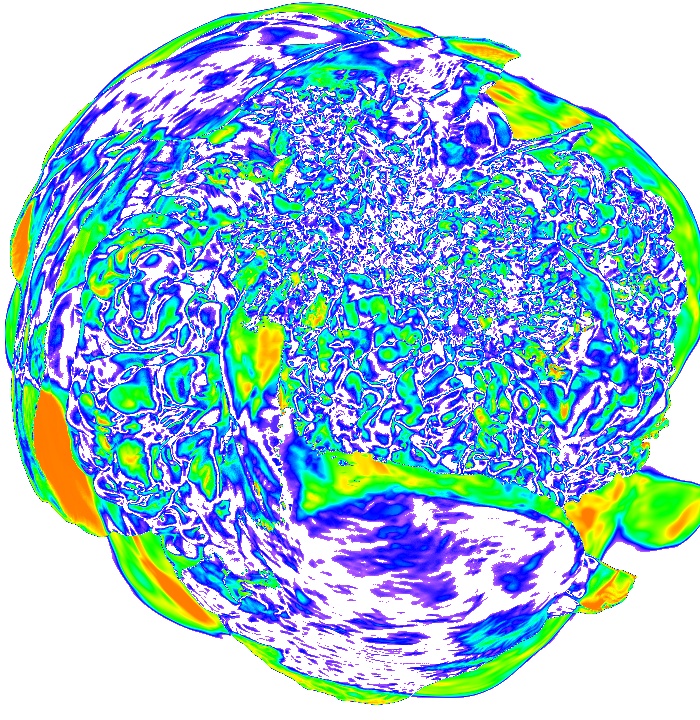}&
    \includegraphics[width=0.095\linewidth]{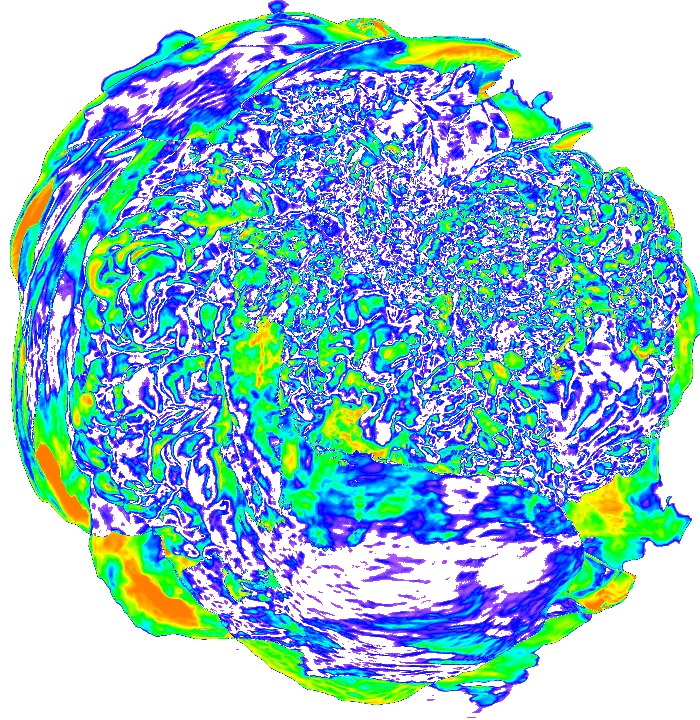}&
    \includegraphics[width=0.095\linewidth]{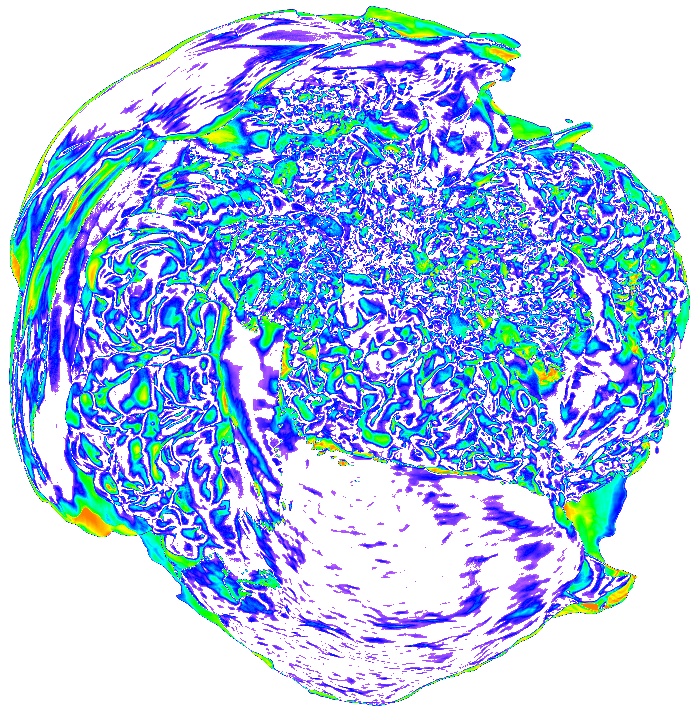}&
    \includegraphics[width=0.095\linewidth]{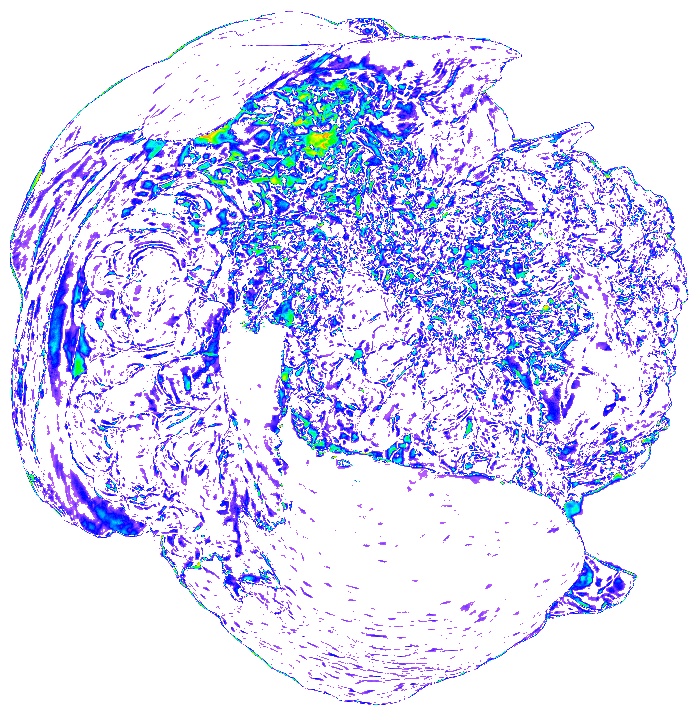}&
    \includegraphics[width=0.095\linewidth]{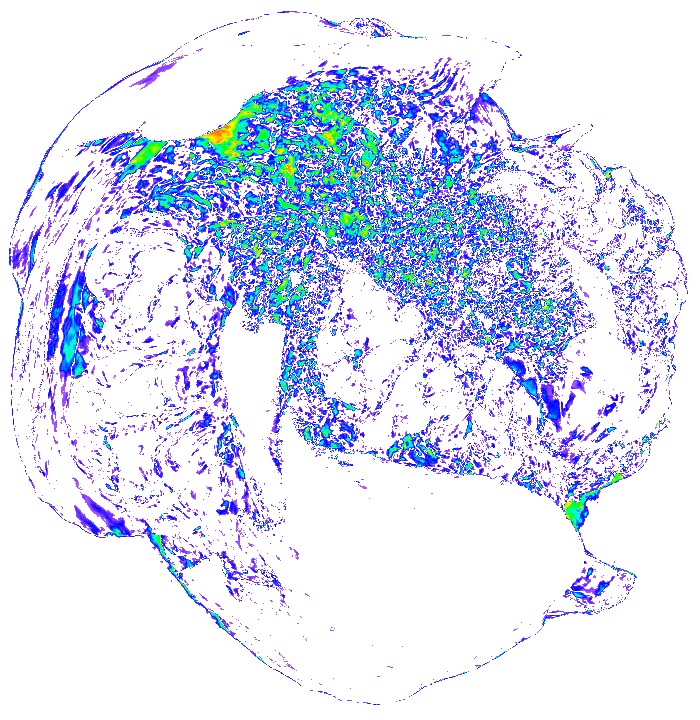}&
    \includegraphics[width=0.095\linewidth]{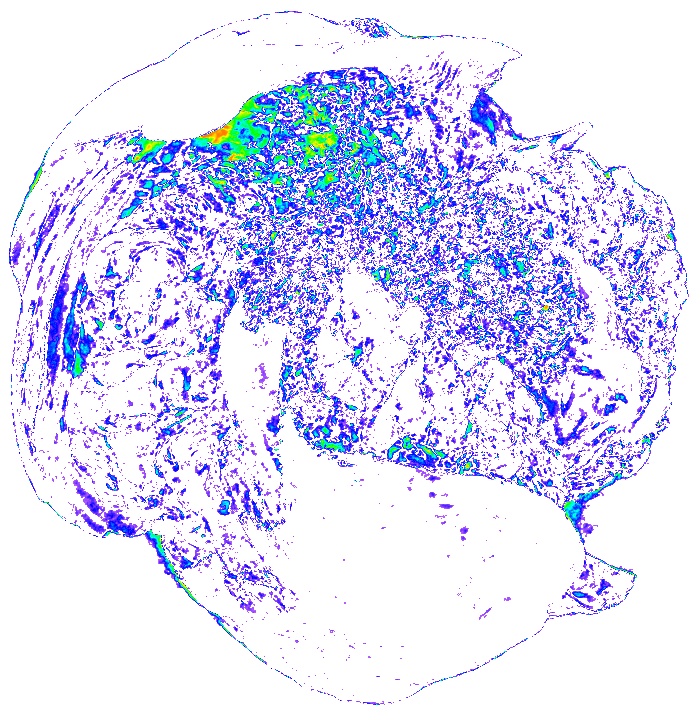}&
    \includegraphics[width=0.095\linewidth]{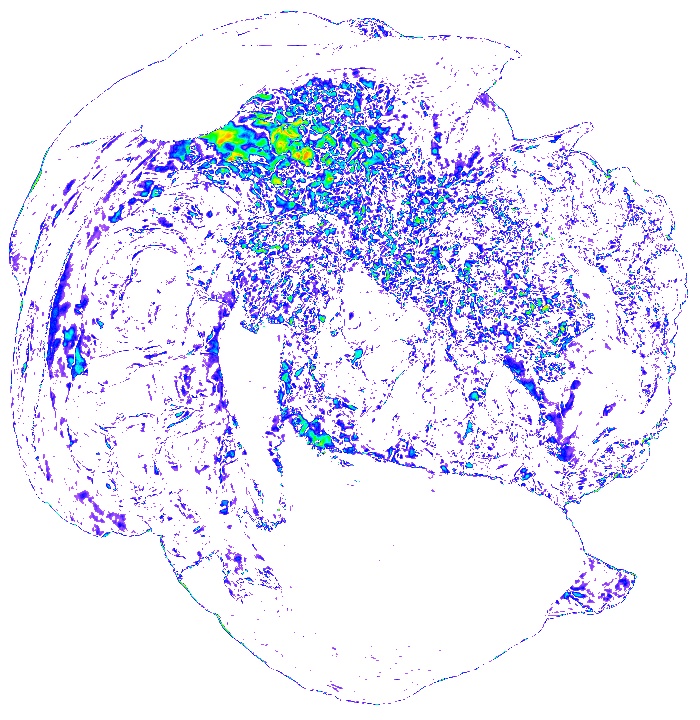}&
    \includegraphics[width=0.095\linewidth]{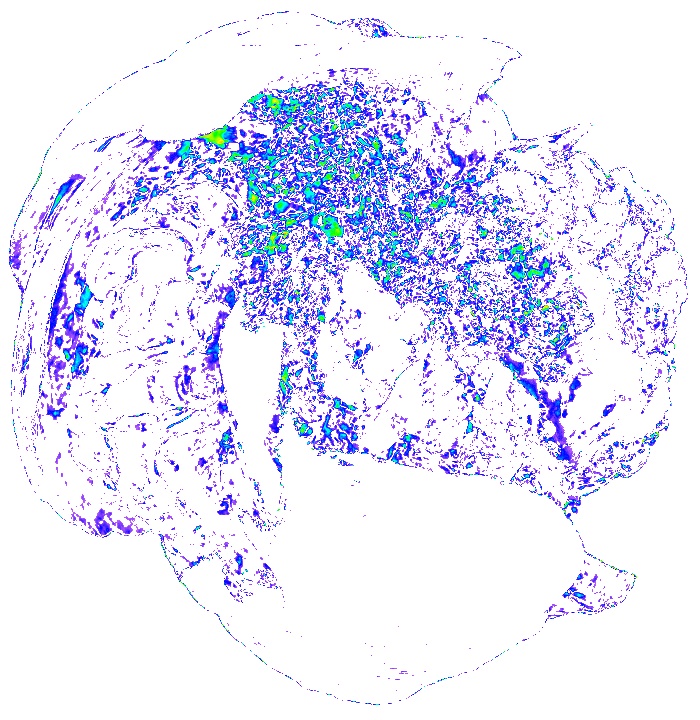}&
    \includegraphics[width=0.095\linewidth]{images/baselines/supernova_ir/GT-029-crop.png}\\
    \mbox{\footnotesize (a)} & \mbox{\footnotesize (b)} & \mbox{\footnotesize (c)} & \mbox{\footnotesize (d)} & \mbox{\footnotesize (e)} & \mbox{\footnotesize \prevhot{(f)}} & \mbox{\footnotesize \prevhot{(g)}} & \mbox{\footnotesize (h)} & \mbox{\footnotesize \prevhot{(i)}} & \mbox{\footnotesize (j)}
  \end{array}$
 \end{center}
\vspace{-.25in} 
 \caption{Difference images of novel view synthesis of IR images for static scenes. 
 (a) to (h): InSituNet, CoordNet, StyleGAN2, EG3D, NeRF, 3DGS, Instant-NGP, \prevhot{TensoRF}, \prevhot{ViSNeRF}, and GT. 
 Top to bottom: vortex, five jets, Tangaroa, and supernova.} 
 \label{fig:comp-nvs-ir-diff}
\end{figure}

\begin{table}[t!]
  \caption{Novel view synthesis of static scenes: average PSNR (dB), SSIM, and LPIPS values across all 181 synthesized views. 
  }
  \vspace{-0.1in}
  \centering
  \resizebox{\columnwidth}{!}{
  \begin{tabular}{c|c|rrr|rrr}
  	& & \multicolumn{3}{c|}{DVR}  & \multicolumn{3}{c}{IR} \\ \hline
  dataset       & method & PSNR$\uparrow$ & SSIM$\uparrow$ & LPIPS$\downarrow$ & PSNR$\uparrow$ & SSIM$\uparrow$ & LPIPS$\downarrow$ \\ \hline
                & InSituNet & 17.15 & 0.714 & 0.189 & 16.09 & 0.658 & 0.222 \\
                & CoordNet  & 17.42 & 0.722 & 0.210 & 16.26 & 0.656 & 0.259 \\
          & StyleGAN2 & 17.35 & 0.727 & 0.199 & 16.42 & 0.669 & 0.232 \\
 vortex 	           & EG3D      & 21.47    & 0.857 & 0.109 & 17.00 & 0.717 & 0.221 \\
 (256$\times$256)               & NeRF      & 28.78  & 0.963 & 0.023 & 24.51 & 0.926 & 0.063 \\
                & 3DGS & 33.30 & 0.990 & 0.009 & 28.60 & 0.966 & 0.028 \\
                & Instant-NGP & 32.00 & 0.986 & 0.014 & 26.03 & 0.933 & 0.092 \\
                & \prevhot{TensoRF} & \prevhot{35.43} & \prevhot{0.993} & \prevhot{0.005} & \prevhot{29.30} & \prevhot{0.968} & \prevhot{0.027} \\
                & \prevhot{ViSNeRF}   & \prevhot{\bf{37.32}} & \prevhot{\bf{0.996}} & \prevhot{\bf{0.003}} & \prevhot{\bf{29.46}} & \prevhot{\bf{0.970}} & \prevhot{\bf{0.025}} \\ \hline
                & InSituNet & 17.19 &  0.794 & 0.152 & 19.93 & 0.822 & 0.140 \\
                & CoordNet  & 18.30 & 0.820 & 0.119 & 20.70 & 0.849 & 0.119 \\
          & StyleGAN2 & 17.26 & 0.806 & 0.163 & 20.50 & 0.836 & 0.137  \\
  five jets	           & EG3D      & 24.13 & 0.911 & 0.044 & 26.79 & 0.936 & 0.046 \\
  (256$\times$256)		& NeRF      & 28.48 & 0.957 & 0.023 & 26.32 & 0.952 & 0.051 \\
                & 3DGS & 31.57 & 0.979 & 0.014 & 35.96 & 0.992 & 0.008 \\
                & Instant-NGP & 31.02 & 0.974 & 0.025 & 32.68 & 0.977 & 0.036 \\
                & \prevhot{TensoRF} & \prevhot{34.82} & \prevhot{0.990} & \prevhot{0.007} & \prevhot{38.63} & \prevhot{\bf{0.995}} & \prevhot{\bf{0.005}} \\
                & \prevhot{ViSNeRF}   & \prevhot{\bf{35.80}} & \prevhot{\bf{0.992}} & \prevhot{\bf{0.005}} &  \prevhot{\bf{38.66}} & \prevhot{\bf{0.995}} & \prevhot{\bf{0.005}} \\ \hline
                & InSituNet & 17.54 & 0.826 & 0.197 & 17.65 & 0.845 & 0.184 \\
                & CoordNet  & 18.79 & 0.835 & 0.212  & 18.34 & 0.851 & 0.192 \\
          & StyleGAN2 & 17.87 & 0.823 & 0.211 & 17.75 & 0.838 & 0.197 \\
  Tangaroa	           & EG3D      & 22.82 & 0.873 & 0.089 & 20.61 & 0.884 & 0.102 \\
  (1024$\times$1024)		& NeRF      & 29.64 & 0.954 & 0.045 & 23.95 & 0.928 & 0.089 \\
                & 3DGS & 32.36 & 0.979 & 0.026 & 30.47 & 0.980 & 0.026 \\
                & Instant-NGP & 35.19 & 0.986 & 0.013 & 30.77 & 0.976 & 0.030 \\
                & TensoRF   & 33.24 & 0.980 & 0.022 & 31.68 & 0.982 & 0.019 \\
                & \prevhot{ViSNeRF}   & \prevhot{\bf{37.25}} & \prevhot{\bf{0.991}} & \prevhot{\bf{0.007}} & \prevhot{\bf{32.23}} & \prevhot{\bf{0.984}} & \prevhot{\bf{0.016}} \\ \hline
                & InSituNet & 15.99 & 0.626 & 0.352 & 16.74 & 0.726 & 0.269 \\
                & CoordNet  & 18.05 & 0.674 & 0.381 & 17.76 & 0.744 & 0.277 \\
          & StyleGAN2 & 17.38 & 0.669 & 0.352 & 17.16 & 0.736 & 0.249 \\
  supernova	           & EG3D      & 20.00 & 0.697 & 0.346 & 20.22 & 0.774 & 0.226 \\
  (1024$\times$1024)		& NeRF      & 23.82  & 0.771 & 0.221 & 25.15 & 0.873 & 0.129 \\
                & 3DGS & 24.21 & 0.815 & 0.129 & 27.49 & 0.931 & 0.051 \\
                & Instant-NGP & 25.99 & 0.830 & 0.148 & 27.58 & 0.914 & 0.082 \\
                & TensoRF & 25.86 & 0.828 & 0.181 & 29.32 & 0.938 & 0.071 \\
                & ViSNeRF & \bf{27.01} & \bf{0.859}  & \bf{0.120} & \bf{29.71} & \bf{0.946} & \bf{0.049} \\ 
  \end{tabular}
  }
  \label{tab:psnr-lpips-nvs}
\end{table}

\begin{table}[t!]
  \caption{Novel view synthesis of static scenes: MS (in MB), TT (in hours), and IT (in minutes) across all 181 synthesized views.}
  \vspace{-0.1in}
  \centering
  \resizebox{\columnwidth}{!}{
  \begin{tabular}{c|c|rrr|rrr}
  	& & \multicolumn{3}{c|}{DVR}  & \multicolumn{3}{c}{IR} \\ \hline
  dataset       & method & MS$\downarrow$ & TT$\downarrow$ & IT$\downarrow$ & MS$\downarrow$ & TT$\downarrow$ & IT$\downarrow$ \\ \hline
                & InSituNet & 166.78 & 1.92 & 0.11 & 166.78 & 1.92 & \bf{0.08} \\
                & CoordNet  & 5.71 & 9.36 & 0.64 & 5.71 & 9.22 & 0.78 \\
                & StyleGAN2 & 102.62 & 1.52 & 0.08 & 102.62 & 1.50 & \bf{0.08} \\
  vortex	      & EG3D      & 145.24 & 14.73 & 0.37 & 145.24 & 14.70 & 0.37 \\
  (256$\times$256)              & NeRF      & \bf{2.27} & 27.26 & 9.40 & \bf{2.27} & 26.07 & 9.26 \\
                & 3DGS & 21.64 & 0.06 & \bf{0.07} & 37.49 & 0.07 & \bf{0.08} \\
                & Instant-NGP & 14.31 & \bf{0.05} & 0.23 & 14.49 & \bf{0.05} & 0.23 \\
                & \prevhot{TensoRF}   & \prevhot{12.41} & \prevhot{0.41} & \prevhot{0.97} & \prevhot{12.51} & \prevhot{0.42} & \prevhot{1.47} \\                
                & \prevhot{ViSNeRF}   & \prevhot{12.57} & 0.22 & \prevhot{0.83} & \prevhot{12.60} & 0.25 & \prevhot{0.88} \\ \hline
                & InSituNet & 166.78 & 2.02 & 0.08 & 166.78 & 1.91 & 0.07 \\
                & CoordNet  & 5.71 & 10.08 & 0.66 & 5.71 & 9.48 & 0.71 \\
                & StyleGAN2 & 102.62 & 1.49 & 0.10 & 102.62 & 1.50 & 0.10 \\
  five jets	    & EG3D      & 145.24 & 14.64 & 0.37 & 145.24 & 14.59 & 0.32 \\
  (256$\times$256)              & NeRF      & \bf{2.27} & 25.80 & 9.06 & \bf{2.27} & 25.66 & 9.26 \\
                & 3DGS  & 18.39 & \bf{0.05} & \bf{0.07} & 14.22 & \bf{0.05} & \bf{0.05} \\
                & Instant-NGP & 14.26 & \bf{0.05} & 0.23 & 14.29 & 0.06 & 0.20 \\
                & \prevhot{TensoRF}   & \prevhot{12.48} & \prevhot{0.42} & \prevhot{0.73} & \prevhot{12.59} & \prevhot{0.40} & \prevhot{0.77} \\                
                & \prevhot{ViSNeRF}   & \prevhot{12.62} & 0.26 & \prevhot{0.83} & \prevhot{12.68} & 0.26 & \prevhot{0.77} \\ \hline
                & InSituNet & 318.73 & 17.82 & \bf{0.37} & 318.73 & 17.80 & \bf{0.35} \\
                & CoordNet  & \bf{5.71} & 44.60 & 12.43 & \bf{5.71} & 43.20 & 11.16 \\
                & StyleGAN2 & 157.68 & 18.19 & 1.07 & 157.68 & 18.13 & 1.12 \\
  Tangaroa	    & EG3D      & 149.76 & 26.15 & 1.49 & 149.75 & 26.25 & 1.49 \\
  (1024$\times$1024)             & NeRF      & 16.36 & 27.89 & 148.99 & 16.36 & 27.23 & 148.06 \\
                & 3DGS  & 97.88 & 0.17 & 0.53 & 105.33 & 0.19 & 0.53 \\
                & Instant-NGP & 55.87 & \bf{0.07} & 2.57 & 57.35 & \bf{0.08} & 2.03 \\               
                & TensoRF   & 68.25 & 0.94 & 11.57 & 68.50 & 0.73 & 14.00 \\                
                & ViSNeRF   & 69.22 & 0.35 & 14.24 & 68.63 & 0.41 & 13.61 \\ \hline
                & InSituNet & 318.73 & 17.92 & \bf{0.53} & 318.73 & 17.75 & \bf{0.41} \\
                & CoordNet  & \bf{5.71} & 43.17 & 10.32 & \bf{5.71} & 44.00 & 10.47 \\
                & StyleGAN2 & 157.68 & 18.22 & 1.91 & 157.68 & 18.12 & 1.59 \\
  supernova	    & EG3D      & 149.75 & 26.38 & 1.35 & 149.75 & 26.27 & 1.32 \\
  (1024$\times$1024)              & NeRF      & 16.36 & 27.24 & 146.05 & 16.36 & 27.23 & 146.46 \\
                & 3DGS  & 149.41 & 0.32 & 1.22 & 120.55 & 0.26 & 1.07 \\
                & Instant-NGP & 57.06 & \bf{0.08} & 6.90 & 57.17 & \bf{0.08} & 4.57 \\                       
                & TensoRF   & 67.16 & 0.90 & 23.38 & 67.22 & 0.61 & 17.45 \\                
                & ViSNeRF   & 67.29 & 0.57 & 35.50 & 67.22 & 0.49 & 24.85 \\
  \end{tabular}
  }
  \label{tab:time-size-nvs}
\end{table}

\vspace{-0.05in}
\subsection{Quantitative Analysis}

As shown in Table~\ref{tab:psnr-lpips-nvs}, ViSNeRF outperforms all baselines in terms of PSNR, SSIM, and LPIPS for novel view synthesis in static scenes, using the same training datasets.
This performance advantage is consistent across resolutions, from lower resolution (256$\times$256) to higher resolution (1024$\times$1024), and applies to both DVR and IR scenes.
These results demonstrate ViSNeRF's ability to deliver consistently higher-quality image generation than the baselines in all tested scenarios.

According to Table~\ref{tab:time-size-nvs}, while ViSNeRF and TensoRF are significantly faster than NeRF, their training and rendering speeds remain slower than 3DGS and Instant-NGP. 
This is largely due to their implementation in PyTorch rather than CUDA and differences in NeRF representations.
While not suitable for real-time rendering, ViSNeRF is still practical for interactive applications.
Transitioning to the CUDA framework could be a future improvement, but the current efficiency of ViSNeRF is acceptable.
ViSNeRF has relatively close model sizes to 3DGS, Instant-NGP, and TensoRF, as their training configurations are carefully adjusted to ensure fair comparisons with equivalent model capabilities.

From Tables~\ref{tab:psnr-lpips-nvs}~and~\ref{tab:time-size-nvs}, we can draw similar conclusions as Table 3 in the paper for dynamic scenes when comparing ViSNeRF to InSituNet, CoordNet, StyleGAN2, and EG3D.

\vspace{-0.05in}
\subsection{Qualitative Analysis}

Figures~\ref{fig:comp-nvs-dvr}~and~\ref{fig:comp-nvs-ir} show DVR and IR images synthesized by ViSNeRF and baseline methods for comparison, where the difference images are shown in Figures~\ref{fig:comp-nvs-dvr-diff}~and~\ref{fig:comp-nvs-ir-diff}.
The visual quality of images synthesized by ViSNeRF is the best among all these methods.
Similar to the results observed in experiments with dynamic scenes, traditional 2D-based methods such as InSituNet, CoordNet, and StyleGAN2 struggle to reconstruct visualizations accurately.
In contrast, NeRF-based methods, including NeRF, 3DGS, Instant-NGP, TensoRF, and ViSNeRF, synthesize visualization images with substantially better visual quality.
Among these, ViSNeRF demonstrates finer details in the synthesized images as well as the lowest errors in difference images for both DVR and IR cases due to its ability to handle transparency and lighting with better accuracy and consistency in novel views.
This capability also ensures higher fidelity in synthesizing visualizations for dynamic scenes.

\vspace{-0.075in}
\section{Baseline Training Details}

Like ViSNeRF, all the baselines are trained using a single NVIDIA Tesla V100 graphic card with 32 GB of video memory.
While 3DGS and Instant-NGP leverage the CUDA framework to accelerate training and rendering, other baseline methods, including InSituNet, CoordNet, StyleGAN2, 3DGS, NeRF, and TensoRF, are implemented in PyTorch. 
In Table~\ref{tab:hyperparam}, we provide hyperparameters of the baselines for network training. 
InSituNet, StyleGAN2, and EG3D are trained with a batch of images per iteration, whereas CoordNet, NeRF, 3DGS, Instant-NGP, TensoRF, and ViSNeRF are trained with a batch of pixels in an image per iteration.

As shown in Tables~\ref{tab:iteration-256}~and~\ref{tab:iteration-1024}, in experiments of dynamic scenes, for a single parameter input (e.g., timestep), we found that tripling the number of training iterations compared to the static scene was enough for the models to converge adequately. 
When two or three input parameters are used (e.g., TF-1 or simulation parameters), the number of training iterations should be 6$\times$ or 7.5$\times$ for the static scene to train the baselines fully. 
In contrast, ViSNeRF can still be sufficiently trained for the tripled number of iterations.
Other hyperparameters and model configurations of baseline methods, such as the number of channels in each layer and positional encoding computation, follow the original papers and default values provided in the code.

\begin{table}[t!]
\caption{Hyperparameters of baseline training. For 1024$\times$1024 resolution, the batch size is reduced due to GPU memory constraint (shown in parentheses).}
\vspace{-0.1in}
\centering
{\fontsize{6}{7.2}\selectfont
\begin{tabular}{c|cccc}
method   & batch size & initial learning rate & $\beta_1$ & $\beta_2$ \\ \hline
InSituNet   & 4 (2) & $5 \times 10^{-5}$ & 0 & 0.999 \\
CoordNet    & 32,000 & $10^{-5}$ & 0.9 & 0.999 \\
StyleGAN2    & 16 (2) & $10^{-3}$ & 0 & 0.99 \\
EG3D        & 4 (1) & $2.5 \times 10^{-3}$ & 0 & 0.99 \\
NeRF        & 4096 & $5 \times 10^{-4}$ & 0.9 & 0.999 \\
3DGS  & 256 & $1.6 \times 10^{-4}$ & 0.9 & 0.999 \\
Instant-NGP & 262144 & $10^{-2}$ & 0.9 & 0.99 \\
TensoRF     & 4096 & 0.02 & 0.9 & 0.99 
\end{tabular}
}
\label{tab:hyperparam}
\end{table}

\begin{table}[t!]
\caption{Training iterations for different numbers of training images with 256$\times$256 resolution. The number of parameters is shown in parentheses (refer to Table 2 in the paper).}
\vspace{-0.1in}
\centering
{\fontsize{6}{7.2}\selectfont
\begin{tabular}{c|cccc}
method  & 42 (1) & 462 (1) & 1512 (2) & \prevhot{1890 (3)} \\ \hline
InSituNet  & 150,000 & 450,000 & 900,000 & \prevhot{1,125,000} \\
CoordNet   & 150,000 & 450,000 & 900,000 & \prevhot{1,125,000}   \\
StyleGAN2   & 150,000 & 450,000 & 900,000 & \prevhot{1,125,000}  \\
EG3D       & 150,000 & 450,000 & 900,000 & \prevhot{1,125,000} \\
NeRF       & 100,000 & --- & ---  & \prevhot{---} \\
3DGS & 30,000 & --- & --- & --- \\
Instant-NGP & 35,000 & --- & --- & --- \\
TensoRF    & 30,000 & --- & --- & \prevhot{---} \\
ViSNeRF    & 30,000 & 90,000 & 90,000 & \prevhot{90,000}
\end{tabular}
}
\label{tab:iteration-256}
\end{table}

\begin{table}[t!]
  \caption{Training iterations for different numbers of training images with 1024$\times$1024 resolution. The number of parameters is shown in parentheses (refer to Table 2 in the paper).}
  \vspace{-0.1in}
  \centering
  {\fontsize{6}{7.2}\selectfont  
  \begin{tabular}{c|cc}
    method  & 42 (1) & 546 (1) \\ \hline
  InSituNet  & 150,000 & 450,000 \\
  CoordNet   & 600,000 & 1,800,000  \\
  StyleGAN2   & 150,000 & 450,000  \\
  EG3D       & 150,000 & 450,000  \\
  NeRF       & 100,000 & ---  \\
  3DGS & 30,000 & --- \\
  Instant-NGP & 35,000 & --- \\
  TensoRF    & 30,000 & --- \\
  ViSNeRF    & 30,000 & 90,000 
  \end{tabular}
  }
  \label{tab:iteration-1024}
  \end{table}

\begin{table}[t!]
\caption{\prevhot{Hyperparameter study: number of training images using Tangaroa IR images.}}
\vspace{-0.1in}
\centering
{\fontsize{6}{7.2}\selectfont
\begin{tabular}{c|ccc}
\# training images & 12 & 42 & 92 \\ \hline
PSNR$\uparrow$ & \prevhot{24.16} & \prevhot{32.23} & \prevhot{{\bf 33.72}} \\
SSIM$\uparrow$ & \prevhot{0.937} & \prevhot{0.984} & \prevhot{{\bf 0.988}} \\
LPIPS$\downarrow$ & \prevhot{0.053} & \prevhot{0.016} & \prevhot{{\bf 0.013}}
\end{tabular}
}
\label{tab:param-img}
\end{table}
 
\begin{figure}[t!]
  \begin{center}
  $\begin{array}{c@{\hspace{0.025in}}c@{\hspace{0.025in}}c@{\hspace{0.025in}}c}
  \includegraphics[height=1.15in]{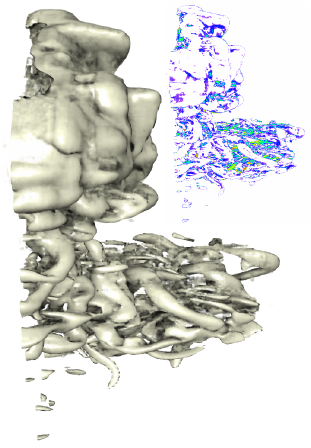}&
  \includegraphics[height=1.15in]{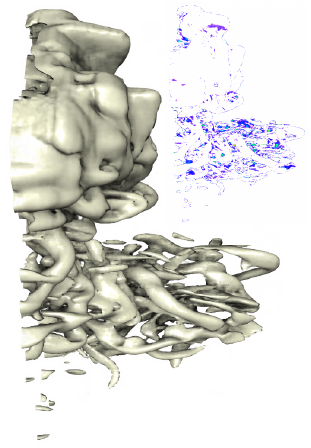}&
  \includegraphics[height=1.15in]{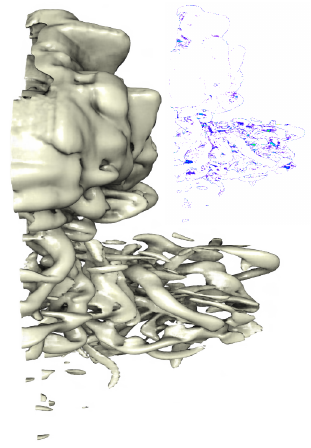}&
  \includegraphics[height=1.15in]{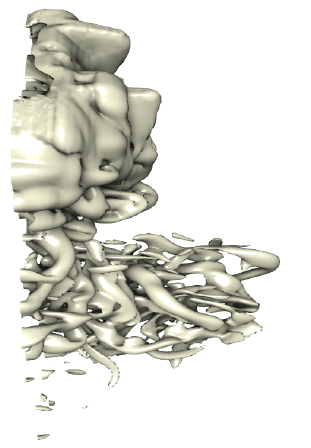}\\
 \mbox{\footnotesize (a) 12} & \mbox{\footnotesize (b) 42} & \mbox{\footnotesize (c) 92} & \mbox{\footnotesize (d) GT} \\
 \end{array}$
 \end{center}
 \vspace{-.25in} 
 \caption{\prevhot{Inferred Tangaroa IR images using different numbers of training images.}} 
 \label{fig:comp-img}
 \end{figure}

\begin{table}[t!]
\caption{\prevhot{Hyperparameter study: number of training iterations using five jets DVR images.}}
\vspace{-0.1in}
\centering
{\fontsize{6}{7.2}\selectfont
\begin{tabular}{c|cccc}
\# iterations               & 10,000 & 20,000 & 30,000 & 40,000 \\ \hline
PSNR$\uparrow$      & \prevhot{33.02} & \prevhot{34.83} & \prevhot{35.80} & \prevhot{{\bf 35.98}} \\
SSIM$\uparrow$       & \prevhot{0.985} & \prevhot{0.990} & \prevhot{{\bf 0.992}} & \prevhot{{\bf 0.992}} \\
LPIPS$\downarrow$ & \prevhot{0.010} & \prevhot{0.007} & \prevhot{{\bf 0.005}} & \prevhot{{\bf 0.005}}
\end{tabular}
}
\label{tab:param-iter}
\end{table}

\begin{figure}[t!]
  \begin{center}
   
  $\begin{array}{c@{\hspace{0.01in}}c@{\hspace{0.01in}}c@{\hspace{0.01in}}c@{\hspace{0.01in}}c}
  \includegraphics[width=0.185\linewidth]{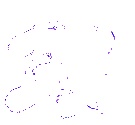}&
  \includegraphics[width=0.185\linewidth]{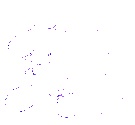}&
  \includegraphics[width=0.185\linewidth]{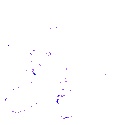}&
  \includegraphics[width=0.185\linewidth]{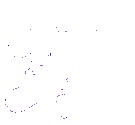}&
\\
  \includegraphics[width=0.185\linewidth]{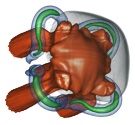}&
  \includegraphics[width=0.185\linewidth]{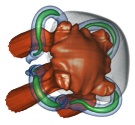}&
  \includegraphics[width=0.185\linewidth]{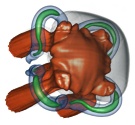}&
  \includegraphics[width=0.185\linewidth]{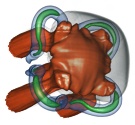}&
  \includegraphics[width=0.185\linewidth]{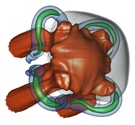}\\
  \mbox{\footnotesize (a) 10,000} & \mbox{\footnotesize (b) 20,000} & \mbox{\footnotesize (c) 30,000} & \mbox{\footnotesize (d) 40,000} & \mbox{\footnotesize (e) GT}
  \end{array}$
 \end{center}
 \vspace{-.25in} 
 \caption{\prevhot{Inferred five jets DVR images using different numbers of training iterations.}} 
 \label{fig:comp-iter}
 \end{figure}

 \begin{table}[t!]
  \caption{\prevhot{Hyperparameter study: number of feature grid voxels using supernova DVR images.}}
  \vspace{-0.1in}
  \centering
  {\fontsize{6}{7.2}\selectfont
  \begin{tabular}{c|cccc}
  \# voxels              & $200^3$ & $300^3$ & $400^3$ & $500^3$ \\ \hline
  PSNR$\uparrow$      & \prevhot{26.01} & \prevhot{27.01} & \prevhot{27.17} & \prevhot{{\bf 27.22}} \\
  SSIM$\uparrow$       & \prevhot{0.821} & \prevhot{0.859} & \prevhot{0.864} & \prevhot{{\bf 0.865}} \\
  LPIPS$\downarrow$ & \prevhot{0.183} & \prevhot{0.120} & \prevhot{0.113} & \prevhot{{\bf 0.108}}
  \end{tabular}
  }
  \label{tab:param-voxel}
  \end{table}

\begin{figure}[t!]
\begin{center}

$\begin{array}{c@{\hspace{0.01in}}c@{\hspace{0.01in}}c@{\hspace{0.01in}}c@{\hspace{0.01in}}c}
\includegraphics[width=0.185\linewidth]{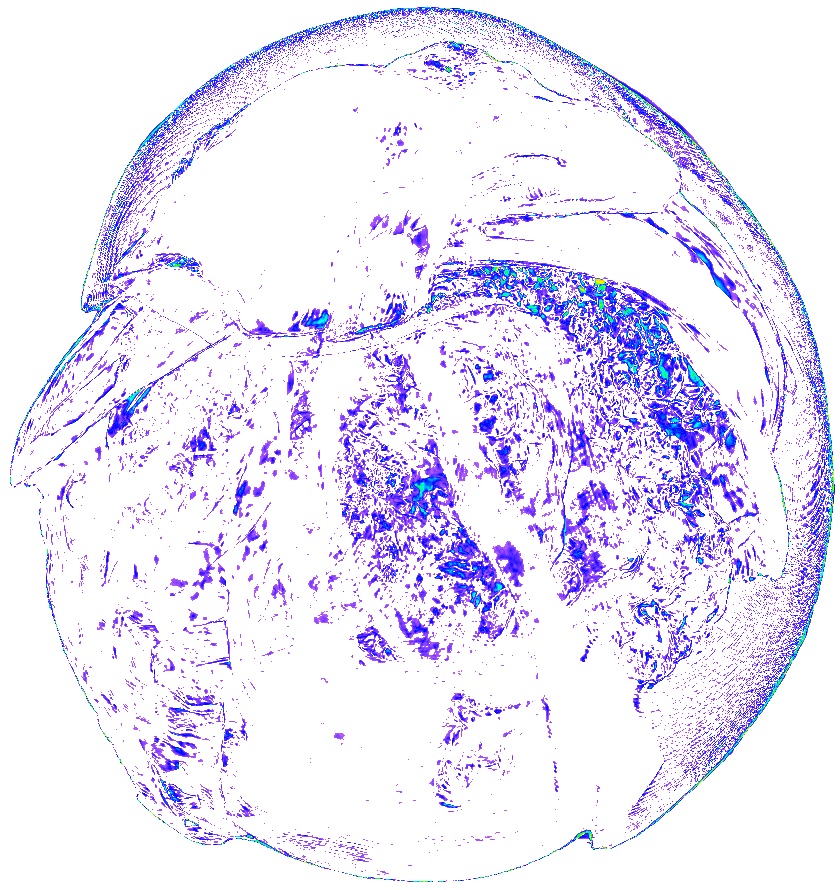}&
\includegraphics[width=0.185\linewidth]{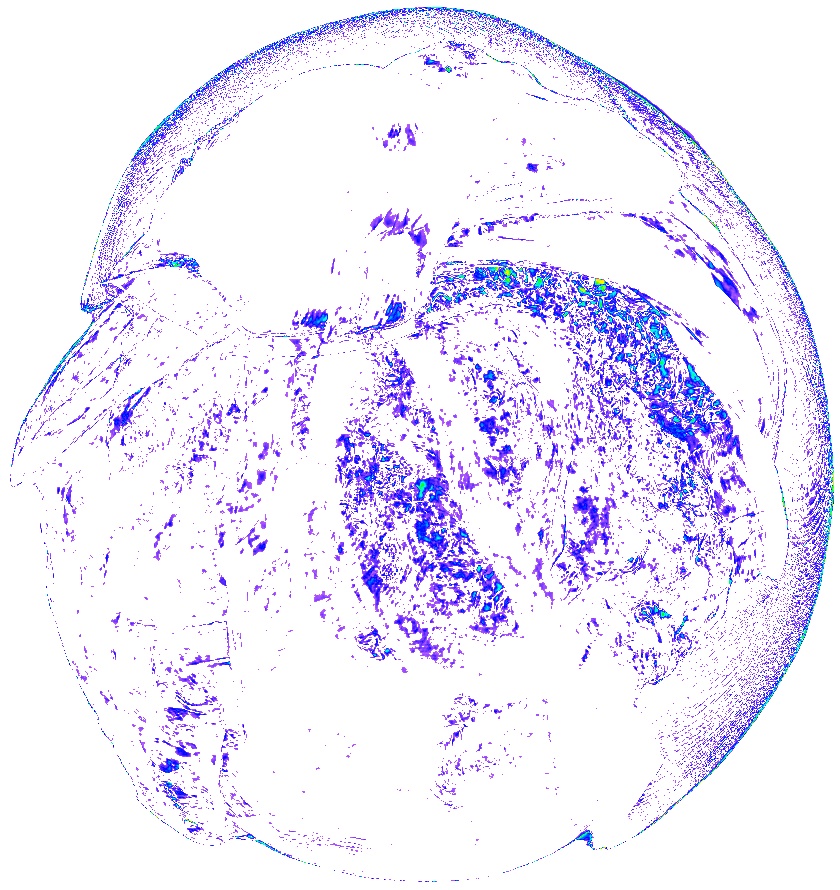}&
\includegraphics[width=0.185\linewidth]{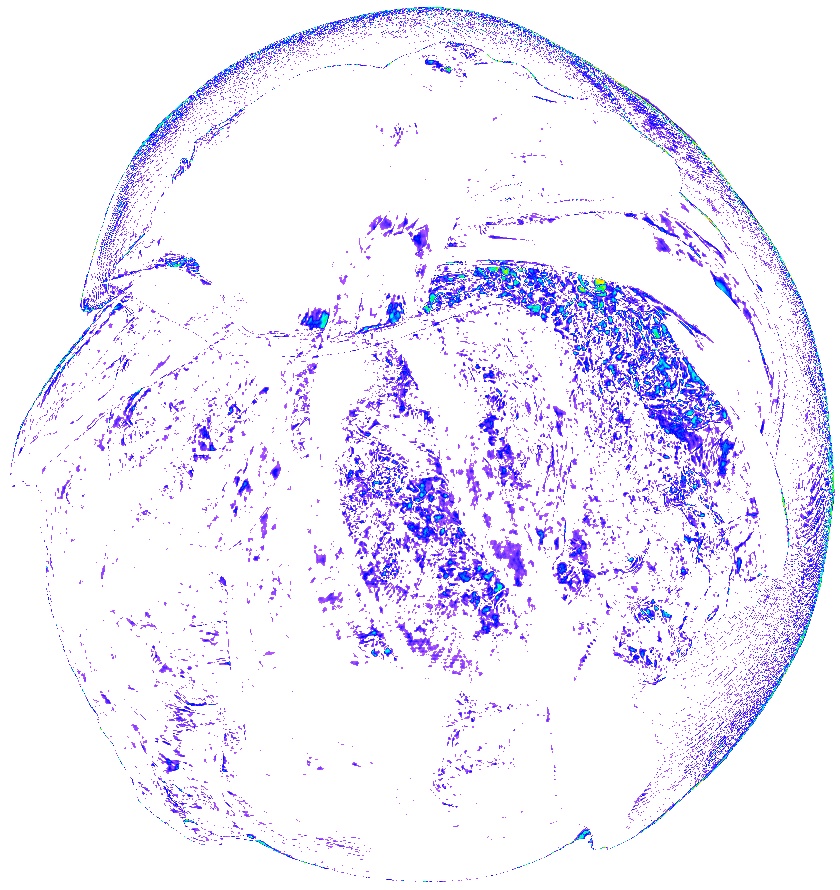}&
\includegraphics[width=0.185\linewidth]{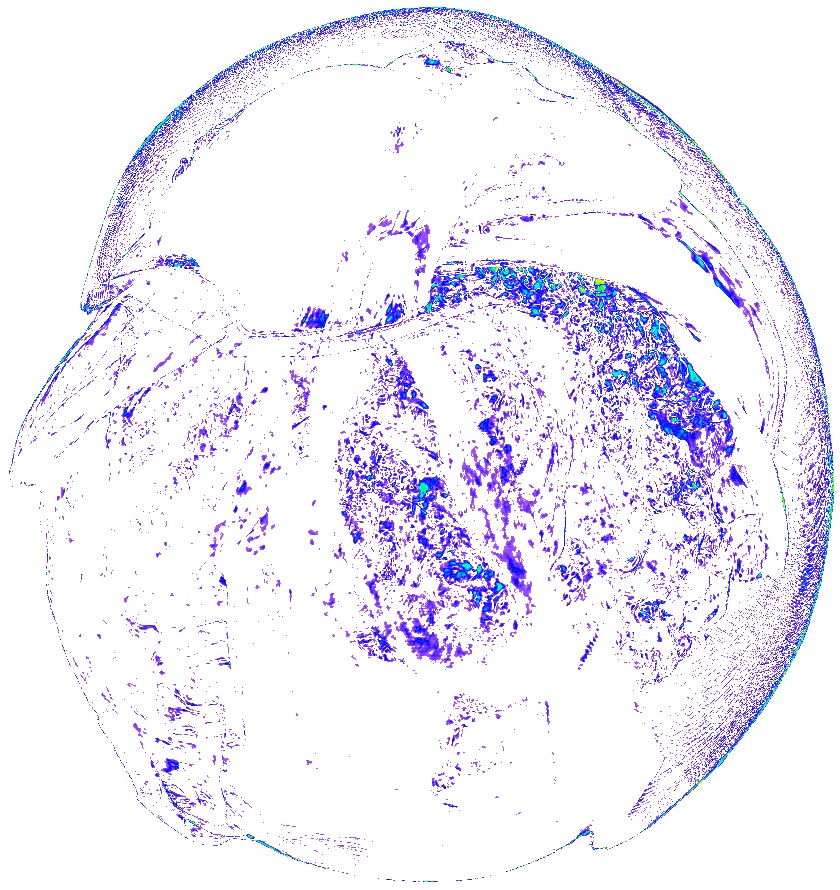}&
\\
\includegraphics[width=0.185\linewidth]{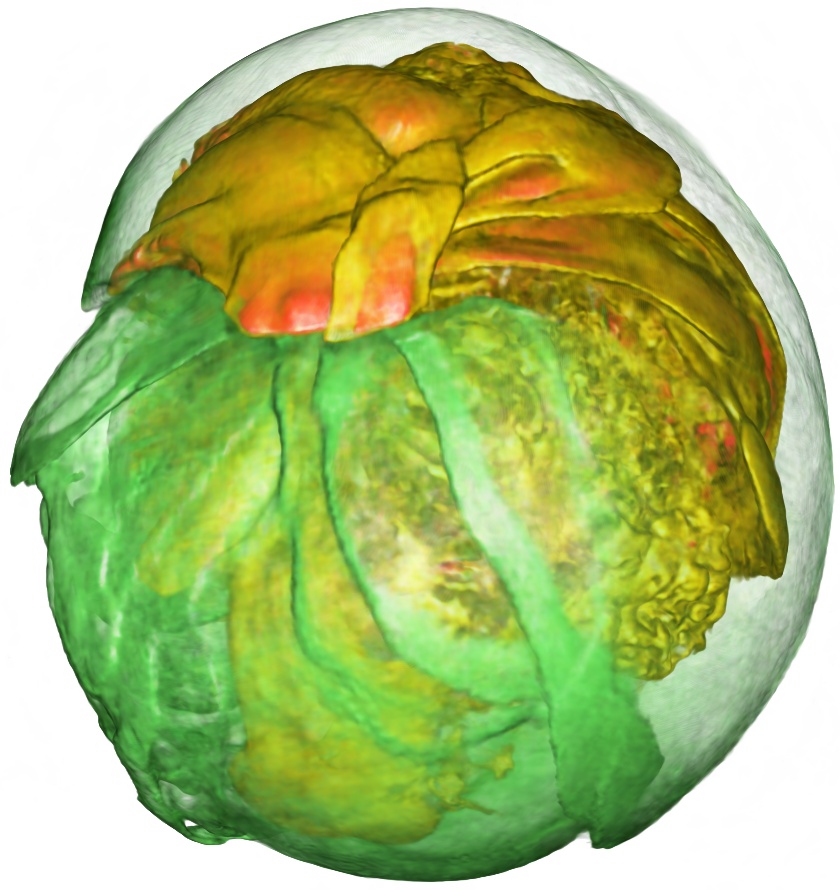}&
\includegraphics[width=0.185\linewidth]{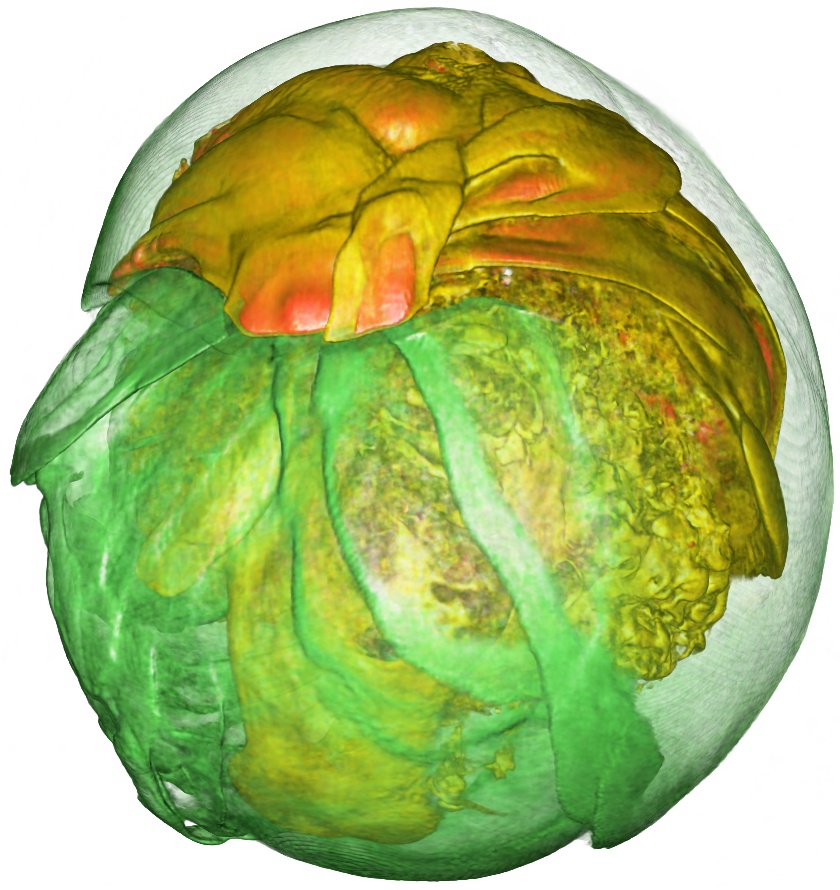}&
\includegraphics[width=0.185\linewidth]{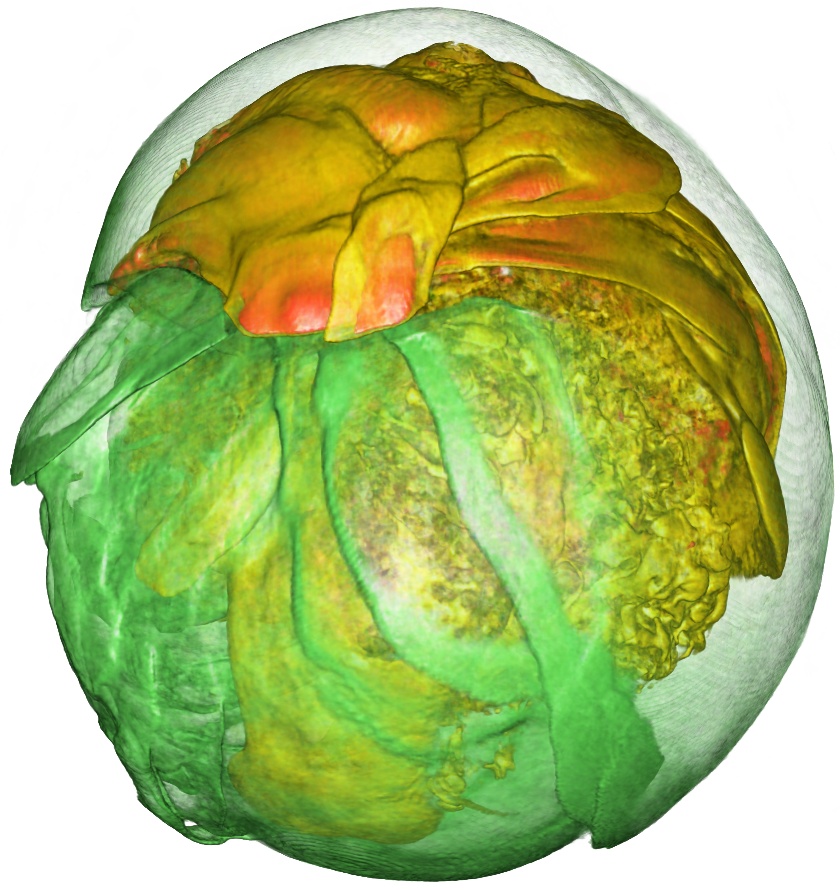}&
\includegraphics[width=0.185\linewidth]{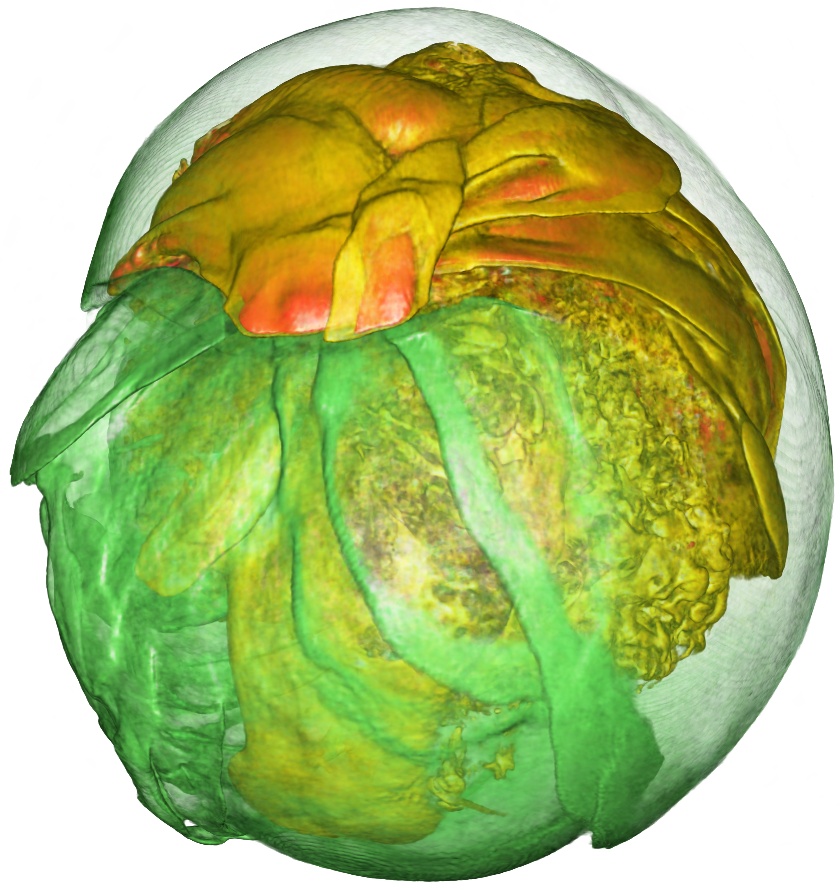}&
\includegraphics[width=0.185\linewidth]{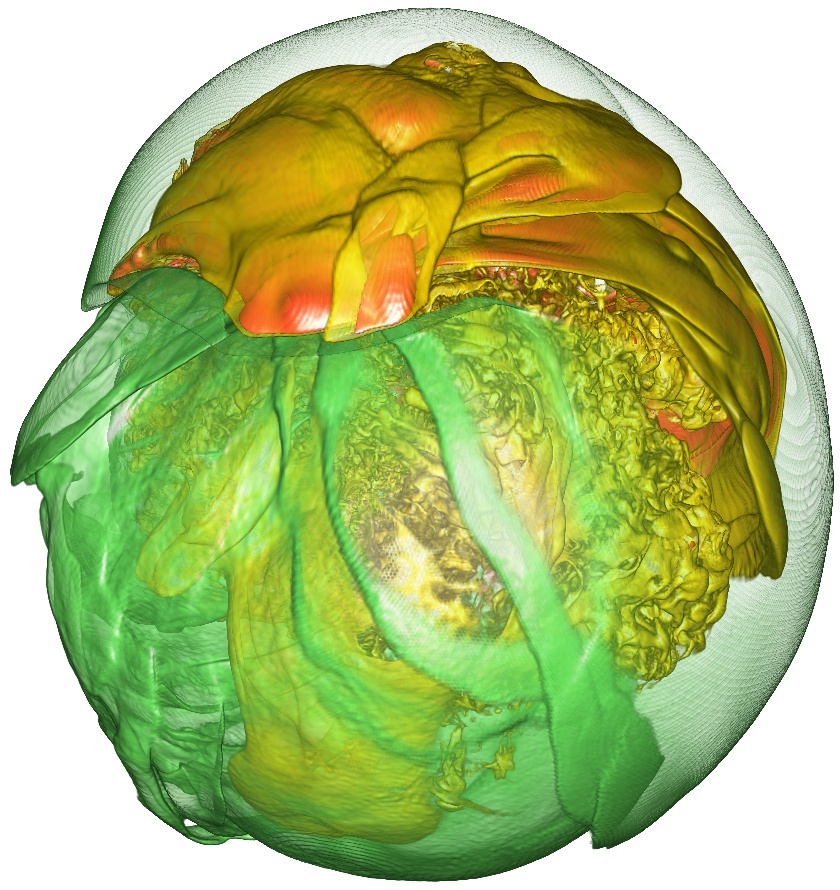}\\
\mbox{\footnotesize (a) $200^3$} & \mbox{\footnotesize (b) $300^3$} & \mbox{\footnotesize (c) $400^3$} & \mbox{\footnotesize (d) $500^3$} & \mbox{\footnotesize (e) GT}
\end{array}$
\end{center}
\vspace{-.25in} 
\caption{\prevhot{Inferred supernova DVR images using different numbers of feature grid voxels.}} 
\label{fig:comp-voxel}
\end{figure}

 \begin{table}[t!]
  \caption{\prevhot{Hyperparameter study: depth of density feature using supernova IR images.}}
  \vspace{-0.1in}
  \centering
  {\fontsize{6}{7.2}\selectfont
  \begin{tabular}{c|cccc}
  depth               & 8 & 16 & 24 & 32 \\ \hline
  PSNR$\uparrow$      & \prevhot{29.08} & \prevhot{29.71} & \prevhot{\bf{29.79}} & \prevhot{29.77} \\
  SSIM$\uparrow$       & \prevhot{0.937} & \prevhot{0.946} & \prevhot{{\bf 0.947}} & \prevhot{{\bf 0.947}} \\
  LPIPS$\downarrow$ & \prevhot{0.072} & \prevhot{\bf{0.049}} & \prevhot{0.051} & \prevhot{0.050}
  \end{tabular}
  }
  \label{tab:param-sigma}
  \end{table}

  \begin{figure}[t!]
  \begin{center}
  
  $\begin{array}{c@{\hspace{0.01in}}c@{\hspace{0.01in}}c@{\hspace{0.01in}}c@{\hspace{0.01in}}c}
  \includegraphics[width=0.185\linewidth]{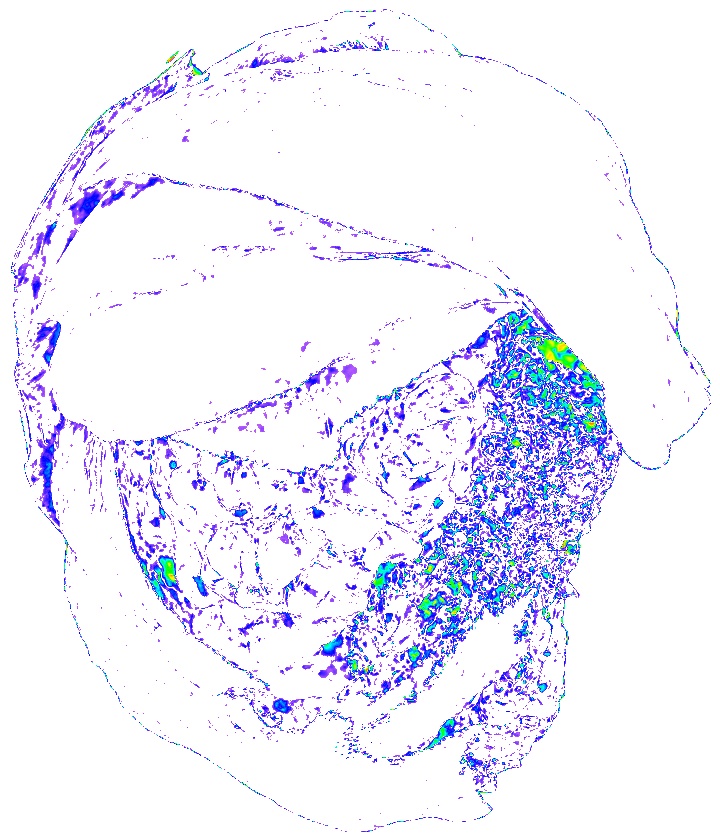}&
  \includegraphics[width=0.185\linewidth]{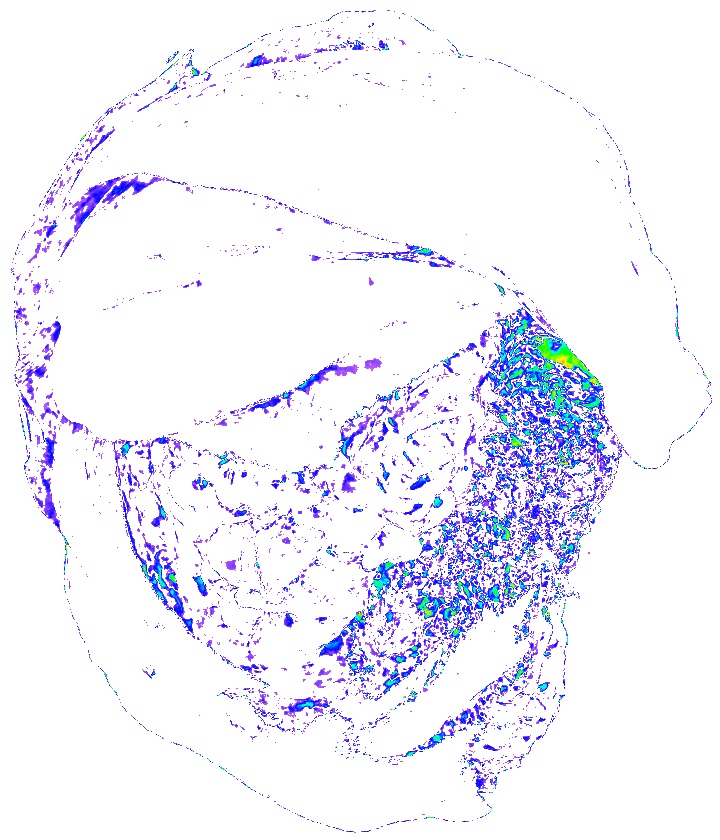}&
  \includegraphics[width=0.185\linewidth]{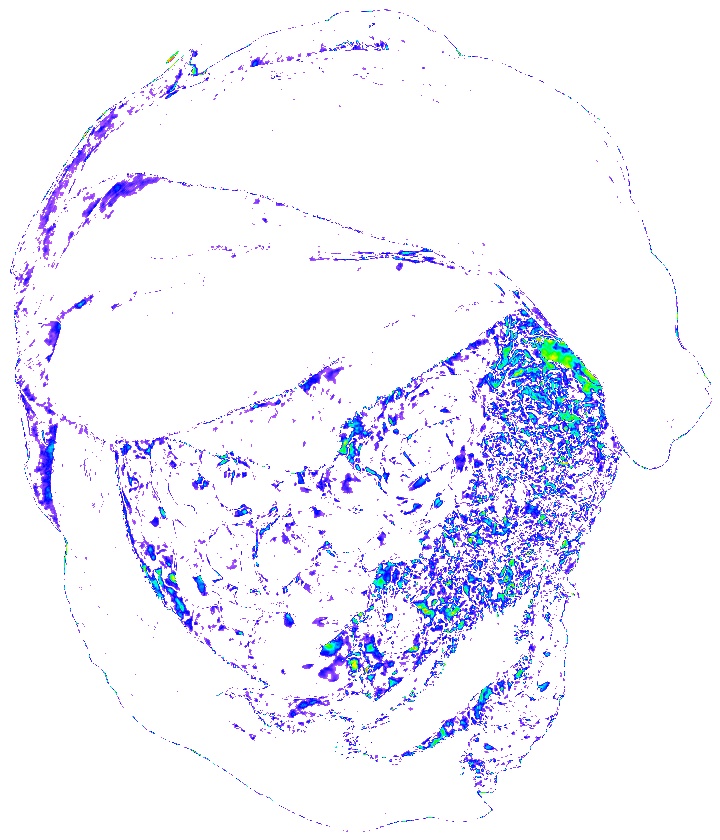}&
  \includegraphics[width=0.185\linewidth]{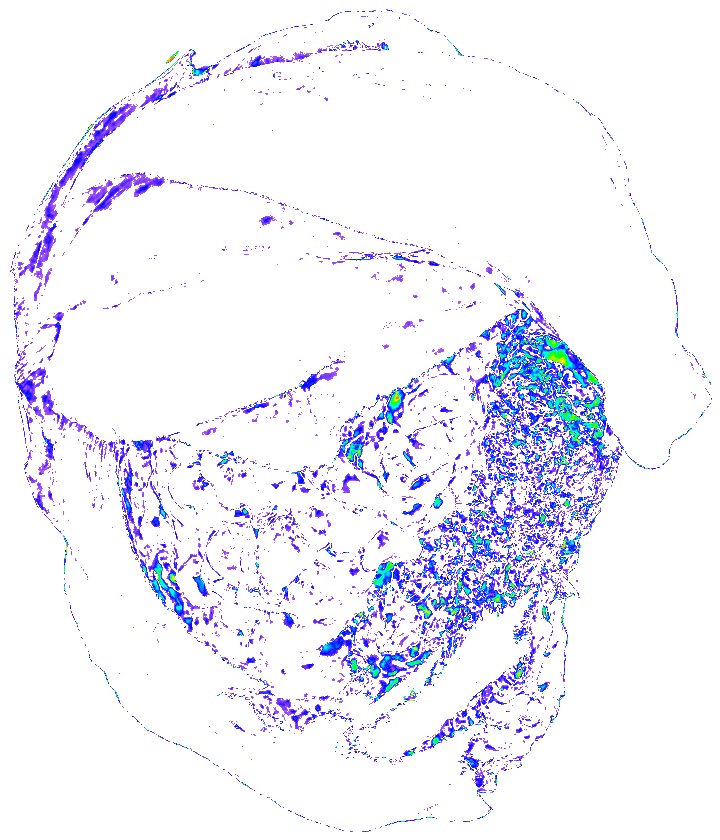}&
\\
  \includegraphics[width=0.185\linewidth]{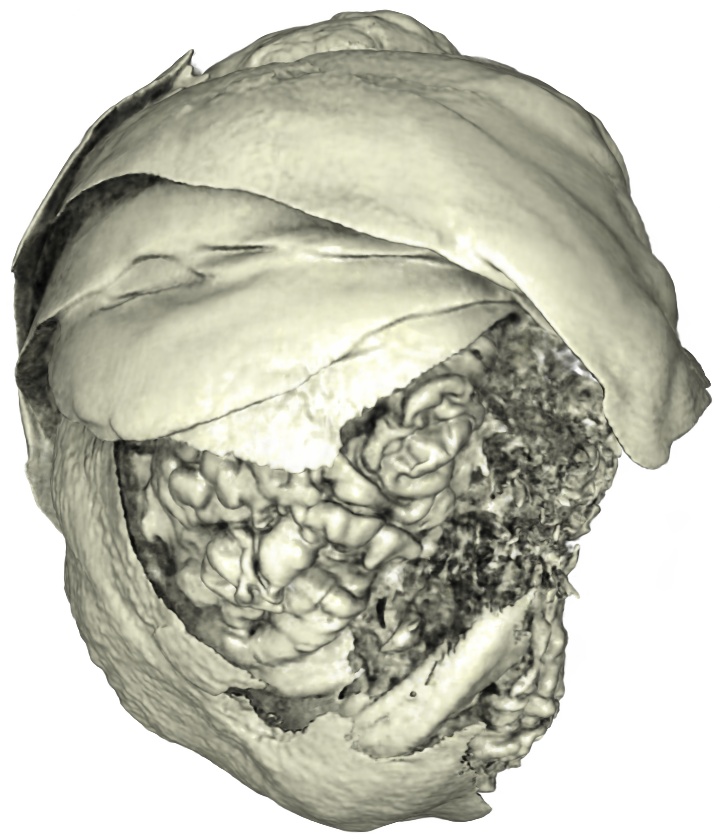}&
  \includegraphics[width=0.185\linewidth]{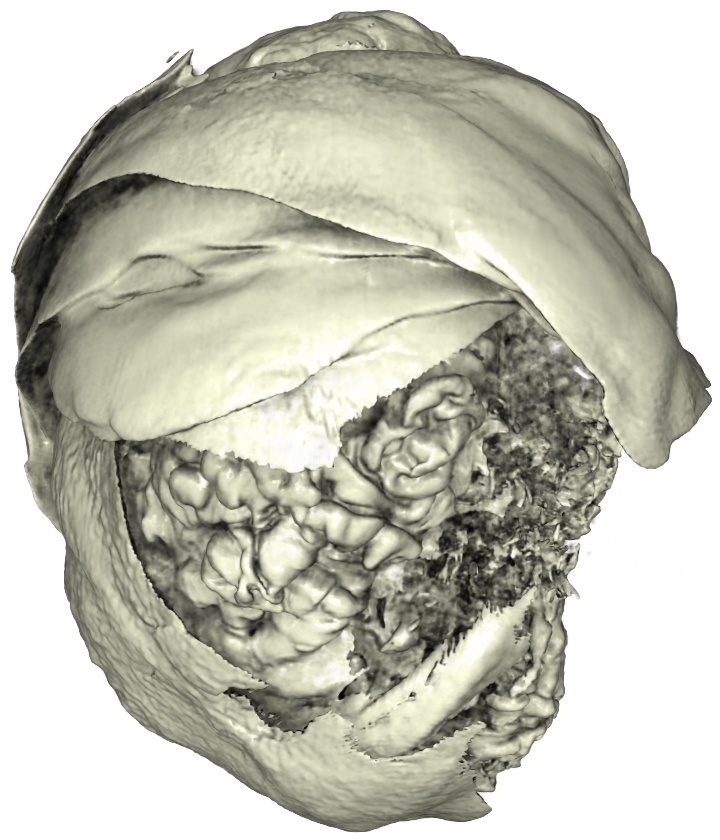}&
  \includegraphics[width=0.185\linewidth]{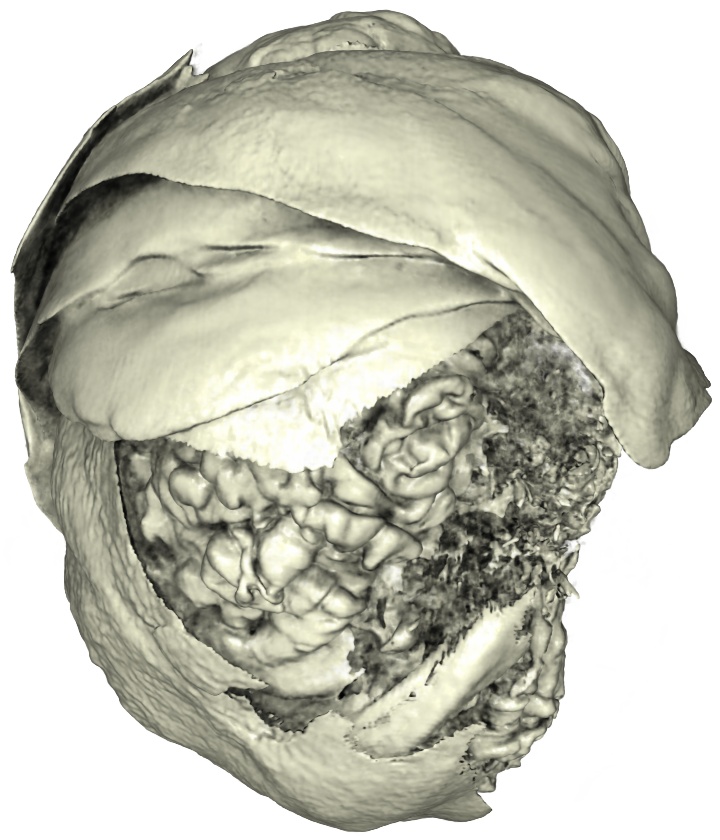}&
  \includegraphics[width=0.185\linewidth]{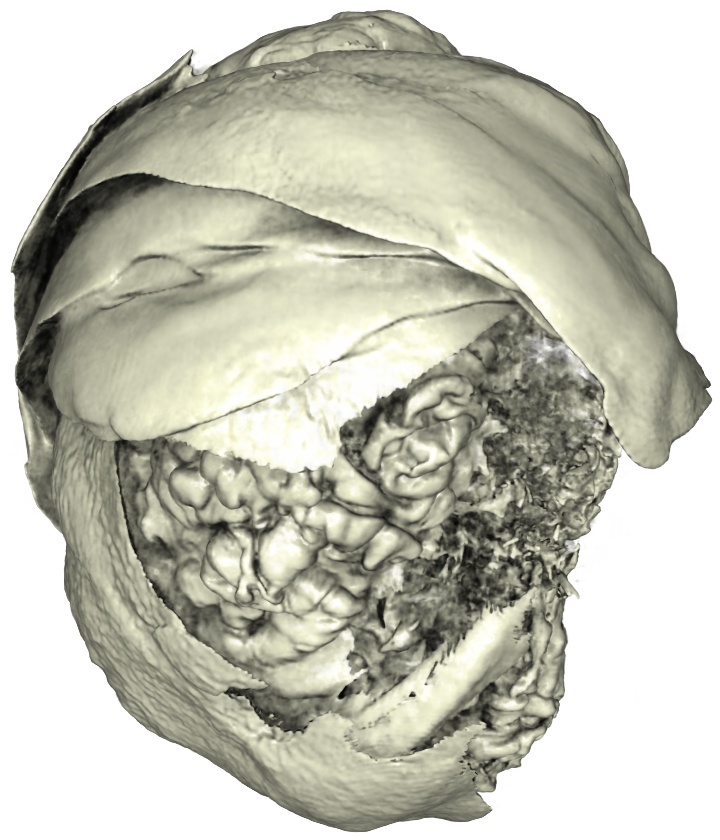}&
  \includegraphics[width=0.185\linewidth]{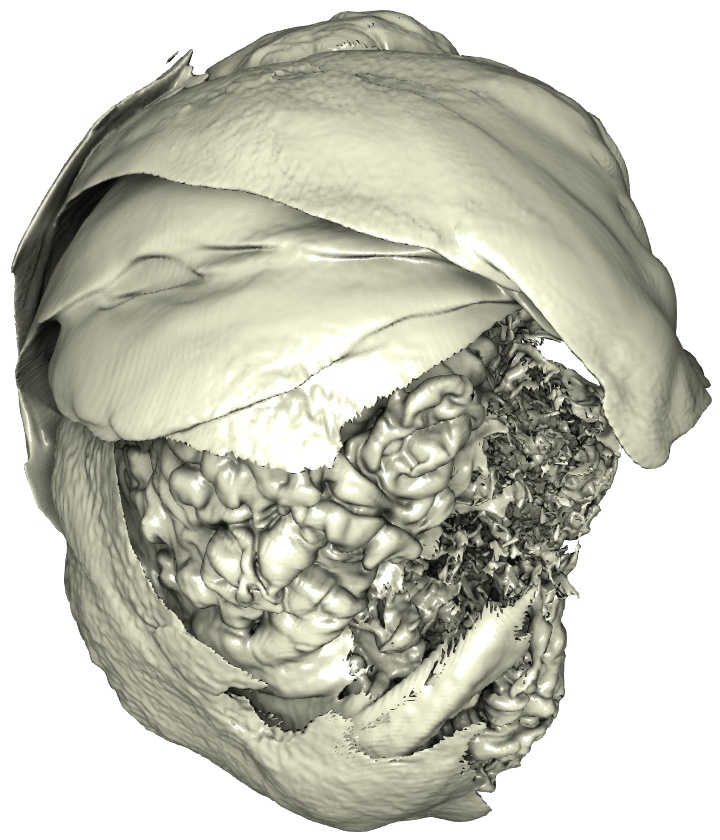}\\
  \mbox{\footnotesize (a) 8} & \mbox{\footnotesize (b) 16} & \mbox{\footnotesize (c) 24} & \mbox{\footnotesize (d) 32} & \mbox{\footnotesize (e) GT}
  \end{array}$
  \end{center}
  \vspace{-.25in} 
  \caption{\prevhot{Inferred supernova IR images using different depths of density feature.}} 
  \label{fig:comp-sigma}
  \end{figure}
 
  \begin{table}[t!]
    \caption{\prevhot{Hyperparameter study: depth of appearance feature using vortex DVR images.}}
    \vspace{-0.1in}
    \centering
    {\fontsize{6}{7.2}\selectfont
    \begin{tabular}{c|cccc}
    depth               & 32 & 40 & 48 & 56 \\ \hline
    PSNR$\uparrow$      & \prevhot{35.25} & \prevhot{36.83} & \prevhot{\prevhot{37.32}} & \prevhot{{\bf 38.04}} \\
    SSIM$\uparrow$       & \prevhot{0.993} & \prevhot{0.995} & \prevhot{0.996} & \prevhot{{\bf 0.997}} \\
    LPIPS$\downarrow$ & \prevhot{0.005} & \prevhot{0.004} & \prevhot{{\bf 0.003}} & \prevhot{{\bf 0.003}}
    \end{tabular}
    }
    \label{tab:param-app}
  \end{table}

\begin{figure}[t!]
  \begin{center}
  
  $\begin{array}{c@{\hspace{0.01in}}c@{\hspace{0.01in}}c@{\hspace{0.01in}}c@{\hspace{0.01in}}c}
  \includegraphics[width=0.185\linewidth]{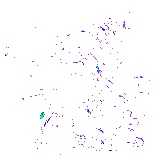}&
  \includegraphics[width=0.185\linewidth]{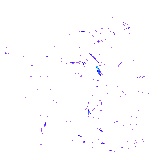}&
  \includegraphics[width=0.185\linewidth]{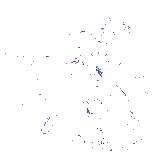}&
  \includegraphics[width=0.185\linewidth]{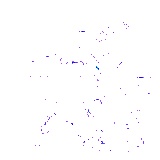}&
 \\
  \includegraphics[width=0.185\linewidth]{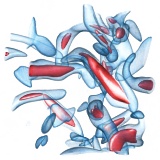}&
  \includegraphics[width=0.185\linewidth]{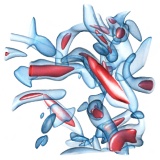}&
  \includegraphics[width=0.185\linewidth]{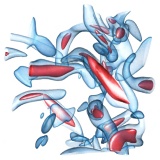}&
  \includegraphics[width=0.185\linewidth]{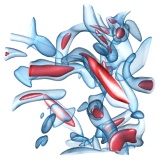}&
  \includegraphics[width=0.185\linewidth]{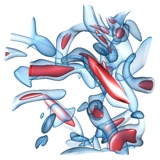}\\
  \mbox{\footnotesize (a) 32} & \mbox{\footnotesize (b) 40} & \mbox{\footnotesize (c) 48} & \mbox{\footnotesize (d) 56} & \mbox{\footnotesize (e) GT}
  \end{array}$
  \end{center}
  \vspace{-.25in} 
  \caption{\prevhot{Inferred vortex DVR images using different depths of appearance feature.}} 
  \label{fig:comp-app}
  \end{figure}

\vspace{-0.125in}
\section{Hyperparameter Study}

We investigate several key hyperparameters for ViSNeRF, including the number of training images, training iterations, feature grid voxels, and depths of density feature and appearance feature.

{\bf Number of training images.}
Reconstruction accuracy largely depends on the number of training images available to ViSNeRF. 
As shown in Table~\ref{tab:param-img}, from 12 to 42 images, the additional 30 images boost PSNR \prevhot{from 24.16 dB to 32.23 dB}.
Adding another 50 extra images improves PSNR \prevhot{only by 1.49 dB} with slightly fewer errors shown in Figure~\ref{fig:comp-img}. 
We conclude that 42 images are sufficient for ViSNeRF to build knowledge of the entire visualization for most cases.
However, more images will help refine the reconstructed visualization's details, such as highlights, shadows, and ambient occlusion.

{\bf Number of training iterations.}
As Table~\ref{tab:param-iter} and Figure~\ref{fig:comp-iter} suggest, ViSNeRF reaches satisfactory convergence after 30,000 iterations for a single static visualization.
While increasing the number of iterations may slightly improve the performance, we set the training iterations to 30,000 for all static visualizations to standardize the training process.

{\bf Number of feature grid voxels.}
As indicated by Table~\ref{tab:param-voxel} and Figure~\ref{fig:comp-voxel}, increasing the number of feature grid voxels may help restore fine details in the visualization. In contrast, a sparse feature grid could result in visualizations with reduced clarity. However, increasing the size of the feature grid will significantly increase the model size and training time. 
For example, if we increase the feature grid from $300^3$ to $400^3$ for the supernova dataset, the mode size increases from 67 MB to 119 MB, and the training time increases from \prevhot{34 to 59 minutes}. 
To strike a balance, ViSNeRF has $300^3$ voxels in the finest feature grid for all the cases.

{\bf Depths of density feature and appearance feature.}
From Tables~\ref{tab:param-sigma} and \ref{tab:param-app} and Figures~\ref{fig:comp-sigma} and \ref{fig:comp-app}, we find that the depths of density feature and appearance feature are minor factors to the reconstruction quality. 
The results suggest that the optimal depth for the density feature is 16, while that for the appearance feature is 48. 
Increasing the depth beyond these values yields little quality improvement.


\begin{table}[t!]
\caption{\prevhot{Unified vs.\ split feature grids: average PSNR (dB), SSIM, LPIPS, MS (in MB), TT (in hours), and IT (in minutes) across all 181 synthesized supernova IR views.}}
\vspace{-0.1in}
\centering
{\fontsize{6}{7.2}\selectfont
\begin{tabular}{c|ccc|ccc}
feature grid               & PSNR$\uparrow$ & SSIM$\uparrow$ & LPIPS$\downarrow$ & MS$\downarrow$ & TT$\downarrow$ & IT$\downarrow$\\ \hline
unified            & \prevhot{29.64} & \prevhot{0.943} & \prevhot{0.056} & \prevhot{\bf{67.22}} & \prevhot{0.98} & \prevhot{33.22} \\
split    & \prevhot{\bf{29.71}} & \prevhot{\bf{0.946}} & \prevhot{\bf{0.049}} & \prevhot{\bf{67.22}} & \prevhot{\bf{0.62}} & \prevhot{\bf{24.13}}
\end{tabular}
}
\label{tab:ablation-split}
\end{table}

\begin{figure}[t!]
\begin{center}
$\begin{array}{c@{\hspace{0.025in}}c@{\hspace{0.025in}}c}
\includegraphics[height=1.15in]{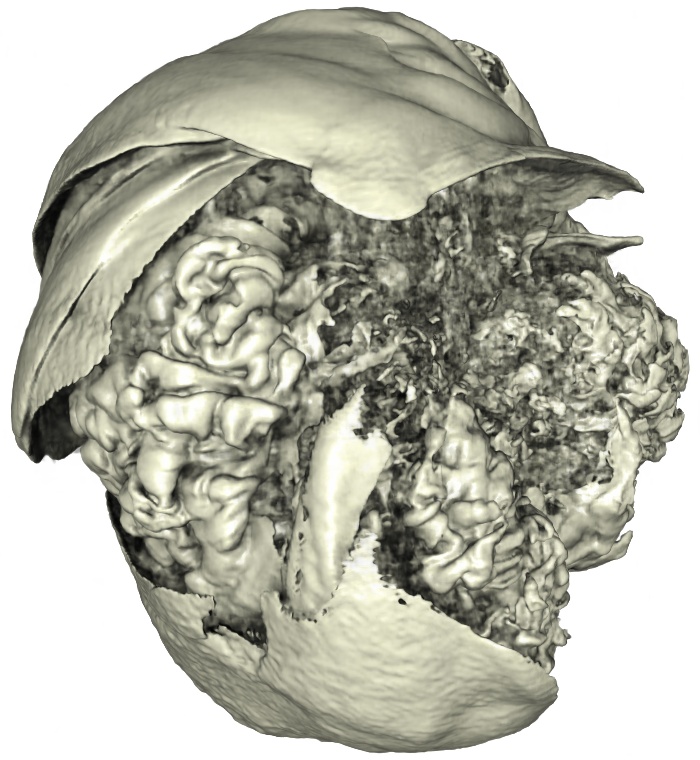}&
\includegraphics[height=1.15in]{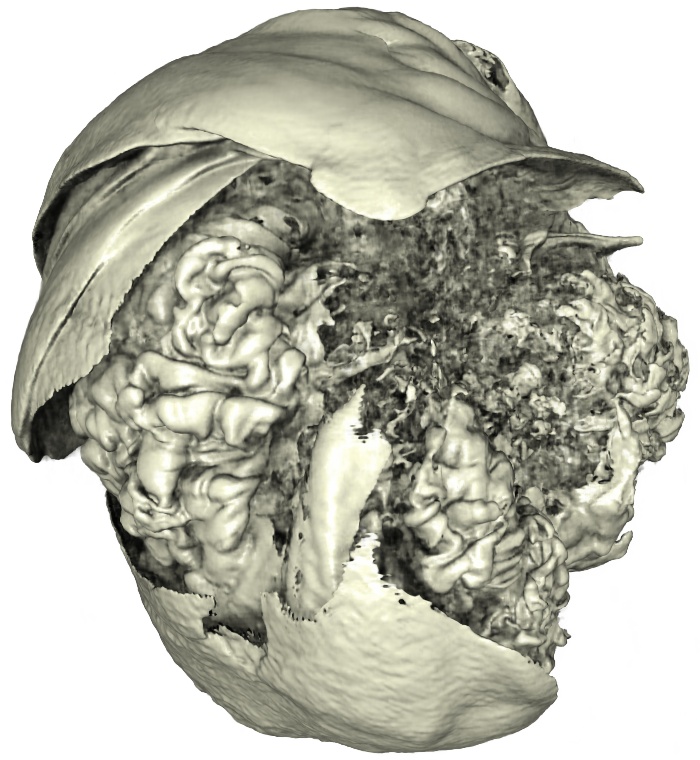}&
\includegraphics[height=1.15in]{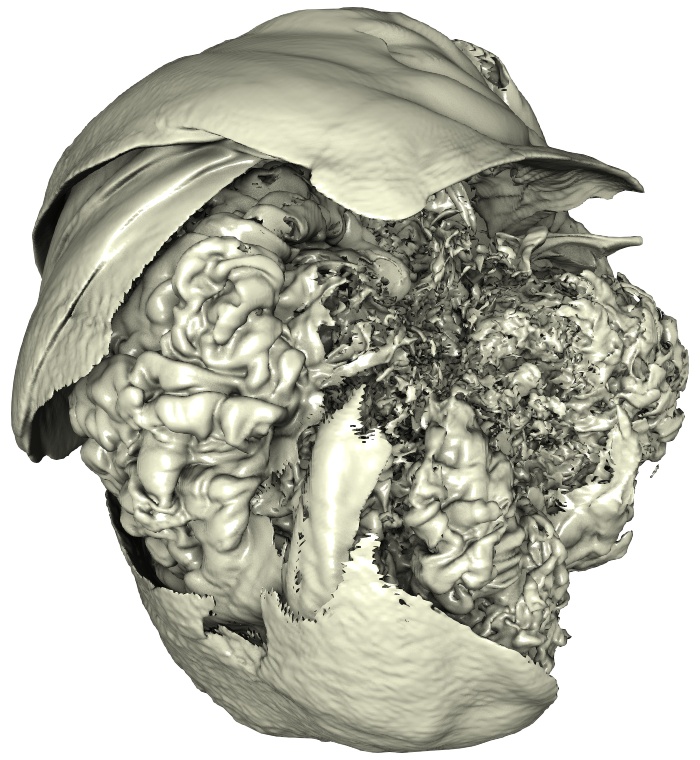}\\
\mbox{\footnotesize (a) unified} & \mbox{\footnotesize (b) split} & \mbox{\footnotesize (c) GT}
\end{array}$
\end{center}
\vspace{-.25in} 
\caption{\prevhot{Unified vs.\ split feature grids: ViSNeRF-synthesized supernova IR images.}} 
\label{fig:ablation-split}
\end{figure}

\begin{table}[t!]
\caption{\prevhot{Tensor decomposition: average PSNR (dB), SSIM, and LPIPS across all 181 synthesized supernova DVR views.}}
\vspace{-0.1in}
\centering
{\fontsize{6}{7.2}\selectfont
\begin{tabular}{c|ccc}
decomposition            & PSNR$\uparrow$ & SSIM$\uparrow$ & LPIPS$\downarrow$ \\ \hline
vectors                     & \prevhot{22.53} & \prevhot{0.717} & \prevhot{0.313}  \\
matrices                    & \prevhot{25.83} & \prevhot{0.827} & \prevhot{0.155} \\
vectors+matrices & \prevhot{{\bf 27.01}} & \prevhot{{\bf 0.859}} & \prevhot{{\bf 0.120}}
\end{tabular}
}
\label{tab:ablation-vm}
\end{table}

\begin{figure}[t!]
\begin{center}
$\begin{array}{c@{\hspace{0.05in}}c}
\includegraphics[width=0.45\linewidth]{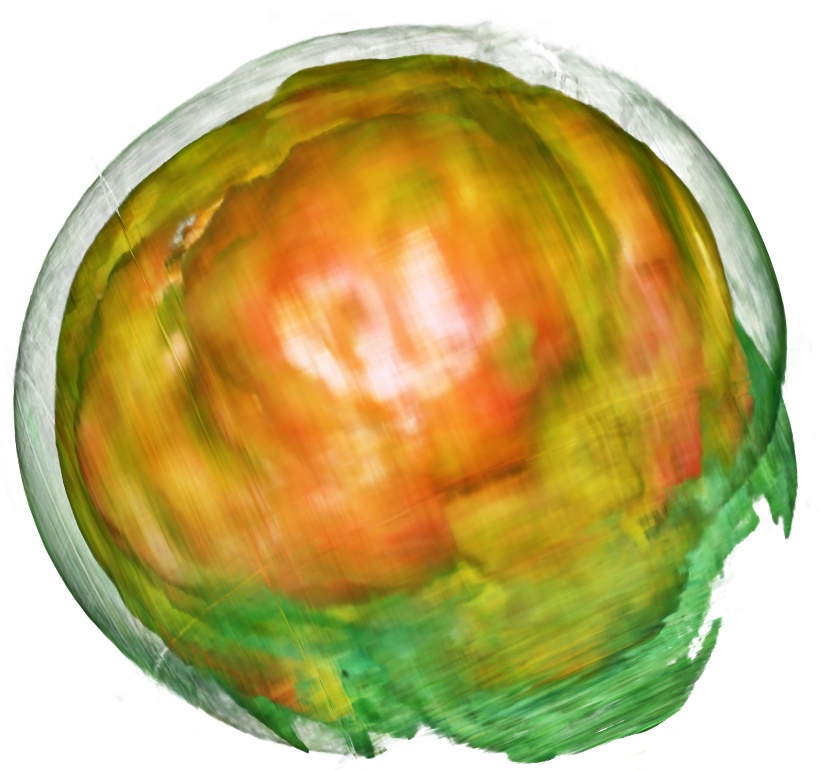}&
\includegraphics[width=0.45\linewidth]{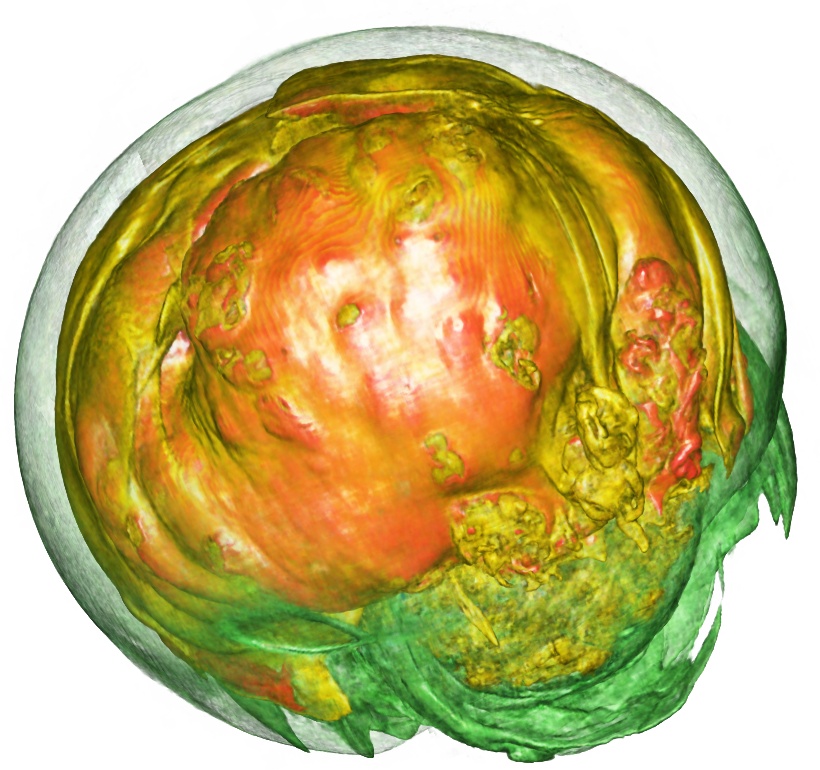}\\
\mbox{\footnotesize (a) vectors} & \mbox{\footnotesize (b) matrices} \\
\includegraphics[width=0.45\linewidth]{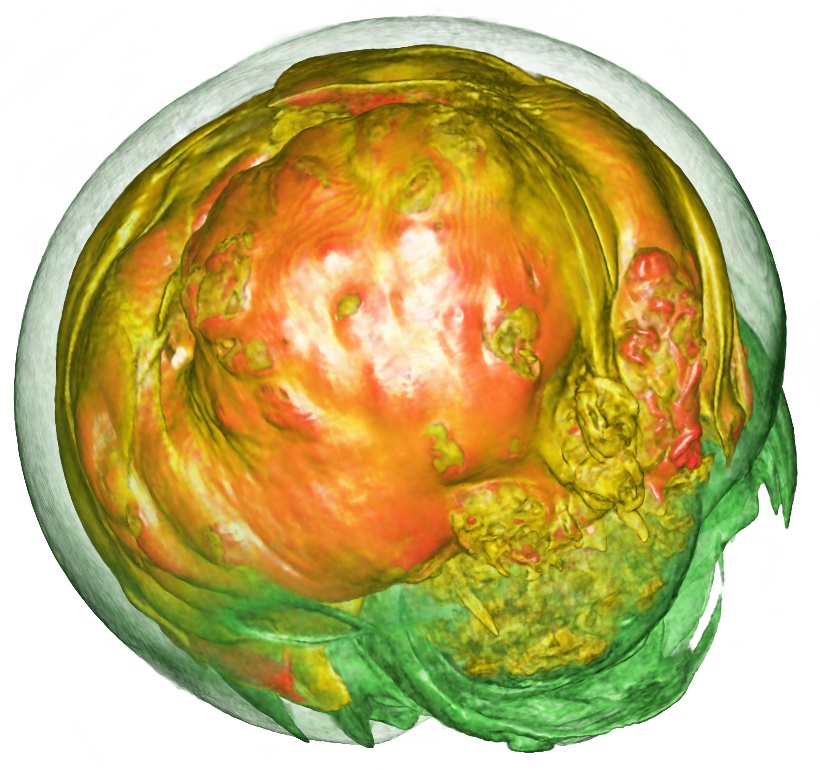}&
\includegraphics[width=0.45\linewidth]{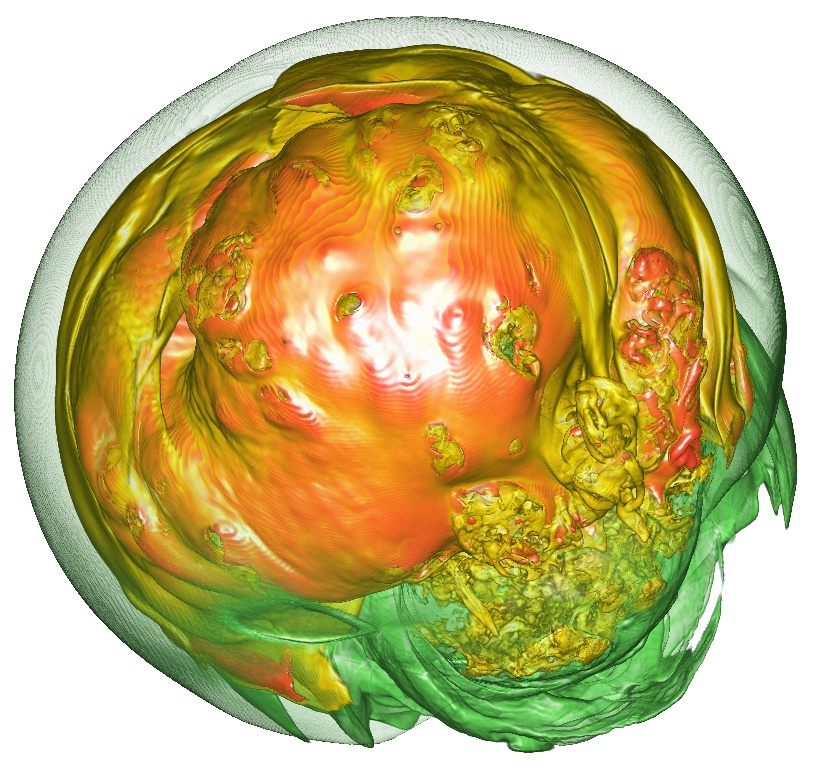}\\
\mbox{\footnotesize (c) vectors+matrices} & \mbox{\footnotesize (d) GT}
\end{array}$
\end{center}
\vspace{-.25in} 
\caption{\prevhot{Tensor decomposition: ViSNeRF-synthesized supernova DVR images.}} 
\label{fig:ablation-vm}
\end{figure}

\begin{figure}[t!]
\begin{center}
$\begin{array}{c@{\hspace{0.05in}}c}
\includegraphics[width=0.45\linewidth]{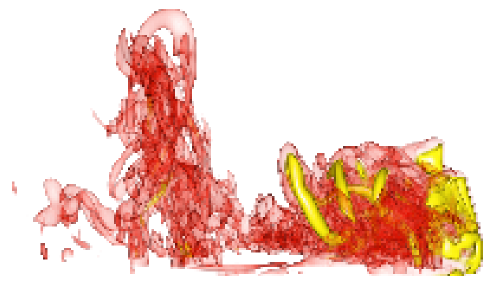}&
\includegraphics[width=0.45\linewidth]{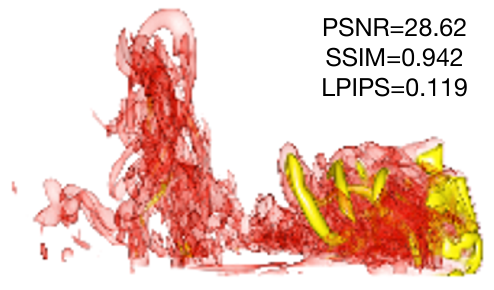}\\
\mbox{\footnotesize (a) input (256$\times$256)} & \mbox{\footnotesize (b) bicubic (1024$\times$1024)} \\
\includegraphics[width=0.45\linewidth]{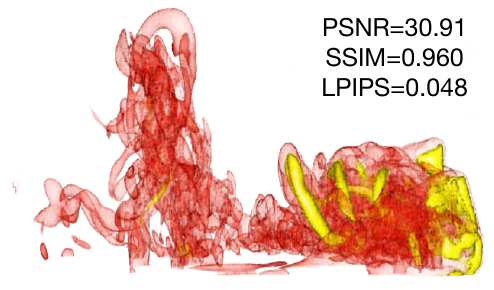}&
\includegraphics[width=0.45\linewidth]{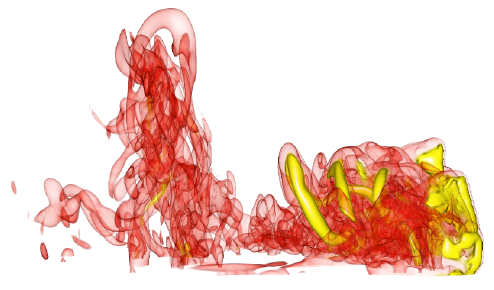}\\
\mbox{\footnotesize (c) ViSNeRF (1024$\times$1024)} & \mbox{\footnotesize (d) GT (1024$\times$1024)}
\end{array}$
\end{center}
\vspace{-.25in} 
\caption{\prevhot{Super-resolution synthesis: low-resolution input and high-resolution synthesized Tangaroa DVR images.}} 
\label{fig:comp-ss}
\end{figure}

\begin{table*}[t!]
  \caption{\prevhot{DVR or IR vs.\ ViSNeRF for dynamic scenes. All timing numbers reported are in minutes, and data/image sizes or memory consumptions reported are in GB.}}
  \vspace{-0.1in}
  \centering
  \prevhot{
  {\fontsize{6}{7.2}\selectfont
  \begin{tabular}{c|ccc|cccc}
                                &  \multicolumn{3}{c|}{DVR or IR}            &  \multicolumn{4}{c}{ViSNeRF}\\ \hline
  dataset                  & data          & processing   & RAM/VRAM    & image     & training & inference & RAM/VRAM \\ 
  (scenario)              & size           & time         & consumption &  size        & time    & time     & consumption\\ \hline
  five jets (timestep) & 1.414  &  14.06 & 2.10/1.10 &  \hot{0.137}    & \hot{85.80} & \hot{21.20} & \hot{27.94/5.83} \\
  Tangaroa (isovalue) &   0.877		  &  1.51  & 1.00/0.27 & 0.117 & 149.68 & 31.42 & 27.77/9.91 \\
  vortex (TF-2)          &   	0.008	  &  0.23 & 1.01/1.26  & 0.010 & 84.92 & 1.45 & 3.87/5.24 \\
  Nyx-DVR (simulation parameters)   &  8.188 		  & 14.90  & 2.90/1.15 & 0.119 & 99.53 & 1.62 & 7.97/5.58 \\
  \end{tabular}
  }}
  \label{tab:vs-gpu}
  \end{table*}

\vspace{-0.05in}
\section{Additional Results and Discussion}
 
{\bf Split feature grid.}
By splitting the feature grid, as shown in Table~\ref{tab:ablation-split}, \prevhot{ViSNeRF achieves reduced training and inference times while maintaining the quality of the synthesized images.}
Note that, to maintain the model size, we set the depth of features in the unified grid to 64, the sum of the depths of split feature grids.  
As shown in Figure~\ref{fig:ablation-split}, with split feature grids, \prevhot{ViSNeRF generates synthesized images of similar quality compared to those produced using a unified approach.}

{\bf Tensor decomposition.}
We investigate the effectiveness of vector-matrix decomposition by ablating ViSNeRF regarding the factorization of the 3D volume representing the radiance fields.
As shown in Figure~\ref{fig:ablation-vm}, with vector decomposition, reconstructed visualization is blurry.
Although matrix decomposition is sufficient to restore the visualization details, thinner areas, especially the shell colored in green, are not correctly synthesized.
By employing vector-matrix decomposition, as shown in Table~\ref{tab:ablation-vm}, the performance of ViSNeRF is further improved.
  
{\bf Super-resolution synthesis.}
A by-product of VisReRF is the ability to synthesize super-resolution rendering images from low-resolution training images. 
With an optimized NeRF, visualizations of any resolutions can be rendered by the volume renderer of ViSNeRF.
As shown in Figure~\ref{fig:comp-ss}, leveraging the knowledge acquired from training 256$\times$256 resolution images, ViSNeRF demonstrates its capability to restore images at a higher resolution of 1024$\times$1024.
Most of the details in the high-resolution images can be recovered by ViSNeRF despite noticeable artifacts due to a limited number of 42 training images.

\prevhot{{\bf Traditional rendering vs.\ ViSNeRF.} 
In Table~\ref{tab:vs-gpu}, we compare traditional DVR and IR with ViSNeRF regarding memory consumption and rendering time using the same GPU.
Our results indicate that traditional DVR and IR necessitate access to the original volume data or isosurfaces, which can be sizeable, often reaching multiple gigabytes in the case of time-varying or ensemble datasets. An exception is observed in the vortex dataset, where only a small volume of 8 MB is necessary due to the interpolation across transfer functions. 
In contrast, ViSNeRF operates primarily on rendered images (PNG format), which typically occupy less storage than volume data or isosurfaces. To illustrate, around 100 MB is sufficient for processing 546 images of Tangaroa and 1890 images of Nyx, and notably, \hot{only 10 MB is needed for 462 images of vortex}. 
This storage efficiency is significant, especially considering ViSNeRF's ability to produce high-quality visualization synthesis from interpolated parameters without the actual data. 

Regarding generation time to produce all rendering or inference images, although traditional DVR and IR include the file reading time, they are still more efficient than ViSNeRF, which requires training.
\hot{However, ViSNeRF exhibits a faster generation time for Nyx datasets of low image resolution (256$\times$256). }
Regarding memory usage, ViSNeRF inherently demands more RAM and VRAM, a common trait of machine learning methods. Before training commences, ViSNeRF must load training and inference images, converting them into a data structure of rays and associated attributes. Each ray consists of ray origin and direction (six floats), RGB values (three floats), and parameters (one float per parameter), totaling approximately ten floats per pixel. Given that each 256$\times$256 image comprises 65,536 pixels, the memory requirement for one image can be calculated as $65,536\times10\times4=2,621,440$ bytes, equating to roughly 2.5 MB per image.
For a dataset like \hot{vortex}, which includes 462 training images and 181 inference images of 256$\times$256 resolution, the total memory footprint for storing these images as rays would be approximately 1.57 GB.

Considering the Tangaroa dataset, which utilizes images of resolution 1024$\times$1024, the initial memory requirement for storing image data in ViSNeRF is substantial. The calculation for this dataset would be $1024\times1024\times10\times4\times(546+181)=30,492,590,080$ bytes, amounting to approximately 29 GB.
However, ViSNeRF employs an optimization by filtering out rays that do not intersect with the bounding box, effectively reducing the long-term runtime RAM usage below 29 GB for the Tangaroa dataset.
Most of the remaining memory consumption, particularly VRAM, is attributed to the model training process. This includes the storage of the density feature grid, color feature grid, parameter feature grid, two small MLPs, and their gradient graphs, which are essential for model optimization. 
Given these requirements, we recommend a minimum of 32 GB of RAM and 10 GB of VRAM for running ViSNeRF efficiently. A configuration with at least 64 GB of RAM and 16 GB of VRAM is ideal for optimal performance, especially with high-resolution datasets.}

\vspace{-0.05in}
\bibliographystyle{abbrv-doi}

\bibliography{template}
\end{document}